\documentclass[11pt]{article}
\usepackage{amsmath,amssymb,amsfonts,graphicx}

\usepackage{hyperref}
\usepackage[nosort]{cite}
\usepackage{verbatim}
\usepackage{multicol,color,longtable}

\DeclareFontFamily{U}{matha}{\hyphenchar\font45}
\DeclareFontShape{U}{matha}{m}{n}{
      <5> <6> <7> <8> <9> <10> gen * matha
      <10.95> matha10 <12> <14.4> <17.28> <20.74> <24.88> matha12
      }{}
\DeclareSymbolFont{matha}{U}{matha}{m}{n}

\DeclareMathSymbol{\oleft}{2}{matha}{"68}
\DeclareMathSymbol{\oright}{2}{matha}{"69}

\usepackage[small,bf,hang]{caption}
\usepackage{slashed}
\usepackage{latexsym,epsfig}

\definecolor{darkred}{rgb}{0.65,0.15,0}
\definecolor{darkgreen}{rgb}{.05,.5,.05}
\hypersetup{pdfborder={0 0 0},colorlinks=true,urlcolor=darkred,citecolor=blue,linkcolor=darkred,linktocpage=true}

\usepackage{mathrsfs}
\usepackage{dsfont}
\usepackage[Symbol]{upgreek}

\makeatletter
\g@addto@macro\bfseries{\boldmath}
\makeatother

\newcommand{\eprint}[1]{{\href{http://arxiv.org/abs/#1}{\texttt{[#1}]}}}
\newcommand{\eprintN}[1]{{\href{http://arxiv.org/abs/#1}{\texttt{#1 [hep-th]}}}}
\newcommand{\doi}[2]{{\hypersetup{urlcolor=blue}\href{http://dx.doi.org/#2}{#1}\hypersetup{urlcolor=darkred}}}

\newcommand{\nn}{\nonumber}

\newcommand{\ints}{\mathds{Z}}
\newcommand{\reals}{\mathds{R}}

\newcommand{\cM}{{\mathcal{M}}}
\newcommand{\cV}{{\mathcal{V}}}
\newcommand{\cJ}{{\mathcal{J}}}
\newcommand{\cS}{{\mathcal{S}}}

\newcommand{\be}{\begin{align}}
\newcommand{\ee}{\end{align}}

\newcommand{\dL}{{\mathsf{d}}}
\newcommand{\dK}{{\mathsf{K}}}

\newcommand{\lb}{\left[}
\newcommand{\rb}{\right]}
\newcommand{\id}{\mathds{1}}

\newcommand{\mf}[1]{{\mathfrak{#1}}}

\makeatletter

\@addtoreset{equation}{section}
\makeatother

\newcommand{\Tr}{{\mathrm{Tr}}}

\newcommand{\ket}[1]{ |#1 \rangle}
\newcommand{\bra}[1]{ \langle #1 |}
\newcommand{\ad}{\mathrm{ad}}
\newcommand{\pot}{V}

\def\be{\begin{equation}}\def\ee{\end{equation}}
\def\bea{\begin{eqnarray}}\def\eea{\end{eqnarray}}
\newcommand{\CR}{\nonumber \\*}
\def\cL{{\cal L}}
\def\cP{{\cal P}}
\def\uA{{\underline{A}}}
\def\uB{{\underline{B}}}
\def\uC{{\underline{C}}}
\def\uD{{\underline{D}}}
\def\uE{{\underline{E}}}
\def\uF{{\underline{F}}}
\def\uG{{\underline{G}}}
\def\uH{{\underline{H}}}
\def\tr{{\rm tr}}

\begin{document}

{\flushright {CPHT-RR104.11201}\\
{QMUL-PH-18-28}\\[5mm]}

\begin{center}
  {\LARGE \sc  $E_9$ exceptional field theory\\[2mm] I. The potential}
    \\[15mm]

{\large
Guillaume Bossard${}^{1}$, Franz Ciceri${}^2$, Gianluca Inverso${}^3$,\\[1ex] Axel Kleinschmidt${}^{2,4}$ and  Henning Samtleben${}^5$}

\vspace{10mm}
${}^1${\it Centre de Physique Th\'eorique, Ecole Polytechnique, CNRS\\
Universit\'e Paris-Saclay FR-91128 Palaiseau cedex, France}
\vskip 1 em
${}^2${\it Max-Planck-Institut f\"{u}r Gravitationsphysik (Albert-Einstein-Institut)\\
Am M\"{u}hlenberg 1, DE-14476 Potsdam, Germany}
\vskip 1 em
${}^3${\it Centre for Research in String Theory, School of Physics and Astronomy, \\
Queen Mary University of London, 327 Mile End Road, London, E1 4NS, United Kingdom}
\vskip 1 em
${}^4${\it International Solvay Institutes\\
ULB-Campus Plaine CP231, BE-1050 Brussels, Belgium}
\vskip 1 em
${}^5${\it Univ Lyon, Ens de Lyon, Univ Claude Bernard, CNRS,\\
Laboratoire de Physique, FR-69342 Lyon, France}

\end{center}

\vspace{5mm}

\begin{center} 
\hrule

\vspace{10mm}

\begin{tabular}{p{12cm}}
{\small
We construct the scalar potential for the exceptional field theory based on the affine symmetry group $E_9$. The fields appearing in this potential live formally on an infinite-dimensional extended spacetime and transform under $E_9$ generalised diffeomorphisms. In addition to the scalar fields expected from $D=2$ maximal supergravity, the invariance of the potential requires the introduction of new constrained scalar fields. Other essential ingredients in the construction include the Virasoro algebra and indecomposable representations of $E_9$.
Upon solving the section constraint, the potential reproduces the dynamics of either eleven-dimensional or type IIB supergravity in the presence of two isometries.
}
\end{tabular}
\vspace{7mm}
\hrule
\end{center}

\thispagestyle{empty}

\newpage
\setcounter{page}{1}


\tableofcontents

\section{Introduction}


Exceptional geometry is a way of unifying the local symmetries of supergravity theories by combining geometric diffeomorphisms with matter gauge transformations into a single so-called generalised Lie derivative~\cite{Hull:2007zu,Pacheco:2008ps,Hillmann:2009pp,Berman:2010is,Berman:2011pe,Coimbra:2011ky,Berman:2012vc,Coimbra:2012af,Park:2013gaj,Cederwall:2013naa,Cederwall:2013oaa,Aldazabal:2013mya,Bossard:2017aae,Cederwall:2017fjm}. This generalised Lie derivative generates generalised diffeomorphisms acting on the fields of the theory and requires the introduction of an extended space beyond the usual space-time geometry of gravity. The generalised Lie derivative forms a closed gauge algebra only when the so-called section condition is imposed on the fields, restricting their dependence on the extended space. Upon solving the section constraint explicitly one recovers the standard supergravity theories. However, exceptional geometry also offers the possibility of describing more complicated (local or global) situations that have been named non-geometric backgrounds.

The symmetry groups of maximal supergravities in $D=11-n$ dimensions belong to the (split real) exceptional series $E_n$~\cite{Cremmer:1978ds,Julia:1982gx,Nicolai:1987kz,Cremmer:1997ct,Cremmer:1998px}. For each exceptional symmetry group $E_n$ one can construct an extended space that is described locally by a set of coordinates $Y^M$ where $M$ labels a representation of $E_n$~\cite{Berman:2012vc}. Adjoining to these coordinates the `external' $D$-dimensional space with coordinates $x^\mu$ and an appropriate notion of external diffeomorphisms one obtains a total space with coordinates $(x^\mu, Y^M)$. Taking the fields from $D$-dimensional maximal supergravity, as prescribed by the tensor hierarchy~\cite{deWit:2005hv,deWit:2008ta}, one may try to construct an action invariant under generalised and external diffeomorphisms. 
As it turns out, closure of the gauge algebra of $p$-forms and invariance of the action require the introduction of additional $p$-forms of rank $p\ge D-2$ beyond those of $D$-dimensional supergravity. These extra p-forms are covariantly constrained in the sense that they obey algebraic constraints analogous to those satisfied by the internal partial derivatives $\partial_M = \frac{\partial}{\partial Y^M}$ by virtue of the section constraint. Moreover, these fields do not constitute additional
degrees of freedom, but are related by first order equations to the propagating fields of the theory. 
Combining all these ingredients leads to a unique theory called $E_n$ exceptional field theory
and that has been explicitly constructed for $E_n$ with $n\leq 8$ \cite{Hohm:2013vpa,Hohm:2013uia,Hohm:2014fxa}.
Imposing a solution to the section condition relates exceptional field theory to maximal supergravity in eleven space-time dimensions, to type IIB supergravity or to their dimensional reductions depending on the choice of solution to the section condition. 

The invariant `actions' of exceptional field theories combine various terms. They carry an Einstein--Hilbert-type term, kinetic terms for the various matter fields including a non-linear sigma model for the scalars, a topological term for the $p$-forms, and a `potential' term for the scalar fields. The scalar fields belong to the coset space $E_n/K(E_n)$, where $K(E_n)$ denotes the maximal compact subgroup of $E_n$, and may be parameterised by a symmetric matrix $\cM_{MN}$ which determines the internal generalised metric on the extended space.
The `potential' $V(\cM)$ is bilinear in the internal derivatives $\partial_M$ with respect to the extended coordinates but does not carry derivatives $\partial_\mu$ with respect to the `external' coordinates. Under generalised Scherk--Schwarz reduction~\cite{Berman:2012uy,Musaev:2013rq,Godazgar:2013oba,Lee:2014mla,Hohm:2014qga} in the extended space, it is $V(\cM)$ that generates the scalar potential term of gauged supergravity. From the point of view of exceptional geometry, $V(\cM)$ is invariant under generalised diffeomorphisms up to a total derivative and plays to some extent the role of the curvature scalar on the extended space. Its structure has been worked out up to and including $E_8$ and it is tied to the remaining terms in the exceptional field theory Lagrangian by external diffeomorphisms.

The first infinite-dimensional group in the $E_n$ series is the affine symmetry group $E_9$ that is known to be a rigid symmetry of $D=2$ ungauged maximal supergravity \cite{Julia:1981wc}. The associated exceptional field theory has not yet been constructed and the aim of the present paper is to begin filling this gap. The $E_9$ generalised Lie derivative was recently introduced in~\cite{Bossard:2017aae} and it acts on fields that depend on infinitely many coordinates $Y^M$. The latter transform under $E_9$ in the basic lowest weight representation.
Closure of the algebra requires a section constraint of the generic form
\begin{align}
\label{eq:SCintro}
Y^{MN}{}_{PQ} \,\partial_M \otimes \partial_N = 0
\end{align}
with the internal derivatives acting on any pair of fields. Here, $Y^{MN}{}_{PQ}$ is a specific 
$E_9$ invariant tensor that can be expressed most easily in terms of quadratic combinations of the $E_9$ generators as we shall review in Section~\ref{sec:GL}.

The purpose of the present article is to construct the potential $V$ for $E_9$ exceptional field theory. 
In $D=2$, the scalar fields appearing in maximal supergravity parameterise the coset space
\begin{equation}
\frac{ \hat E_8 \rtimes\left( \mathds{R}^+_{\sf d}\ltimes\mathds{R}_{L_{-1}} \right) }
 {K(E_9)}\,,
 \label{intro:coset}
\end{equation}
where $\hat{E}_8$ denotes the centrally extended loop group over $E_8$. 
Its quotient by the maximal `compact' subgroup%
\footnote{More precisely, $K(E_9)$ is the maximal $E_9$ subgroup acting unitarily  on the representation of the extended space coordinates. We shall henceforth refer to it as the maximal unitary subgroup.}
 $K(E_9)=K(\hat{E}_8)$ contains the infinite tower of dual scalar fields from $D=2$ maximal supergravity together with the conformal factor of the external metric. The factor $\mathds{R}^+_{\sf d}\ltimes\mathds{R}_{L_{-1}}$ is parameterised by two more scalar fields $\{\rho,\tilde\rho\}$ which, in $D=2$ ungauged supergravity, are related by a free duality equation. The generator $\dL$ associated with the dilaton $\rho$ enhances $\hat{E}_8$ to 
 \begin{equation}
 E_9 = \hat E_8 \rtimes\mathds R^+_\dL\,,
 \end{equation}
 while the (Virasoro) generator $L_{-1}$ associated with the axion $\tilde\rho$ acts as a translation generator on the loop parameter of the loop group $\hat{E}_8$. The generator $L_{-1}$ also appears in the $E_9$ generalised Lie derivative~\cite{Bossard:2017aae}.

A novel feature of $E_9$ exceptional field theory, compared to $E_n$ for $n\leq 8$, is that the scalar sector~\eqref{intro:coset} of maximal supergravity is not sufficient to define the theory. This can be seen by 
extrapolating the generic field content of exceptional field theories down to two external dimensions and noting that the covariantly constrained additional $p$-forms mentioned above already start from $p=0$ forms for $D=2$ external dimensions. Therefore one has to enhance the scalar sector \eqref{intro:coset} 
by additional constrained fields $\chi_M$ transforming under $E_9$ and generalised diffeomorphisms. 
Specifically, these additional scalar fields obey the constraints
\begin{align}
Y^{MN}{}_{PQ} \,\chi_M \otimes \partial_N  = Y^{MN}{}_{PQ} \,\chi_M \otimes \chi_N = 0
\;,
\end{align}
with the tensor $Y^{MN}{}_{PQ}$ from \eqref{eq:SCintro},
and similar relations with other constrained objects. 
For the M-theory solution of the section constraint which makes 
all fields independent of all but nine of the extended directions $Y^M$,
the field $\chi_M$ thus has at most nine truly independent components.

As is already the case for $E_8$ \cite{Hohm:2014fxa}, the closure of the algebra of generalised diffeomorphisms requires to not only consider the generalised Lie derivative along a gauge parameter  $\Lambda^M$ in the basic representation, but to also include an additional set of gauge transformations with a covariantly constrained parameter $\Sigma^M{}_N$~\cite{Bossard:2017aae}. The Lie derivative along the generalised vector $\Lambda^M$ only involves $E_9$, while the constrained parameter also induces a gauging of the $\reals_{L_{-1}}$ symmetry when $\Sigma^M{}_M\neq0$.

Our construction of the $E_9$ potential is guided by the following requirements. Firstly, 
the different terms must transform as scalar densities under rigid $E_9$  and rigid $\mathds{R}_{L_{-1}}$ transformations. Secondly, the combination of these terms must be such that the 
potential is invariant under generalised diffeomorphisms, up to a total derivative. Finally, the potential must
reproduce parts of the known Lagrangian of $E_8$ exceptional field theory upon truncation to a suitable subset of coordinates.  
These requirements allow us to uniquely pinpoint the $E_9$ potential. 

The $\mathds{R}_{L_{-1}}$ gauge transformations mentioned above can be gauge-fixed by setting $\tilde\rho=0$ without breaking the rest of the generalised diffeomorphisms (\textit{i.e.} those satisfying $\Sigma^M{}_M= 0$).
This choice also breaks the rigid $\mathds{R}_{L_{-1}}$ invariance. For simplicity, we now present the resulting potential for $\tilde\rho=0$ such that $\cM_{MN}$ is valued in $E_9$,
\begin{align}
\label{pot_intro}
V (\cM,\chi) = \frac14 \rho^{-1}\,\eta^{\alpha\beta}\,{\cal M}^{MN}\,{\cal J}_{M,\alpha}\,{\cal J}_{N,\beta} - \frac12 \rho^{-1}\, {\cal M}^{PQ}\,({\cal J}_{M}){}^N{}_P \,({\cal J}_{N})^M{}_Q\\
+\frac12 \rho\,{\cal M}^{PQ}\,({\cal J}^-_{M}){}^N{}_P \,({\cal J}^-_{N})^M{}_Q + \rho^{-2}\,\partial_M\rho\,\partial_N \cM^{MN}   \,.\nn
\end{align}
The scalar current ${\cal J}_M$ is defined as
\begin{align}
\cM^{PS} \partial_M \cM_{SQ} = \left(\cJ_M\right){}^P{}_Q = \cJ_{M,\alpha} \, (T^\alpha)^P{}_Q\,,
\end{align}
in terms of the generators $T^\alpha$ of $\mathfrak{e}_9$ written in the  representation $R(\Lambda_0)_0$ that we define below. The (inverse) invariant bilinear form on $\mf e_9$ is denoted $\eta^{\alpha\beta}$, and the \textit{shifted} current ${\cal J}_M^-$ is defined as
\begin{align}
{\cal J}^-_M = {\cal S}_{-1}\left({\cal J}_M\right) + \chi_M\,\dK
\;,
\label{intro:shift}
\end{align}
where the operator ${\cal S}_{-1}$ defined in (\ref{eq:Sm}) shifts the mode number of the $\hat{E}_8$ loop generators and the Virasoro generators. 
The additional scalar field $\chi_M$ appears as the component along the $\mf e_9$ central element $\dK$, and is necessary to ensure covariance of the shifted current under rigid $E_9$ transformations.
The first two terms in~\eqref{pot_intro} are the $E_9$ version of the generic terms that appear in the potential of all exceptional field theories~\cite{Cederwall:2017fjm}.
The third term contains the new constrained scalar field $\chi_M$ via \eqref{intro:shift}
and generalises a structure which has so far only occurred in
the potential of the $E_8$ exceptional field theory \cite{Hohm:2014fxa}.
Each term in (\ref{pot_intro}) is separately invariant under rigid $\hat E_8$ and scales with the same weight under rigid $\mathds R^+_\dL$, which is analogous to the homogeneous scaling of higher-dimensional exceptional field theory Lagrangians under the trombone symmetry.
The relative coefficients of the various terms in (\ref{pot_intro}) are fixed by generalised diffeomorphism invariance.

In the bulk of the paper we shall also derive the potential away from $\tilde\rho=0$ and in this way restore the full invariance under generalised diffeomorphisms and rigid $\mathds{R}_{L_{-1}}$ transformations. All the terms in~\eqref{pot_intro} then become functions of $\tilde\rho$ and its derivatives, such that they are invariant under rigid $\hat E_8\rtimes (\mathds R^+_\dL\ltimes \reals_{L_{-1}})$ transformations up to homogeneous scaling under $\mathds R^+_\dL$. Moreover, the same relative combination as in~\eqref{pot_intro} becomes invariant under all generalised diffeomorphisms as we shall demonstrate in detail.

The construction of $E_9$ exceptional field theory is interesting for several reasons. To begin with,
it yields the first example of an exceptional field theory based on an infinite-dimensional duality group
with fields and coordinates transforming in infinite-dimensional representations.
As an immediate application, the $E_9$ potential can provide a prediction for the yet elusive scalar potential of gauged maximal $D=2$ supergravity~\cite{Samtleben:2007an} by performing a generalised Scherk--Schwarz reduction. The $D=2$ potential seems at present inaccessible by standard supersymmetry considerations because of the intricacies of $K(E_9)$ representation theory. As two-dimensional gauged supergravities generically involve a gauging of the $\reals_{L_{-1}}$ symmetry~\cite{Samtleben:2007an}, it is crucial to construct the $E_9$ potential at $\tilde\rho\neq 0$, which is invariant under all generalised diffeomorphisms. Another possible application is the study of non-geometric backgrounds~\cite{Hull:2007zu,Pacheco:2008ps,Coimbra:2011ky,Coimbra:2012af}. Moreover, $D=2$ supergravity is the arena for exotic branes of co-dimension two (or lower)~\cite{deBoer:2010ud,deBoer:2012ma} for which $E_9$ exceptional field theory may provide the appropriate framework~\cite{Bakhmatov:2017les,Berman:2018okd}.

Our construction does not depend on the details of the group $E_8$ and in fact the expressions we give will be valid for any simple group $G$ and its affine extension $\hat{G}$. This provides the potential for extended field theories with coordinates in the basic representation of $\hat{G}$ that are invariant under rigid $\hat{G}\rtimes (\reals_{\dL}^+ \ltimes \reals_{L_{-1}})$ and $\hat{G}$ generalised diffeomorphisms. 

The rest of this paper is organised as follows. In Section~\ref{sec:not} we review some basic facts and properties of the exceptional algebra $\mathfrak{e}_9$ and its representations. Section~\ref{sec:build} introduces the building blocks for $E_9$ exceptional field theory by reviewing the field content of $D=2$ maximal supergravity and the $E_9$ generalised 
Lie derivative from~\cite{Bossard:2017aae}. We furthermore introduce the covariantly constrained scalar fields $\chi_M$.
Section~\ref{sec:pot} presents the main result of this paper, the construction of the $E_9$ potential $V(\cM,\chi)$
invariant under generalised diffeomorphisms.
Finally, in Section~\ref{sec:redE8} we consider the branching under $E_8$ and show that the $E_9$ potential  reproduces all the terms of the $E_8$ exceptional field theory that do not depend on the two-dimensional external derivatives. This shows that after solving the section condition, the $E_9$ potential $V(\cM,\chi)$ reproduces $D=11$ and type IIB supergravity for field configurations constant along the two-dimensional external spacetime.
We finish with conclusions in Section~\ref{sec:conc} and two appendices that contain some technical details and identities.

\section{$E_9$ basic representation and Virasoro algebra}
\label{sec:not}

In this section, we introduce some notions we require from $E_9$ along with our notation to be used throughout the paper.

\subsection{$E_9$ preliminaries}
\label{sec:e9prelim}

At the Lie algebra level, $E_9$ is an infinite-dimensional Kac--Moody algebra that we call $\mf{e}_9$. The core part of $\mf{e}_9$ is the centrally extended loop algebra $\hat{\mathfrak{e}}_8$ over the Lie algebra $\mf{e}_8$ and we only work with the split real forms. The Lie algebra $\mf{e}_8$ has dimension $248$ and we denote its generators by $T^A$ with $A=1,\ldots,248$ and $\mf{e}_8$ structure constants
\begin{align}
\mf{e}_8\,:\quad\quad \lb T^A, T^B \rb = f^{AB}{}_C T^C\,.
\end{align}
The $\mf{e}_8$-invariant and non-degenerate metric is $\eta^{AB}$ with inverse $\eta_{AB}$. The loop generators of $\hat{\mf{e}}_8$ are denoted by $T^A_m$ with mode number $m\in \ints$ and commutation relations
\begin{align}
\lb T^A_m , T^B_n \rb = f^{AB}{}_C T^C_{m+n} + m \, \eta^{AB} \delta_{m,-n} \dK\,,
\end{align}
where $\dK$ denotes the central extension of the loop algebra with $\lb \dK, T^A_m\rb =0$. In order to define the affine algebra $\mf{e}_9$ one also has to adjoin the derivation operator $\dL$ that satisfies
\begin{align}
\lb \dL , T^A_m \rb = -m\, T^A_m\,,\quad 
\lb \dL, \dK \rb =0\,.
\end{align}
As a vector space, $\mf{e}_9= \hat{\mf{e}}_8 \oplus \langle\dL\rangle$. 
There is an isomorphic copy of $\mf{e}_8$ embedded in $\mf{e}_9$ by considering the generators $T^A_0$ at mode number zero. 
In terms of a loop parameter $w$, the loop generators can be realised in the adjoint representation as $T_m^A \sim w^m T^A $, $\dL \sim - w \partial_w$  \cite{Goddard:1986bp}.

The above defines the adjoint representation of $\mf{e}_9$ and we will also require other representations. Irreducible highest or lowest weight representations can be constructed in a Fock space manner as reviewed for example in~\cite{Goddard:1986bp,Bossard:2017aae}. Here, we focus on the so-called basic representation that starts from an $\mf{e}_8$ invariant ground state and we shall employ a Fock space notation. The ground state $\ket{0}$ satisfies
\begin{align}
T^A_0 \ket{0} = 0 \,, \quad \dK  \ket{0}&=\ket{0} \,,\quad \dL \ket{0} = h\ket{0}\\
\textrm{and}\quad T^A_n \ket{0} &=0 \quad\textrm{for $n>0$}\,.\nn
\end{align}
While the eigenvalue of $\dK$ is fixed to one by unitarity, the eigenvalue of $\dL$ is a priori not determined. There is a one-parameter family of basic representations labelled by $h$ that appears in $\dL\ket{0} = h \ket{0}$, and we shall denote these representations by $R(\Lambda_0)_h$.\footnote{We have changed conventions with respect to \cite{Bossard:2017aae}, where such representations were denoted $R(\Lambda_0)_{-h}$. } 
In general, the eigenvalue of the central element $\dK$ on an irreducible module is an integer called the affine level.

Note that $\mf{e}_9$ is not simple as $\dK$ is central and $\dL$ never appears on the right-hand side of any commutator. Thus $\mf{e}_9$ admits a one-dimensional representation $\rho^h$ on which $\hat{\mf{e}}_8$ acts trivially and $\dL$ acts with eigenvalue $-h/2$. The module $R(\Lambda_0)_h$ can therefore be defined as the product representation $R(\Lambda_0)_h = \rho^{-2h} R(\Lambda_0)_0$. 

General elements in the basic representation will be denoted by ket-vectors $\ket{V}$ and can be expanded as
\begin{align}
\label{eq:Vdec1}
\ket{V} = \left( V^0 +  \sum_{n=1}^\infty V_{A_1\,\ldots\, A_n} \,T_{-1}^{A_1}\cdots T_{-1}^{A_n} \right)\ket{0}\,.
\end{align}
This representation of $\mf{e}_9$ is made irreducible by removing all singular vectors (submodules) that arise when acting on $\ket{0}$. As a consequence, each coefficient $V_{A_1\,\ldots\, A_n}$ is constrained to live in the subset of irreducible $\mf{e}_8$ representations contained in $\otimes^n{\bf 248}$ according to the graded decomposition of $R(\Lambda_0)_h$ under $\mf{e}_8$
\begin{align}
\label{eq:R0e8}
R(\Lambda_0)_h = {\bf 1}_h \oplus {\bf 248}_{h+1} \oplus \left({\bf 1} \oplus {\bf 248} \oplus {\bf 3875}\right)_{h+2} \oplus\ldots
\end{align}
The subscript on the $\mf{e}_8$ representations denotes their $\dL$ eigenvalue. This representation of the Lie algebra $\mf{e}_9$ is integrable and can be lifted to a representation of the affine Kac--Moody group $E_9=\hat{E}_8 \rtimes \reals_\dL^+$. Subtleties in defining this infinite-dimensional group will be discussed in Section~\ref{BorelGauge}.

At some places we shall also utilise an index notation for elements of the basic representation rather than a Fock space notation. Choosing an infinite countable basis of the Fock space module $\ket{e_M}$ with $M\in\{ 0; A; \ldots\}$ a collection of $\mf{e}_8$ indices reproducing the decomposition \eqref{eq:R0e8}, equation \eqref{eq:Vdec1} becomes
\begin{align}
\label{eq:R0ind}
\ket{V} = V^M \ket{e_M}\,,
\end{align}
so that components of vectors are $V^M$ and we will refer to $M$ as a `fundamental index'. We will use the bases $\ket{e_M}$ and indices $M,N,\ldots$ to label the components of $R(\Lambda_0)_h$ for all $h$, understanding that they characterise the $\hat{E}_8$ basic representation, whereas $h$ labels the representation under $\mathds{R}^+_\dL$. This convention is defined such that all the $E_9$ group elements $g$ are understood to be defined in the $R(\Lambda_0)_0$ representation, and the additional factor of $\rho(g)^{-2h}$ will be written explicitly. Note that for finite dimensional groups $E_n$ with $n<9$,  the symmetry of exceptional field theories is $E_n\times \reals^+$, and one writes various $E_n$ tensors of different weight with respect to  $\mathds{R}^+$. For $E_9$ the structure is very similar except that $E_9$ is only a semi-product $E_9 =\hat{E}_8\rtimes  \mathds{R}^+_\dL $.

We shall also require the representation $\overline{R(\Lambda_0)_h}$ conjugate to $R(\Lambda_0)_h$. Elements of the conjugate representation will be denoted by bra-vectors. As we shall review below, coordinates of the $E_9$ exceptional geometry belong to the $R(\Lambda_0)$ representation and derivatives to its conjugate.
To describe objects in $\overline{R(\Lambda_0)_h}$ in index notation, we introduce a basis $\bra{e^M}$ dual to $\ket{e_M}$ so that 
\begin{align}
\label{eq:barR0ind}
\bra{W} = W_M \bra{e^M}\,.
\end{align}
Again, we use the same notation for any value of $h$, which will be specified separately.

\subsection{Virasoro algebra}
\label{sec:Vir}

For the Fock space representation of the basic module (at affine level $1$) we define, following Sugawara~\cite{Sugawara:1967rw}, Virasoro generators in the enveloping algebra by
\begin{align}
\label{eq:Vir}
L_m = \frac{1}{2(1+g^\vee)} \sum_{n\in\ints} \eta_{AB} : T_n^A T_{m-n}^B:  \,,
\end{align}
where the colon denotes normal ordering such that the positive modes appear on the right. For $\mf{e}_8$ the dual Coxeter number $g^\vee=30$. The Virasoro generators~\eqref{eq:Vir} acting on the lowest weight basic representation satisfy the Virasoro algebra
\begin{align}
\label{eq:Viralg}
\lb L_m , L_n  \rb = (m-n) L_{m+n} + \frac{c}{12} m (m^2-1) \delta_{m,-n} \dK
\end{align}
with $c= \frac{\dim\mf{e}_8 }{1+g^\vee}=8$ and $\dK=1$ is the identity operator. The central charge $c=8$ comes from the fact that the module $R(\Lambda_0)_0$ can be realised as the Hilbert space of the two-dimensional conformal field theory of eight free chiral bosons parameterising the $E_8$ torus on which $\hat{\mathfrak{e}}_8$ acts as the current algebra.\footnote{For general extended loop groups $\hat{G}$,  $c$ is the rank of the group $G$, corresponding to the fact that the basic representation of $\hat{\mathfrak{g}}$ can be realised as the Hilbert space of $c$ chiral bosons on the torus $\mathds{R}^c / \Lambda_\mathfrak{g}$ with $\Lambda_\mathfrak{g}$ the even lattice generated by the simple roots of the simply-laced Lie algebra $\mathfrak{g}$.}
The Virasoro algebra is valid on $R(\Lambda_0)_h$ for any $h$.

We will denote the span of all Virasoro generators by
\begin{align}
\mf{vir} = \left\langle L_m\,\middle|\, m\in\ints\right\rangle\,.
\end{align}
We note that this space is not closed under commutation~\eqref{eq:Viralg} due to the central extension that we identify with $\dK$ and that is already contained in $\hat{\mf{e}}_8$. The maximal algebra that we shall consider in this paper is 
\begin{align}
\label{eq:fullalg}
\mf{f} = \hat{\mf{e}}_8 \oleft \mf{vir}\,,
\end{align}
which is the loop algebra extended by $\dK$ and \textit{all} Virasoro generators $L_m$. The Virasoro generators~\eqref{eq:Vir} act on any lowest weight $\mf{e}_9$ representation which therefore is automatically a representation of $\hat{\mf{e}}_8\oleft \mf{vir} $ where the sum is semi-direct according to
\begin{align}
\label{eq:VirE9}
\left[ L_m, T_n^A \right] = - n T_{m+n}^A\,,\quad \lb L_m ,\dK \rb=0\,.
\end{align}
We shall use more generally the notation $\oleft$ to denote an indecomposable $\hat{E}_8$ (or $E_9$) representation $X\oleft Y$ such that $X$ and the quotient $X\oleft Y / \{ X \sim 0\}$ are submodules of $\hat{E}_8$ but $Y\subset X\oleft Y$ is not, because $T_n^A Y \nsubseteq Y$.

In terms of a loop parameter $w$, the loop generators can be realised in the adjoint representation as $L_m = -w^{m+1}\partial_m$.

From~\eqref{eq:VirE9} we see that $\dL$ acting on the module $R(\Lambda_0)_h$ can be identified with $L_0+h$. Both $\dL$ and $L_0$ satisfy therefore the same commutation relations with the loop algebra, but $L_0\ket{0}=0$ for any $h$. In the basic representation, $\mf{e}_9$ is the span of the generators $\{ T_M^A, \dK\}$ of $\hat{\mf{e}}_8$ and $L_0$ such that 
\begin{align}
\label{eq:hate8e9}
\mf{e}_9 = \langle T_m^A , \dK , \dL \rangle = \langle T_m^A , \dK , L_0 \rangle\,.
\end{align}

As is well-known, the Virasoro algebra~\eqref{eq:Viralg} has an $\mf{sl}(2)$ subalgebra given by $\langle  L_{-1}, L_0, L_1 \rangle$. The group generated by $\hat{\mf{e}}_8 \oleft \langle  L_{-1}, L_0, L_1 \rangle$ through the exponential map is 
\begin{align}
\label{eq:gp}
\hat{E}_8 \rtimes SL(2)\, .
\end{align}
This group was identified in \cite{Julia:1996nu} as a symmetry of ungauged maximal supergravity in two dimensions. The symmetry group of the $E_9$ exceptional field theory will turn out to be its subgroup generated by ${\mf{e}}_9 \oleft \langle  L_{-1} \rangle$,
\begin{align}
\label{eq:symmetry}
\hat{E}_8 \rtimes ( \mathds{R}^+_{\dL} \!\! \ltimes \mathds{R}_{L_{-1}}) \, .
\end{align}
This group acts on $\mf{f}$ in an indecomposable representation.

We will denote collectively the generators of $\mf{f}$ in the basic representation $R(\Lambda_0)_h$ by $\langle T^\alpha \rangle = \langle T^A_n, \dK, L_n \rangle$. By construction they do not depend on $h$, and $\dK=1$ on the module, although we choose to write it explicitly for clarity.  

We can define a collection of $\hat{E}_8$ invariant symmetric bilinear forms $\eta_{m}$ by
\begin{align}
\label{eq:etam}
\eta_{m\, \alpha\beta} T^\alpha \otimes T^\beta = \sum_{n\in \ints} \eta_{AB} T^A_n \otimes T^B_{m-n} - L_m \otimes \dK - \dK\otimes L_m\,.
\end{align}
For the value $m=0$ the range of the  generators $T^\alpha$ is restricted to  $\langle T^A_n, \dK, L_0 \rangle$ and this form defined in the basic representation $R(\Lambda_0)_0$ (for which $\dL=L_0$) is the inverse of the standard invariant Killing form $\eta^{\alpha\beta}$ on $\mf{e}_9$. We shall also denote this form by just $\eta_{\alpha\beta}\equiv \eta_{0\,\alpha\beta}$. Similarly for  $\eta_m$  the range of the  generators $T^\alpha$ is restricted to $\langle T^A_n, \dK, L_m \rangle$ (for all $n\in \mathds{Z} $ but only one $m$), which also defines an algebra for which  $\eta_m$ is an invariant non-degenerate bilinear form.

It will be very convenient to also introduce shift operators $\cS_m$ (for $m\in\ints$), acting on $\mf{f}=\hat{\mf{e}}_8\oleft \mf{vir}$ according to
\begin{align}
\label{eq:Sm}
\cS_m(\dK) =\delta_{m,0}\, \dK \,,\quad \cS_m(L_n) = L_{m+n}\,,\quad \cS_m(T_n^A) = T_{m+n}^A\,.
\end{align}
$\cS_0$ is the identity. Combining this definition with \eqref{eq:etam} we find a useful identity for $m\ne 0$
\begin{equation}
\label{eq:etashiftid}
\eta_{(n+m)\,\alpha\beta}T^\alpha\otimes T^\beta = \eta_{n\,\alpha\beta}T^\alpha\otimes\mathcal{S}_{m}(T^\beta) - L_{n+m}\otimes\dK\,.
\end{equation}
It follows that the shift operators are not invariant under $\hat{E}_8$. Their transformation properties are discussed in appendix~\ref{app:cocycle}.

Finally, the Hermitian conjugate $T^{\alpha \dagger}$ in the representation $R(\Lambda_0)_h$ is defined as
\begin{align}
\label{eq:HC}
 L_n^\dagger  = L_{-n}\,,\quad \dK^\dagger = \dK\,,\quad \dL = \dL^\dagger\,,\quad T_n^{A\,\dagger} = \eta_{AB} T_{-n}^{B}\, . 
\end{align}
It acts on the shift operators as $\cS_m(T^\alpha)^\dagger=\,\cS_{-m}(T^{\alpha\,\dagger})$. We use the Hermitian conjugate to also define the maximal unitary subgroup $K(E_9)\subset E_9$ which consists of those elements $k\in E_9$ satisfying $k^\dagger k=k k^\dagger=1$ when acting on $R(\Lambda_0)_h$.\footnote{For finite-dimensional groups, the notion of maximal unitary subgroup coincides with that of maximal compact subgroup.}

 The representation of $\mf{f}$ on $\overline{R(\Lambda_0)_h}$ in terms of the generator $T^\alpha$ on the Hilbert space of bra vectors can be realised equivalently in terms of the generators $T^{\alpha \dagger}$ on the Hilbert space of ket vectors, using 
\be \bigl( \langle W| X \bigr)^\dagger = X^\dagger | W \rangle \label{eq:clarify}\ee
where we will write $\langle W|^\dagger= | W\rangle$ and it will be clear from the context that $|W\rangle \in \overline{R(\Lambda_0)_h}$. One consequence of this is that $\bra{W} \dL = \bra{W} (L_0+h)$.
Note that the representation of $\mf{f}$ on ${R(\Lambda_0)_h}$ in terms of the generator $-T^\alpha$ on the Hilbert space of ket vectors only agrees with the conjugate representation for anti-Hermitian elements. In particular, the two modules are isomorphic modules of the maximal unitary subgroup $K(E_9) \subset E_9$.

\section{Building blocks of $E_9$ exceptional field theory}
\label{sec:build}

Many of the variables of $E_9$ exceptional field theory can be extracted from the knowledge of $D=2$ maximal supergravity, which we review first in a reduction from $D=3$. Then we discuss the additional structures that enter the exceptional field theory, in particular the generalised Lie derivative, the section constraint and the presence of constrained fields.

\subsection{$D=2$ maximal supergravity fields}
\label{sec:2dmax}

An affine symmetry of $D=2$ gravity coupled to matter arises on-shell when it is obtained by dimensional reduction of a $D=3$ system with scalars taking values in a symmetric space~\cite{Geroch:1970nt,Julia:1982gx,Breitenlohner:1986um,Breitenlohner:1987dg}. In principle, all  propagating bosonic fields in $D=3$ can be dualized to scalars coupled minimally to a non-dynamical three-dimensional metric.  Assuming an additional space-like isometry with Killing vector $\partial_3\equiv \partial_\varphi$ in the three-dimensional space-time leads to a metric of the form
\begin{align}
\label{eq:met3D}
ds_3^2 = e^{2\sigma}(-dt^2+ dx^2) + \rho^2(d\varphi + A^{\scriptscriptstyle (3)}_\mu dx^\mu)^2\,,
\end{align}
where coordinates have been chosen to make the two-dimensional metric conformally flat with scale factor $e^{2\sigma}$. The variable $\rho$ measures the size of the Killing direction. The index $\mu=0,1$ labels the two coordinates $t$ and $x$ on which all the fields depend.

The field $A^{\scriptscriptstyle (3)}_\mu$ is the Kaluza--Klein vector arising in the reduction from $3$ to $2$ dimensions.
Vector fields in $D=2$ are not propagating and can be eliminated locally by a gauge transformation. 
In the usual formulation of ungauged $D=2$ supergravity with rigid $E_9$ symmetry (and its associated linear systems), this vector field is set to zero~\cite{Breitenlohner:1986um,Nicolai:1987kz}. 
However, the investigation of gauged supergravity in various dimensions has shown the importance of the hierarchy of tensor fields including the non-propagating ones~\cite{deWit:2005hv,deWit:2008ta}. In particular, the gauging of $D=2$ supergravity requires the introduction of an infinity of vector fields (including $A^{\scriptscriptstyle (3)}_\mu$) transforming in the basic representation of $E_9$~\cite{Samtleben:2007an}.
A similar requirement is expected to hold in exceptional field theory, 
but our goal in the present paper is to construct only the potential of $E_9$ exceptional field theory, which does not depend on vectors. Hence, we will postpone their analysis (and possibly that of higher rank forms) to future work.

\bigskip

The propagating scalar fields descend from $D=3$ and can be arranged in a representative $V_0$ of the coset space $E_8/(Spin(16)/\ints_2)$, where $Spin(16)/\ints_2$ is chosen to act on $V_0$ from the left, or alternatively in a Hermitian $E_8$ matrix $M_0=V_0^\dagger V_0^{\phantom{\dagger}\!}$. Written in the adjoint representation of $E_8$, the matrix takes the form $M_{0\, AB}$ with  $A,B\in\{1,\ldots,248\}$  of $\mf{e}_8$. The (bosonic) dynamics of the maximal $D=2$ supergravity theory is entirely described in terms of $V_0$ together with the scalars $\sigma$ and $\rho$ arising from the metric~\eqref{eq:met3D}. 

However, this does not make the infinite-dimensional affine symmetry and the associated integrability of the $D=2$ theory manifest. In order to exhibit this symmetry one has to use an infinite set of dual scalar fields (a.k.a. cascade of dual potentials) that are related to the original scalar fields by non-linear duality relations that are consistent with the equations of motion. These are manifested in a so-called linear system for a generating function of a spectral parameter $w$ that should be identified with the loop parameter of the loop algebra $\hat{\mf{e}}_8$ discussed in section~\ref{sec:e9prelim}.\footnote{We note that there are two spectral parameters that are relevant in gravity reduced to $D=2$; one that is called the `constant spectral parameter' and that we denote by $w$ and another one, often called `space-time dependent spectral parameter' that also depends on the $D=2$ coordinates and that we denote by $\gamma$. The two are related by $\gamma+\gamma^{-1} =2(w+\tilde\rho)/\rho$, so that $\gamma$ is a double cover of $w$. The `axion' $\tilde\rho$ in this relation is defined in equation~\eqref{eq:tilderho}. One can define $K(E_9)$ groups with respect to both choices of spectral parameter and the one that is commonly used in the linear system is $\gamma$. The one that we are using when writing the representative~\eqref{eq:coshatE8} is the constant spectral parameter $w$.} 

The dualisation of the scalar fields $V_0$ leads to scalar fields that parameterise the coset space $\hat{E}_8/K(E_9)$ where $K(E_9)$ denotes the maximal unitary subgroup of $E_9$~\cite{Breitenlohner:1986um,Nicolai:2004nv} which agrees with the maximal unitary subgroup of $\hat{E}_8$ in the basic representation $R(\Lambda_0)_0$.
Expanding around $w=\infty$ we can parameterise a coset representative of the centrally extended loop group $\hat{E}_8$  in the representation $R(\Lambda_0)_0$ as
\begin{align}
\label{eq:coshatE8}
\hat{V} = e^{-\sigma \dK} V_0 \exp( \eta_{AB}  Y_1^A T_{-1}^B ) \exp( \eta_{AB} Y_2^A T_{-2}^B ) \cdots\,.
\end{align}
$V_0$ here is the $E_8/(Spin(16)/\ints_2)$ coset representative containing the original $E_8$ scalars of the $D=3$ theory while the $Y_n^A$ are scalar fields corresponding to the $\hat{\mf{e}}_8$ generators $T_{-n}^A $ for $n>0$, and $\eta_{AB}$ is the Cartan--Killing form over $\mf{e}_8$. The local group $K(E_9)$ acts on $\hat V$ from the left while the rigid $\hat{E}_8$ acts from the right. In writing the coset representative of the centrally extended loop group $\hat{E}_8$ we have fixed a Borel gauge, meaning that only the negative mode loop generators $T^A_{-n}$ appear. This corresponds to fixing the action of $K(E_9)$. 

The fields $Y_n^A$ are on-shell dual to the propagating $E_8$ scalar fields parameterising $V_0$. The first duality relation is
\begin{align}
\label{eq:dual1}
\partial_\mu Y_1^A \eta_{AB} T^B = \rho\, \varepsilon_{\mu\nu} V_0^{-1} P^\nu V_0\,,
\end{align}
where the integrability of this equation is guaranteed by the equations of motion for $V_0$. In the above equation, $P^\nu=\tfrac12 (\partial^\nu V_0 V_0^{-1} + (\partial^\nu V_0 V_0^{-1})^\dagger)$ denotes the coset component of the Maurer--Cartan form, {\it i.e.}, the projection of $\partial^\nu V_0 V_0^{-1}$ to the $128$ non-compact generators of $E_8$. 

Equation~\eqref{eq:dual1} and similar equations for the other $Y_n^A$ are summarised in the linear system of the $D=2$ maximal supergravity~\cite{Nicolai:1987kz} whose precise form we do not require here. These infinitely many fields are required to realise the on-shell $\hat{E}_8$ symmetry. 

According to~\eqref{eq:hate8e9}, the full $E_9$ symmetry also requires the inclusion of the generator $\dL$. The scalar field of $D=2$ supergravity associated with this generator is the dilaton $\rho$~\cite{Julia:1982gx,Julia:1996nu}. This scalar field satisfies the free Klein--Gordon equation and is dual to an axion $\tilde\rho$ via
\begin{align}
\label{eq:tilderho}
\partial_\mu \rho = \varepsilon_{\mu\nu} \partial^\nu \tilde\rho\, .
\end{align}
Unlike for the $E_8$ scalars $V_0$, this duality relation is linear and does not give rise to an  infinite sequence of dual scalar fields.

The dilaton $\rho$ and the axion $\tilde\rho$ parameterise the group $\mathds{R}^+_{\sf d}\ltimes\mathds{R}_{L_{-1}}$.
The full coset space of relevance is therefore
\begin{equation}
 \frac{ \hat E_8 \rtimes\left( \mathds{R}^+_{\sf d}\ltimes\mathds{R}_{L_{-1}} \right) }
 {K(E_9)}\,.
\end{equation}
In the $R(\Lambda_0)_0$ representation, where we can identify $\dL$ with $L_0$, we write the $\mathds{R}^+_{\sf d}\ltimes\mathds{R}_{L_{-1}}$ group element as
\begin{align}
\label{eq:cosSL2}
v = \rho^{-L_0} e^{-\tilde\rho L_{-1}}\,,
\end{align}
and the full coset representative becomes 
\begin{align}
\label{eq:paracoset}
\cV = v \hat{V}\,.
\end{align}
It transforms from the left under the denominator group $K(E_9)$ and from the right under the rigid symmetry group $\hat E_8 \rtimes\left( \mathds{R}^+_{\sf d}\ltimes\mathds{R}_{L_{-1}} \right) $, \textit{i.e.} as $\cV \to k \cV g$.

Using the fact that $v$ can be embedded into the $SL(2)$ group generated by $L_{-1},\,L_0$ and $L_{+1}$, we can conveniently work with the Hermitian element
\begin{align}
\label{eq:cM}
\cM = \cV^\dagger\cV = \hat{V}^\dagger m \hat{V} \in \hat{E}_8\rtimes SL(2)
\end{align}
with
\begin{align}
\label{eq:mSL2}
m = v^\dagger v =  e^{-\tilde\rho L_1} \rho^{-2L_0} e^{-\tilde\rho L_{-1}}\, , 
\end{align}
so that $\cM=\cM^\dagger$.
We shall also decompose $\cM$ as follows
\begin{align}
\label{eq:cM_co}
\cM = \hat{V}^\dagger m \hat{V} = m \hat{g}_{\cM} = \hat{g}_{\cM}^\dagger m\,.
\end{align}
Note that while $m\in SL(2)$ satisfies $m^\dagger=m$, the $\hat{E}_8$ element $\hat{g}_{\cM}$ does not.
We stress that $\cM$ is defined as a group element in the $R(\Lambda_0)_0$ representation, in which $\dL=L_0$.

In the fundamental representation ${\bf 2}$ of $SL(2)$, $m$ can be written as the $2\times2$ matrix
\begin{equation}
\label{eq:m2}
m_{\bf 2}=\rho^{-1}\begin{pmatrix} 
\rho^2-\tilde\rho^2 & - \tilde\rho \\
\tilde\rho &1
\end{pmatrix}\,.
\end{equation}
We note that the Hermitian conjugate \eqref{eq:HC} isolates a non-compact unitary $SO(1,1)$ subgroup of $SL(2)$.
This implies that the finite-dimensional $m_{\bf2}$ cannot be a symmetric matrix but rather satisfies $m_{\bf 2}=\sigma_3 m_{\bf 2}^T \sigma_3$. Note that the whole $SL(2)$ is an on-shell symmetry of ungauged supergravity in two dimensions \cite{Julia:1996nu}, but the exceptional field theory potential will only exhibit the parabolic subgroup $\mathds{R}^+_{\sf d}\ltimes\mathds{R}_{L_{-1}}$ as  symmetry.

The advantage of working with $\cM$ instead of $\cV$ is that it only transforms under the rigid $\hat E_8 \rtimes\left( \mathds{R}^+_{\sf d}\ltimes\mathds{R}_{L_{-1}} \right)$ symmetry of the coset space as
\begin{align}
\label{eq:Mtrm}
\cM \to g^\dagger \cM g\,,\qquad g\in \hat E_8 \rtimes\left( \mathds{R}^+_{\sf d}\ltimes\mathds{R}_{L_{-1}} \right)\,.
\end{align}

Formally, for $\tilde\rho=0$, the element $v$ in~\eqref{eq:cosSL2} is simply a dilatation $\mathds{R}^+_{\sf d}$  and $\cM$ becomes an element of the affine $E_9$ group only. As we shall see many formul\ae{} simplify for $\tilde\rho=0$. 
Even though most of our derivations use $\cM$ for simplicity, a proper evaluation of the potential and definition of the dual scalar fields requires descending to $\cV$, a step we shall explain at the end in Section~\ref{BorelGauge}.

In summary, all scalar fields appearing in the $D=2$ maximal supergravity theory can be packaged into the operator $\cM$. It contains the $128$ propagating fields in the $E_8/(Spin(16)/\ints_2)$ coset representative $V_0$ along with all their dual potentials $Y_n^A$ as well as the dilaton $\rho$, the axion $\tilde\rho$ and the scale factor $\sigma$.

To give a more concrete idea of how the $E_8$ scalar fields parameterising $M_0$ and the dual potentials $Y_n^A$ are embedded in $\cM$, we now display some components of the inverse matrix $\cM^{MN}$. One can give formal definitions of the index-notation coefficients $\cM_{MN}$~and~$\cM^{MN}$ using  \eqref{eq:R0ind} and \eqref{eq:barR0ind} as
\begin{align}
\label{eq:Mind}
\cM = \cM_{MN} \bra{e^M}^\dagger\otimes\bra{e^N}\; ,\quad
\cM^{-1} = \cM^{MN}\ket{e_M}\otimes \ket{e_N}^\dagger\,.
\end{align}
Taking $\ket{e_0}=\ket{0}$, $\ket{e_A}=T_{-1\,A}\ket{0}$, the first few matrix components of $\cM^{MN}$ are computed from \eqref{eq:cM} in the parameterisation~\eqref{eq:coshatE8} and~\eqref{eq:mSL2} as follows
\begin{subequations}
\begin{alignat}{2}
\cM^{00} &= \bra{0}\cM^{-1}\ket{0} &&= e^{2\sigma}\,,\\\
\cM^{0A} &=\bra{0}\cM^{-1}T_{1}^{A\,\dagger} \ket{0} &&= - e^{2\sigma} Y_1^A\,,\\
\cM^{AB} &= \bra{0}T_{1}^A\cM^{-1}T_{1}^{B\,\dagger} \ket{0} &&= e^{2\sigma}\left(\rho^2 M_0^{AB}  + Y_1^A Y_1^B\right)\,,
\end{alignat}
\end{subequations}
with $M^{AB}_0$ the matrix components of the inverse $E_8$ matrix $M^{-1}_0= V_0^{-1} (V_0^{\dagger})^{-1}$.
We have also used $M_0^{-1}(T_m^A)^\dagger = M_0^{AB} T_{m\,B}M_0^{-1}$ and the fact that $\ket{0}$ is $SL(2)$ invariant.
The other dual potentials $Y^A_m$, $m>1$, as well as $\tilde\rho$ start appearing at higher levels.
Performing a similar expansion for $\cM_{MN}$ would give infinite divergent series in $Y_m^A$ at each level in the graded decomposition. 
However, the way $\cM$ enters in the potential $V(\cM,\chi)$ is such that the latter is well-defined for $\cV$ in the Borel gauge~\eqref{eq:coshatE8} and only involves finite combinations of terms, as we shall exhibit in Section~\ref{BorelGauge} and also in Section~\ref{sec:redE8} when we discuss the reduction to $E_8$.

\subsection{Generalised diffeomorphisms and scalar fields}

As usual in the construction of exceptional field theory, the supergravity fields are the basic building blocks and are promoted to fields depending on both the `external coordinates' and the `internal coordinates' of the exceptional geometry. The $E_9$ exceptional geometry is characterised by its generalised diffeomorphisms that we shall review first.

\subsubsection{Generalised Lie derivative}
\label{sec:GL}

As shown in~\cite{Bossard:2017aae}, the correct representation for coordinates and canonical generalised vectors in $E_9$ exceptional geometry is the basic representation $R(\Lambda_0)_{-1}$ discussed above. Writing $\ket{V}$ for a vector field in this representation, the action of a generalised diffeomorphism reads
\begin{align}
\label{eq:Lie}
\mathscr{L}_{\Lambda,\Sigma} \ket{V} =\ & \langle \partial_V \ket{\Lambda} \ket{V} 
-\eta_{\alpha\beta}\bra{\partial_\Lambda}T^\alpha\ket{\Lambda} T^\beta\ket{V}
-\langle\partial_\Lambda\ket{\Lambda}\ket{V} -\eta_{-1\,\alpha\beta}\, \Tr{(T^\alpha\Sigma)}\, T^\beta\ket{V}\,.
\end{align}
This very compact expression uses the Fock space notation for $\mf{e}_9$ representations and the bilinear forms \eqref{eq:etam}.
As the coordinates are valued in the $R(\Lambda_0)_{-1}$ representation just like generalised vectors, derivatives are in the dual $\overline{R(\Lambda_0)_{-1}}$ representation with $\sf d$ eigenvalue of the vacuum state $\bra{0}{\sf d}=(-1)\bra{0}$. Derivatives are represented as bra-vectors $\bra{\partial}$ with the subscript indicating which object they act on. 

The generalised Lie derivative~\eqref{eq:Lie} depends on two gauge parameters, $\Lambda$ and $\Sigma$. The first parameter $\Lambda$ is the usual generalised diffeomorphism parameter and is also valued in $R(\Lambda_0)_{-1}$. It is thus written as a ket vector. 
The second parameter $\Sigma$ is an extra constrained parameter that generalises a similar constrained parameter in the $E_8$ exceptional field theory~\cite{Hohm:2014fxa}. For $E_9$, $\Sigma$ belongs to $R(\Lambda_0)_0 \otimes \overline{R(\Lambda_0)_{-1}}$ with a constrained `bra index'.
This constraint will be spelt out below.
The trace $\Tr(T^\alpha\Sigma)$ is guaranteed to be finite due to the constrained nature of $\Sigma$.

Since the Fock space notation in~\eqref{eq:Lie} is different from that used for finite-dimensional symmetry groups, we provide a short translation into index notation using~\eqref{eq:R0ind}~and~\eqref{eq:barR0ind}. Vectors carry an upper fundamental index $M$ and co-vectors have a lower fundamental index. In this component notation, the gauge parameters have index structure $\Lambda^M$ and $\Sigma^N{}_M$. The generalised Lie derivative then takes the index form
\begin{align}
\label{eq:Liecpt}
\mathscr{L}_{\Lambda,\Sigma} V^M =&\ 
\Lambda^N \partial_N V^M 
-\eta_{\alpha\beta}\, (T^\alpha)^P{}_Q(T^\beta)^M{}_N\,\partial_P\Lambda^Q\,  V^N - \partial_N \Lambda^N V^M\nonumber\\
&-\eta_{-1\,\alpha\beta}\, (T^\alpha)^P{}_Q  (T^\beta)^M{}_N  \Sigma^Q{}_P \, V^N \,.
\end{align}
More examples of translating between the Fock space notation and the index notation were given in~\cite{Bossard:2017aae}.

The definition \eqref{eq:Lie} generalises to any field $\Phi$ admitting a well-defined action of $\mf{e}_9\oleft \langle L_{-1}\rangle$, not necessarily in a highest/lowest weight representation 
\begin{align}
\label{eq:Lieformal}
\mathscr{L}_{\Lambda,\Sigma}\Phi =&\ 
\langle\partial_\Phi\ket{\Lambda}\Phi 
+\eta_{\alpha\beta}\bra{\partial_\Lambda}T^\alpha\ket{\Lambda}
\delta^\beta \Phi
+\eta_{-1\,\alpha\beta}\Tr{(T^\alpha\Sigma)}\,\delta^\beta\Phi\,,
\end{align}
where $\delta^\alpha\Phi$ is the variation of the field with respect to the global symmetry algebra $\mf{e}_9 \oleft \langle L_{-1}\rangle$.
We stress that this includes the variation with respect to the derivation $\sf d$ and not $L_0$, thus reproducing the third term in \eqref{eq:Lie} with $\dL | V\rangle = (L_0 -1) |V\rangle$.

From the comparison of~\eqref{eq:Liecpt} with the common form of generalised Lie derivatives~\cite{Berman:2012vc} one can already anticipate the form of the section constraint to be
\begin{subequations}
\label{eq:SC}
\begin{align}
\label{eq:SC1}
\eta_{\alpha\beta}\bra{\partial_1}T^\alpha\otimes \bra{\partial_2}T^\beta
+\bra{\partial_1}\otimes\bra{\partial_2}-\bra{\partial_2}\otimes\bra{\partial_1} = 0\,.
\end{align}
This constraint defines the  tensor $Y^{MN}{}_{PQ}$ in~\eqref{eq:SCintro} in the introduction.
The above constraint  has for consequence the additional constraints
\begin{align}
\eta_{-n\,\alpha\beta}\bra{\partial_1}T^\alpha\otimes \bra{\partial_2}T^\beta  &=0 \quad\quad \textrm{for all $n>0$}\,,\label{eq:SC2}\\
\eta_{1\,\alpha\beta}\big(\bra{\partial_1}T^\alpha\otimes \bra{\partial_2}T^\beta+\bra{\partial_2}T^\alpha\otimes \bra{\partial_1}T^\beta\,\big) &=0\label{eq:SC3} \,.
\end{align}
\end{subequations}
The constraint on the gauge parameter $\Sigma$  is more conveniently written using the notation introduced in \cite{Bossard:2017aae}
\begin{equation}\label{eq:Sigmafact}
\Sigma = \Sigma^M{}_N \ket{e^N}\bra{e_M} = \ket{\Sigma}\bra{\pi_\Sigma} \,.
\end{equation}
The section constraint \eqref{eq:SC} is then also imposed when one derivative is replaced by $\bra{\pi_\Sigma}$, and when both derivatives are replaced with $\bra{\pi_{\Sigma_1}}$, $\bra{\pi_{\Sigma_2}}$, where $\Sigma_1,\,\Sigma_2$ can be the same or two different gauge parameters.
Notice that $\Sigma$ generally does not factorize into a tensor product of a bra-vector $\bra{\pi_\Sigma}$ with a ket-vector $\ket{\Sigma}$.\footnote{To write the constraint \eqref{eq:SC1} for the operator $\Sigma$ itself one must introduce an additional arbitrary vector $|V\rangle$, such that for any $|V\rangle$: $ \eta_{\alpha\beta} \Sigma T^\alpha | V\rangle \langle \partial | T^\beta + \Sigma |V\rangle \langle \partial | - \Sigma \langle \partial | V \rangle =0$ and $ \eta_{\alpha\beta} \Sigma _1T^\alpha | V_1\rangle \otimes   \Sigma _2T^\beta | V_2\rangle + \Sigma _1  | V_1\rangle \otimes   \Sigma _2 | V_2\rangle -\Sigma _1  | V_2\rangle \otimes   \Sigma _2 | V_1\rangle=0$.}

As was shown in~\cite{Bossard:2017aae}, the generalised Lie derivative~\eqref{eq:Lie} provides a closed gauge algebra when the section constraints are imposed, satisfying
\begin{equation}
[\mathscr{L}_{\Lambda_1,\,\Sigma_1},\,\mathscr{L}_{\Lambda_2,\,\Sigma_2}]\Phi= \mathscr{L}_{\Lambda_{12},\,\Sigma_{12}}\Phi \, , 
\end{equation}
with 
\begin{align}
\Lambda_{12} &= \frac12 \left(\mathscr{L}_{\Lambda_1} \Lambda_2 - \mathscr{L}_{\Lambda_2} \Lambda_1\right)\,,\nn\\
\Sigma_{12} &= \mathscr{L}_{\Lambda_1} \Sigma_2 + \frac12 \eta_{-1\alpha\beta} T^\alpha \Sigma_1  T^\beta  \Sigma_2- \frac12 \eta_{1\alpha\beta} \bra{\partial_{\Lambda_2}} T^\alpha \ket{\Lambda_{[1}} T^\beta\ket{\Lambda_{2]}}\bra{\partial_{\Lambda_2}}  - \left(1 \leftrightarrow 2\right) \ .  \label{ClosureGD} 
\end{align}
The parameter $\Sigma_{12}$ depends nontrivially on $\Lambda_1$ and $\Lambda_2$ (as well as on $\Sigma_1$ and $\Sigma_2$), compensating for the fact that the $\Lambda$ part of $E_9$ generalised diffeomorphisms does not close onto itself.
An observation that will be crucial in the following is that closure of the generalised Lie derivative is ensured already by restricting to traceless $\Sigma$ parameters, {\it i.e.} $\Sigma_{12}$ is traceless if $\Sigma_1$ and $\Sigma_2$ are.
As is clear from \eqref{eq:Lie} and \eqref{eq:etam}, the trace component $\Tr{(\Sigma)}$ is the only one generating $L_{-1}$ transformations.
We can then gauge-fix the trace component of $\Sigma$ transformations acting on $\cM$ by setting $\tilde\rho=0$ and then consistently restrict to arbitrary $\Lambda$ and traceless $\Sigma$ transformations.
No compensating gauge transformations are then needed to keep $\tilde\rho=0$ and $E_9$ covariance is preserved.

\subsubsection{Unconstrained scalar fields and currents}

The fields of the theory include the element $\cM\in \hat{E}_8\rtimes SL(2)$ introduced in~\eqref{eq:cM} and it depends on the coordinates of the exceptional geometry that take values in the $R(\Lambda_0)_{-1}$ representation. We reiterate that we always take $\cM$ to be defined as a group element in the representation $R(\Lambda_0)_0$. As a side-remark we note that this discrepancy between the weight $h$ of the coordinate representation and the representation of $\cM$ means that the `generalised metric' of $E_9$ exceptional geometry is $\rho^2 \cM$.

{}From $\cM$ in $R(\Lambda_0)_0$ we can, as usual, construct the current
\begin{align}
\label{eq:currdef}
\cJ_M = \cM^{-1} \partial_M \cM
\end{align}
which is valued in the Lie algebra $\hat{\mf{e}}_8\oleft \mf{sl}(2)$ and carries a constrained derivative index $M$ in the $\overline{R(\Lambda_0)_{-1}}$ representation.
In index notation and using the bases \eqref{eq:R0ind} and \eqref{eq:barR0ind} this reads
\begin{align}
\label{eq:cJdef1}
\cM^{PS} \partial_M \cM_{SQ} = (\cJ_M)^P{}_Q = \cJ_{M,\alpha} (T^\alpha)^P{}_Q\,,
\end{align}
where $T^\alpha\in\mf f$ but $\cJ_{M,\alpha}$ is only non-vanishing along $\hat{\mf e}_{8}\oleft \mf{sl}(2)$. 
It will be more convenient to use the Fock space notation, in which the current is defined such that
\begin{align}
\label{eq:cJdef2}
\bra{\partial_\cM} \otimes \cM = \bra{\cJ_\alpha} \otimes \cM T^\alpha\,.
\end{align}
To see the equivalence of this formula with the definition~\eqref{eq:cJdef1} above one may reintroduce indices as
\begin{align}
\partial_M \cM_{PQ} = \cJ_{M,\alpha} \cM_{PS} (T^\alpha)^S{}_Q
\end{align}
and multiply the equation with $\cM^{-1}$. The current satisfies the following useful identity
\begin{equation}
\label{eq:curid}
\bra{\cJ_\alpha}\otimes\cM^{-1} T^{\alpha\,\dagger}\cM=\,\bra{\cJ_\alpha}\otimes T^\alpha\,,
\end{equation}
which follows from the Hermiticity of $\cM$. It will also be convenient to introduce the matrix $H(\mathcal{M})^\alpha{}_\beta$ defined by
\begin{align}
\label{eq:Hdef}
 \cM^{-1} T^{\alpha\,\dagger} \cM
=  H(\cM)^\alpha{}_\beta T^\beta \,,
\end{align}
in terms of which the identity \eqref{eq:curid} reads  
\begin{equation}
\label{eq:Hid}
H(\cM)^\beta{}_\alpha\, \cJ_\beta =\cJ_\alpha\,.
\end{equation}

We write out the Lie algebra valued current in its components more explicitly as follows 
\begin{align}
\label{eq:curcpt}
\bra{\cJ_\alpha} \otimes T^\alpha= \sum_{n\in\ints} \bra{J^{\,n}_A} \otimes T^A_n + \sum_{\tiny{q=-1,0,1}} \bra{J_q}\otimes L_q + \bra{J_\dK}\otimes \dK\,.
\end{align}
The above expression stresses once again that,  while our conventions are such that $T^\alpha$ are the generators of $\mf{f}=\hat{\mf{e}}_8\oleft \mf{vir}$, the current has only components  along $\hat{\mf{e}}_8\oleft \mf{sl}(2)$ since it is constructed from an element $\cM$ in the group $\hat{E}_8 \ltimes SL(2)$~\eqref{eq:cM}. In other words, we have $\bra{J_q}=0$ for $|q|>1$. 

We also note that the $\mf{sl}(2)$-part of the current is identical to that constructed solely out of the $SL(2)$ element $m$ defined in~\eqref{eq:mSL2},
\begin{align}
\label{eq:SL2cur}
m^{-1} \partial_M m = J_{M,0} L_0 + J_{M,1} L_1 + J_{M,-1} L_{-1}\,,
\end{align}
where, due to the structure of $m$, one has in Fock space notation
\begin{align}
\label{eq:jrelation}
\bra{J_{-1}} + \tilde\rho \bra{J_0} + (\tilde\rho^2-\rho^2)\bra{J_{1}} = 0\,.
\end{align}
This relation can be derived easily from the matrix representation~\eqref{eq:m2} and can be used to solve for $\bra{J_{-1}}$ in terms of the other components.

\subsubsection{Constrained scalar fields and shifted currents}

A crucial ingredient in exceptional field theory is the existence of constrained fields. For $E_{11-D}$ exceptional field theory with a $D$-dimensional external spacetime, the constrained fields appear starting from the $(D-2)$-form sector. For instance, in the $E_8$ exceptional field theory, corresponding to $D=3$, there are constrained vector fields that are crucial in the construction of the theory~\cite{Hohm:2014fxa}. In the present case of $D=2$, the constrained fields appear already in the scalar sector and therefore are essential for the scalar potential. 

As was argued in~\cite{Bossard:2017aae}, the constrained scalar fields carry an index labelling the dual basic representation $\overline{R(\Lambda_0)_{-1}}$ and we write them as $\chi_M$ or $\bra{\chi}$. The fact that they are constrained means that they satisfy a condition analogous to the section constraint~\eqref{eq:SC}, namely
\begin{subequations}
\begin{align}
\eta_{\alpha\beta}\bra{\chi}T^\alpha\otimes \bra{\partial}T^\beta
+\bra{\chi}\otimes\bra{\partial}-\bra{\partial}\otimes\bra{\chi} &= 0\,,\\
\eta_{-n\,\alpha\beta}\bra{\chi}T^\alpha\otimes \bra{\partial}T^\beta  &=0 \quad\quad \textrm{for all $n>0$}\,,\\
\eta_{1\,\alpha\beta}\big(\bra{\chi}T^\alpha\otimes \bra{\partial}T^\beta+\bra{\partial}T^\alpha\otimes \bra{\chi}T^\beta\,\big) &=0\, ,
\end{align}
\end{subequations}
and the same identities bilinear in $\langle \chi|$. Here, $\bra{\partial}$ can be any derivative, as for instance that carried by the current $\bra{\cJ_\alpha}$, or also the constrained index of the generalised diffeomorphism parameter $\Sigma$.

As will become clearer when we discuss the transformation properties of the various fields, it is useful to also define a shifted version of the current $\bra{\cJ_\alpha}$ defined in~\eqref{eq:cJdef2}, by making use of the shift operators defined in \eqref{eq:Sm}
\begin{align}
\label{eq:shiftcur}
\bra{\cJ_\alpha^{-}} \otimes T^\alpha = \sum_{k=0}^\infty \tilde\rho^k \bra{\cJ_\alpha} \otimes \cS_{-1-k}(T^\alpha) + \bra{\chi}\otimes \dK\,.
\end{align}
The constrained scalar field $\chi$ appears in the definition of the shifted current in order to ensure covariance under the rigid symmetries, as we shall explain in detail below. Unlike the unshifted current~\eqref{eq:curcpt}, the shifted current has also non-trivial components along the Virasoro generators $L_q$ for all $q\leq 1$ because these are generated by the shift operators.

\section{The potential}
\label{sec:pot}
In this section, we present the $E_9$ exceptional field theory potential, depending on $\cM$ and the constrained scalar field $\chi$, written as a sum of four terms as
\begin{align}
\label{eq:pot}
\pot(M,\chi)= \tfrac14 \mathcal{L}_1 -\tfrac12 \mathcal{L}_2 + \tfrac12 \mathcal{L}_3 +\tfrac12 \mathcal{L}_4\; . 
\end{align}
The four terms are independently invariant under rigid $\hat E_8$ and $\mathds{R}_{L_{-1}}$ transformations, and transform with the expected homogeneous scaling under $\mathds{R}_\dL^+$. The symmetry $\mathds{R}_\dL^+$ is known to be a symmetry of the equations of motion, whereas the potential itself is not invariant but transforms homogeneously with weight one. The relative coefficients between the four terms are fixed by requiring the potential to transform into a total derivative under generalised diffeomorphisms.

Let us briefly compare the general structure of the potential to that of other $E_n$ exceptional field theories with $n\leq 8$. While the terms $\mathcal{L}_1$, $\mathcal{L}_2$ and $\mathcal{L}_4$ have direct analogues in the other cases~\cite{Hohm:2013vpa,Hohm:2013uia,Hohm:2014fxa,Cederwall:2017fjm}, the term $\mathcal{L}_3$ is a vast generalisation of a term that only appears in $E_8$ exceptional field theory. The main new feature is that the $E_9$ scalar fields $\cM$ and $\chi$ form an \textit{indecomposable} representation, meaning that they cannot be separated into a direct sum of irreducible $E_9$ representations. The term $\mathcal{L}_3$ contains crucially the shifted current~\eqref{eq:shiftcur} that comprises all these scalar fields. Another consequence of this indecomposability of the scalar fields is that the four individual terms of the potential are not all manifestly invariant under the rigid $\hat{E}_8 \rtimes \mathds{R}_{L_{-1}}$ transformations and we shall therefore demonstrate this invariance explicitly.

For the sake of clarity of the presentation we shall first consider a partially gauge-fixed version of the potential in which the axion $\tilde{\rho}=0$ and the rigid symmetry $\mathds{R}_{L_{-1}}$ is broken. As mentioned in section~\ref{sec:GL}, this gauge-fixing preserves the closed subalgebra of generalised diffeomorphisms \eqref{eq:Lie} with $\Tr\,(\Sigma)=0$. In this case, the four terms are manifestly invariant under  $\hat E_8$. The term $\mathcal{L}_3$ still retains the indecomposable structure but now involves only the shift operator $\cS_{-1}$ (instead of all $\cS_n$ with $n<0$). This shift operator and the associated shifted bilinear form $\eta_{-1}$ are also expected on the basis of the structure of maximal gauged supergravity~\cite{Samtleben:2007an} where the embedding tensor couples through $\eta_{-1}$.

In a second step, we reintroduce the $\tilde\rho$ dependence and consequently the full generalised diffeomorphism invariance. Besides generalised diffeomorphism invariance, the full potential presented in this section is invariant under rigid $\hat{E}_8 \rtimes \reals_{L_{-1}}$ and these two requirements uniquely fix the combination of the four individual terms. In the following section~\ref{sec:redE8}, we moreover demonstrate that our potential, upon choosing an appropriate solution to the section condition, reproduces all the terms in the $E_{8(8)}$ exceptional field theory \cite{Hohm:2014fxa} that can contribute to the $E_9$ potential. This provides a final check on the $E_9$ potential.

\subsection{The potential  at $\tilde\rho=\,0$}

In this section we restrict ourselves to the case $\tilde\rho=\,0$ in which $\mathcal{M}$, defined in \eqref{eq:cM}, is an element of $E_9$ in the $R(\Lambda_0)_0$ representation. The construction of the potential is greatly simplified in this setting as one simply requires its invariance under rigid $\hat E_8$ transformations and $\Lambda$ generalised diffeomorphisms. The various terms of the potential then read
\begin{subequations}
\label{eq:potterm0}
\begin{align}
\mathcal{L}_1=&\,\rho^{-1} \eta^{\alpha\beta} \bra{\mathcal{J}_\alpha}\mathcal{M}^{-1}\ket{\mathcal{J}_\beta}\,,\label{eq:pot01}\\[1.5mm]
\mathcal{L}_2=&\,\rho^{-1}\bra{\mathcal{J}_\alpha}T^\beta\mathcal{M}^{-1}T^{\alpha\,\dagger}\ket{\mathcal{J}_{\beta}}\,,\label{eq:pot02}\\[1.5mm]
\mathcal{L}_3=&\,\rho\,\bra{\mathcal{J}_\alpha^-}T^\beta\mathcal{M}^{-1}T^{\alpha\,\dagger}\ket{\mathcal{J}^-_\beta}\,,\label{eq:pot03}\\[1.5mm]
\mathcal{L}_4=&\,\rho^{-1}\bra{J_0}T^\alpha\mathcal{M}^{-1}\ket{\mathcal{J}_\alpha}\,\label{eq:pot04}.
\end{align}
\end{subequations}
Their expression in an index notation was already given in the introduction \eqref{pot_intro}. The currents $\bra{\mathcal{J}_\alpha}$ and $\bra{\mathcal{J}_\alpha^-}$ were defined in \eqref{eq:cJdef2} and \eqref{eq:shiftcur}, and since here $\mathcal{M}\in E_9$, their only non-vanishing components are along $\mathfrak{e}_9$ and $\hat{\mathfrak{e}}_8\oleft L_{-1}$, respectively. Since here $\tilde\rho=0$, the current component along $L_0$ is simply given by
\begin{align}
\label{eq:defJ0}
\bra{J_0}=\,-2\,\rho^{-1}\bra{\partial_\rho}\rho\,.
\end{align} 
We also point out that, while writing some of the currents as a ket in \eqref{eq:potterm0} might seem confusing at first, our notation should be clear from the discussion at the end of section \ref{sec:Vir}.

All the terms in \eqref{eq:potterm0} are manifestly Hermitian and, as we shall see below, invariant under rigid $\hat E_8$ transformations. Moreover, note that in this case the new constrained scalar field $\bra{\chi}$ only appears in the third term. As previously mentioned, $\mathcal{L}_3$ generalises a term that so far only appeared in the $E_{8(8)}$ potential, where it involved two $E_{8(8)}$ currents contracted directly \cite{Hohm:2014fxa}. 

\subsubsection{Rigid $E_9$ symmetry}    
Under $g\in E_9$, we have the following rigid transformations 
\begin{align}
\mathcal{M}&\to g^{\dagger}\mathcal{M}g\,,\label{eq:E9actM}\\
\bra{\partial}&\to\rho(g)^2\bra{\partial}g\,,\label{eq:E9actder}\\\
\ket{Y}&\to\rho(g)^{-2} g^{-1} \ket{Y}\,,
\end{align}
where here and in the following, $g$ will always be a group element in the $R(\Lambda_0)_0$ representation. It acts on the derivative bra in the $\overline{R(\Lambda_0)_{-1}}$ representation by multiplication with $g$ from the right and on the coordinate ket in the $R(\Lambda_0)_{-1}$ by multiplication with $g^{-1}$ from the left. The $\mathds{R}^+$-valued function $\rho(g)$ appearing in the transformation laws is the multiplicative character defined in \eqref{eq:rhogdef}. It occurs for instance in the above second variation to account for the fact that the derivative transforms in the $\overline{R(\Lambda_0)_{-1}}$ representation while $g$ is a group element in the $R(\Lambda_0)_0$ representation.\footnote{We note that $\cM$ as a group element is in $R(\Lambda_0)_0$ which naturally multiplies ket vectors in $R(\Lambda_0)_{-h}$ from the left and produces bra vectors in $\overline{R(\Lambda_0)_{+h}}$. It thus acts as an intertwiner of representations.} In particular, we have $\rho(\mathcal{M})=\rho$.

The variation of the $\mathfrak{e}_9$-valued current \eqref{eq:cJdef2} then reads
\begin{align}
\bra{\cJ_\alpha}\otimes T^\alpha\to\rho(g)^2 \bra{\cJ_\alpha}g\otimes g^{-1}T^\alpha g\,,
\end{align}
while its components $\bra{\mathcal{J_\alpha}}$ transform as
\begin{equation}
\label{eq:E9actrho}
\bra{\cJ_\alpha}\to \rho(g)^2\,R(g)^\beta{}_\alpha\bra{\cJ_\beta}g\,,
\end{equation}
where $R(g)^\alpha{}_\beta$ denotes the representation matrix of $g$ acting by conjugation, which is defined by 
\begin{equation}
\label{eq:Rdef'}
R(g)^\alpha{}_\beta T^\beta=\,g^{-1}T^\alpha g\,.
\end{equation}
To further clarify our notation for the current written as a ket, we also provide explicitly its transformation under $E_9$,
\begin{align}
T^\alpha\otimes\ket{\cJ_\alpha}& \to g^{-1}T^\alpha g\otimes g^\dagger\ket{\cJ_\alpha} \,\rho(g)^2 \,,\nn\\[1.5mm]
\ket{\cJ_\alpha} &\to  g^\dagger \ket{\cJ_\beta}  R(g)^\beta{}_\alpha\, \rho(g)^2\,,
\end{align}
which follows from \eqref{eq:clarify}. The variation of the scalar $\rho=\rho(\mathcal{M})$ is by definition
\begin{equation}
\label{eq:rhoL0var}
\rho\to \rho(g)^2\,\rho\,.
\end{equation}
From \eqref{eq:defJ0}, one then simply finds that $\bra{J_0}\to\rho(g)^2\bra{J_0}g$. Using the above transformations it is straightforward to verify the rigid $\hat E_8$  invariance of \eqref{eq:pot02} and \eqref{eq:pot04}. The $\hat E_8$ invariance of \eqref{eq:pot01} is ensured by the presence of the $\mathfrak{e}_9$ invariant bilinear form $\eta^{\alpha\beta}$, which satisfies \mbox{$R(g)^\alpha{}_\gamma R(g)^\beta{}_\delta \eta^{\gamma\delta}=\,\eta^{\alpha\beta}$.} 

The invariance of \eqref{eq:pot03} is a bit more subtle. Start by considering the variation of the current which has been acted upon by the shift operator defined in \eqref{eq:Sm},
\begin{equation}
\label{eq:shiftvar}
\bra{\mathcal{J}_\alpha}\otimes \mathcal{S}_{-1}(T^\alpha)\to\rho(g)^2\bra{\cJ_\alpha}g\,\otimes\big(\rho(g)^{-2}\,g^{-1}\cS_{-1}(T^\alpha)g-\omega^\alpha_{1}(g^{-1})\dK\big)\,.
\end{equation}
This results follows directly from using the relation \eqref{eq:coad}. It involves an $\mathds{R}$-valued function $\omega_1^\alpha(g)$, which is a group 1-cocycle defined in \eqref{eq:coad}. As explained in detail in Appendix \ref{app:cocycle}, this cocycle and the character $\rho(g)$ define an extension of the adjoint representation of $E_9$ by the generator $L_{-1}$. The new scalar field $\bra{\chi}$ must be chosen to transform as part of the dual of this extended representation \eqref{eq:coadext}, {\it i.e.} as
\begin{equation}
\label{eq:chiglotrans}
\bra{\chi}\to\bra{\chi}g+\rho(g)^2\,\omega_1^\alpha(g^{-1})\bra{\cJ_\alpha}g\,,
\end{equation}
in order for the shifted current to transform covariantly under $E_9$. Indeed, using \eqref{eq:shiftvar} and \eqref{eq:chiglotrans} we find that the shifted current \eqref{eq:shiftcur} and its components transform as
\begin{align}
\bra{\cJ_\alpha^-}\otimes T^\alpha&\to\bra{\cJ_\alpha^-}g\otimes g^{-1}T^\alpha g\,,\label{eq:scurglotrans}\\[1.5mm]
\bra{\cJ_\alpha^-}&\to R(g)^\beta{}_\alpha\bra{\cJ_\beta}g\,.
\end{align}
It is then straightforward to verify that the third term in the potential is invariant under $\hat E_8$. Note that the appearance of fields transforming in indecomposable representations, such as $\bra{\chi}$ in \eqref{eq:chiglotrans}, is a new feature in $E_9$ exceptional field theory. In higher-dimensions, all the fields have to transform individually in irreducible representations of the duality group since $E_n$ is then a finite-dimensional reductive group.

Let us finally remark that the potential scales uniformly under transformations generated by  $\dL$,
\begin{equation}
\label{eq:potscale}
\pot\to\rho(g)^{2}\,\pot\,.
\end{equation}
As mentioned previously, the generator $\dL$ is associated to a symmetry of the equations of motion and not of the Lagrangian itself, as is the so-called trombone symmetry in higher dimensions \cite{Cremmer:1997xj}. According to its original definition, the trombone symmetry in two dimensions shifts the conformal factor $\sigma$ of the metric and is the symmetry of the action generated by the central charge $\dK$. It is instead the symmetry generated by $\dL$ that rescales the dilaton field $\rho$ which is not a symmetry of the action.

\subsubsection{Invariance under generalised diffeomorphisms}

We denote an infinitesimal variation under generalised diffeomorphisms by $\delta_{\Lambda,\Sigma}$. By definition it splits into
\begin{align}
\label{eq:Delta}
\delta_{\Lambda,\Sigma} = \mathscr{L}_{\Lambda,\Sigma} + \Delta_{\Lambda,\Sigma},
\end{align}
where the action of the generalised Lie derivative $\mathscr{L}_{\Lambda,\Sigma}$ on an arbitrary field was defined in \eqref{eq:Lieformal}, and where $\Delta_{\Lambda,\Sigma}$ explicitly collects all the non-covariant pieces in the variation. The latter are those terms involving second derivatives of the gauge parameter $\Lambda$ or a single derivative of the gauge parameter $\Sigma$. In this section, we exclusively focus on variations under $\Lambda$ generalised diffeomorphisms, which are simply denoted by $\delta_\Lambda$. The reason is that in the expression of the generalised Lie derivative, the trace of $\Sigma$ appears as the gauge parameter of an infinitesimal $\mathds{R}_{L_{-1}}$ transformation, which can only be considered at $\tilde\rho\neq 0$ when $\cM\in\hat E_8\rtimes SL(2)\supset \mathds{R}_{L_{-1}}$. While we could already consider traceless $\Sigma$ variations in this section, we postpone this discussion to section \ref{sec:pottilderho} where we will prove the invariance of the full potential under arbitrary $\Sigma$ variations.

We start with $\cM$ which transforms covariantly under generalised diffeomorphisms, {\it i.e.} as
\begin{equation}
\label{eq:gendiffM}
\delta_\Lambda \cM=\mathscr{L}_\Lambda\cM=\,\langle\partial_{\cM}|\Lambda\rangle\cM+\eta_{\alpha\beta}\bra{\partial_\Lambda}T^\alpha\ket{\Lambda}\big(\cM\, T^\beta+T^{\beta\,\dagger}\cM\big)\,,
\end{equation}
The expression of the rotation term follows from the action \eqref{eq:E9actM} of $E_9$ on $\cM$. In particular, there is no density term as $\cM$ transforms (from the right and the left) in the $R(\Lambda_0)_0$ representation and therefore carries no weight. According to \eqref{eq:rhoL0var}, the field $\rho$ is an $E_9$ scalar density of weight one and thus transforms as a total derivative
\begin{equation}
\label{eq:gendiffr}
\delta_\Lambda\,\rho=\mathscr{L}_\Lambda\,\rho=\langle \partial_\rho|\Lambda\rangle\rho+\langle\partial_\Lambda|\Lambda\rangle\rho=\bra{\partial}\Big(\ket{\Lambda}\rho\Big)\,.
\end{equation}
The variation of the current \eqref{eq:cJdef2} follows from that of $\cM$ and takes the form
\begin{align}
\label{eq:vargencurr}
\delta_{\Lambda}\bra{\mathcal{J}_\alpha}\otimes T^\alpha=\,\mathscr{L}_\Lambda\bra{\cJ_\alpha}\otimes T^\alpha+\eta_{\alpha\beta}\bra{\partial_\Lambda}T^\alpha\ket{\Lambda}\bra{\partial_\Lambda}\otimes(T^\beta+\cM^{-1}T^{\beta\,\dagger}\cM\big)\,,
\end{align}
where its Lie derivative is given by 
\begin{equation}
\label{eq:Liecur}
\mathscr{L}_\Lambda\bra{\cJ_\alpha}\otimes T^\alpha=\,\big(\langle \partial_\mathcal{J}|\Lambda\rangle\bra{\mathcal{J}_\alpha}+\langle\mathcal{J}_\alpha|\Lambda\rangle\bra{\partial_\Lambda}\big)\otimes T^\alpha-\eta_{\alpha\beta}\bra{\partial_\Lambda}T^\alpha\ket{\Lambda}\bra{\cJ_\gamma}\otimes [T^\beta,T^\gamma]\,.
\end{equation} 
The variation of the current components then reads
\begin{align}
\label{eq:varcurcomp}
\delta_\Lambda \bra{\cJ_\alpha}=\,\mathscr{L}_\Lambda\bra{\cJ_\alpha}+ \eta_{\alpha\beta}  \bra{\partial_\Lambda} \bigl( T^\beta + \cM^{-1} T^{\beta \dagger} \cM \bigr) \ket{\Lambda}\bra{\partial_\Lambda}\,,
\end{align} 
with 
\begin{align}
\label{eq:Liecurcomp}
\mathscr{L}_\Lambda\bra{\cJ_\alpha}=\,\langle\partial_{\cJ}|\Lambda\rangle\bra{\cJ_\alpha}+\langle\mathcal{J}_\alpha|\Lambda\rangle\bra{\partial_\Lambda}-\eta_{\gamma\delta} \bra{\partial_{\Lambda}}T^\gamma \ket{\Lambda} \,f^{\delta\beta}{}_\alpha\bra{\cJ_\beta }\, . 
\end{align}
where $f^{\alpha\beta}{}_\gamma$ denotes the structure constants of $\mathfrak{f}$. To write the non-covariant terms we also used the identity 
\begin{equation} \label{etanCovE9} 
\eta_{-n\,  \alpha\beta}\,\cM^{-1}T^{\alpha\,\dagger}\cM\otimes T^\beta=\,\rho^{-2n} \eta_{ n\,  \alpha\beta}\, T^\alpha\otimes\cM^{-1}T^{\beta\,\dagger}\cM\,
\end{equation}
which only holds here as $\cM\in E_9$, and follows from the covariance of $\eta_{n \, \alpha\beta}$ under $E_9$. From the non-covariant variation in \eqref{eq:varcurcomp}, one gets in particular
\begin{equation}
\label{eq:varJ0}
\Delta_\Lambda\bra{J_0}=-2\,\langle\partial_\Lambda|\Lambda\rangle\bra{\partial_\Lambda}\,.
\end{equation}

Let us now discuss the variation of the shifted current. Acting with the shift operator $\mathcal{S}_{-1}$ on \eqref{eq:vargencurr} gives
\begin{align}
\label{eq:genshift}
\delta_\Lambda \bra{\cJ_\alpha}&\otimes\mathcal{S}_{-1}(T^\alpha)\nonumber\\[1.5mm]
=\,&\big(\langle \partial_\mathcal{J}|\Lambda\rangle\bra{\mathcal{J}_\alpha}+\langle\mathcal{J}_\alpha|\Lambda\rangle\bra{\partial_\Lambda}-\langle\partial_\Lambda|\Lambda\rangle\bra{\cJ_\alpha}\big)\otimes \mathcal{S}_{-1}(T^\alpha)\nonumber\\[1.5mm]
&\,-\eta_{\alpha\beta}\bra{\partial_\Lambda}T^\alpha\ket{\Lambda}\bra{\cJ_\gamma}\otimes [T^\beta,\mathcal{S}_{-1}(T^\gamma)]-\sum\limits_{n\in\mathds{Z}}(n-1)\bra{\partial_\Lambda}T^A_{n-1}\ket{\Lambda}\bra{J_A^{\,n}}\otimes\dK\nonumber\\
&\,+\eta_{\alpha\beta}\bra{\partial_\Lambda}T^\alpha\ket{\Lambda}\bra{\partial_\Lambda}\otimes(\rho^{-2}\cM^{-1}\mathcal{S}_{1}(T^\beta)^\dagger\cM-\omega_{-1}^\beta(\cM)\dK+\mathcal{S}_{-1}(T^\beta)\big)\,.
\end{align}
The density term and the term involving the explicit sum over the loop algebra are generated by pulling the shift operator inside of the commutator in the second term of the second line, while the appearance of the cocycle is a consequence of using the identity \eqref{eq:coad}. The variation of the constrained field $\bra{\chi}$ is chosen to be
\begin{align}
\label{eq:chigenvar}
\delta_\Lambda\bra{\chi}=\,&\mathscr{L}_\Lambda\bra{\chi}-\bra{\partial_\Lambda}\big(L_{-1}+\rho^{-2}L_{1}-\eta_{\alpha\beta}\,\omega_{-1}^\alpha(\cM) \,T^\beta\big)\ket{\Lambda}\bra{\partial_\Lambda}\nonumber\\[1.5mm]
=\,&\mathscr{L}_\Lambda\bra{\chi}-\bra{\partial_\Lambda}\big(L_{-1}+\cM^{-1}L^\dagger_{-1}\,\cM)\ket{\Lambda}\bra{\partial_\Lambda}\,.
\end{align}
with the Lie derivative 
\begin{align}
\mathscr{L}_\Lambda\bra{\chi}=\,\langle\partial_\chi|\Lambda\rangle\bra{\chi}+\langle\chi|\Lambda\rangle\bra{\partial_\Lambda}-\langle\partial_\Lambda|\Lambda\rangle\bra{\chi}+\sum\limits_{n\in\mathds{Z}}(n-1)\bra{\partial_\Lambda}T^A_{n-1}\ket{\Lambda}\bra{J^{\,n}_A}\, . 
\end{align}
The Lie derivative is determined according to \eqref{eq:Lieformal} and the linearisation of the $E_9$ action \eqref{eq:chiglotrans} on the field $\langle \chi |$,
\begin{align} 
 \delta_X \bra{\chi} \equiv X_\alpha\, \delta^\alpha  \bra{\chi} = \bra{\chi} X_\alpha T^\alpha +\eta^{AB} \sum\limits_{n\in\mathds{Z}}(n-1) X^{1-n}_A \bra{J^{\,n}_B}  \quad\quad (X\in \mf{e}_9\,\textrm{only}) \ ,
\end{align}
which follows from using the section constraint and the linearisation of the cocycle \eqref{eq:cocexp}. Note furthermore that the non-covariant variations in \eqref{eq:chigenvar} are consistent with the property that $\langle \chi|$ transforms as part of the dual of the extended representation \eqref{eq:coadext} which includes $L_{-1}$. Combining \eqref{eq:genshift} and \eqref{eq:chigenvar} yields for the shifted
\begin{align}
\label{eq:varshiftcur}
\delta_{\Lambda} \bra{\cJ^-_\alpha}\otimes T^\alpha &=\,\mathscr{L}_\Lambda\bra{\cJ_\alpha^-}\otimes T^\alpha+ \bra{\partial_\Lambda}T^\alpha\ket{\Lambda}\bra{\partial_\Lambda}\otimes\big(\eta_{-1\,\alpha\beta}\,T^\beta+\eta_{1\,\alpha\beta}\,\rho^{-2}\cM^{-1}T^{\beta\,\dagger}\cM\big)\,, \nn \\[1.5mm]
\delta_{\Lambda} \bra{\cJ^-_\alpha} &=\,\mathscr{L}_\Lambda\bra{\cJ_\alpha^-} +  \eta_{-1\alpha\beta} \bra{\partial_\Lambda}\bigl( T^\beta + \cM^{-1} T^{\beta\,\dagger}\cM\big) \ket{\Lambda}\bra{\partial_\Lambda}\,,
\end{align}
where the  non-covariant terms were recombined using \eqref{eq:etashiftid} and \eqref{etanCovE9} in the first and second line, respectively. Due to the $E_9$ covariance \eqref{eq:scurglotrans} of the shifted current, its Lie derivative simply reads
\begin{align}
\label{eq:Liescur}
\mathscr{L}_\Lambda\bra{\mathcal{J}^-_\alpha}\otimes T^\alpha=\,&\big(\langle \partial_\mathcal{J}|\Lambda\rangle\bra{\cJ^-_\alpha}+\langle\mathcal{J}^-_\alpha|\Lambda\rangle\bra{\partial_\Lambda}-\langle\partial_\Lambda|\Lambda\rangle\bra{\cJ^-_\alpha}\big)\otimes T^\alpha\nonumber\\[1.5mm]
&\,-\eta_{\alpha\beta}\bra{\partial_\Lambda}T^\alpha\ket{\Lambda}\bra{\cJ^-_\gamma}\otimes [T^\beta,T^\gamma]\,,
\end{align}
and matches that of the regular current up to a density term.

Having established the necessary transformation rules of the various fields under generalised diffeomorphisms, we now move on to proving the invariance of the potential \eqref{eq:pot} at $\tilde\rho=0$. Its variation takes the form
\begin{equation}
\delta_{\Lambda} \pot=\,\mathscr{L}_\Lambda\pot+\Delta_\Lambda\pot=\bra{\partial}\big(\ket{\Lambda}\pot\big)+\Delta_{\Lambda}\pot\,.
\end{equation}
As was shown in \eqref{eq:potscale}, the potential is an $E_9$ scalar of weight one. As a result, we immediately deduce that the generalised Lie derivative of the potential is a total derivative. In the following we then exclusively focus on the non-covariant variations $\Delta_\Lambda\pot$. From \eqref{eq:vargencurr}, \eqref{eq:varcurcomp} and \eqref{eq:varJ0} one computes that
\begin{align}
\Delta_{\Lambda}\mathcal{L}_1=&\,4\,\rho^{-1}\bra{\partial_\Lambda}T^\alpha\ket{\Lambda}\bra{\partial_\Lambda}\cM^{-1}\ket{\cJ_\alpha}\,,\\[1.5mm]
\Delta_{\Lambda}\mathcal{L}_4=&\,-2\,\rho^{-1}\langle\partial_\Lambda|\Lambda\rangle\bra{\partial_\Lambda}T^\alpha\cM^{-1}\ket{\cJ_\alpha}\nonumber\\[1.5mm]
&\,-\rho^{-1}\langle\partial_\Lambda|\Lambda\rangle\bra{J_0}\cM^{-1}\ket{\partial_\Lambda}+\rho^{-1}\langle J_0|\Lambda\rangle\bra{\partial_\Lambda}\cM^{-1}\ket{\partial_\Lambda}\,,
\end{align}
where we used \eqref{eq:curid} in the variation of $\mathcal{L}_1$ and the section constraint \eqref{eq:SC1} in the variation of $\mathcal{L}_4$. For the second term in the potential, we find
\begin{align}
\Delta_\Lambda\mathcal{L}_2=&\,2\,\rho^{-1}\eta_{\alpha\beta} \bra{\partial_\Lambda}T^\alpha\ket{\Lambda}\bra{\partial_\Lambda}T^\gamma\cM^{-1}\big(T^{\beta\,\dagger}+\cM\, T^\beta\cM^{-1}\big)\ket{\cJ_\gamma} \nonumber\\[1.5mm]
=&\,2\,\rho^{-1}\langle{\cJ_\alpha}|\Lambda\rangle\bra{\partial_\Lambda}T^\alpha\cM^{-1}\ket{\partial_\Lambda}-2\rho^{-1}\langle{\partial_\Lambda}|\Lambda\rangle\bra{\partial_\Lambda}T^\alpha\cM^{-1}\ket{\cJ_\alpha}\nonumber\\[1.5mm]
&\,+2\,\rho^{-1} \eta_{\alpha\beta} \bra{\partial_\Lambda}T^\alpha\ket{\Lambda}\bra{\partial_\Lambda} [ T^\gamma , T^\beta] \cM^{-1}\ket{\cJ_\gamma} \,.
\end{align}
The section constraint  \eqref{eq:SC1} was used on both terms to simplify the first line. Using \eqref{eq:varshiftcur}, the variation of $\mathcal{L}_3$ reads
\begin{align}
\Delta_\Lambda\mathcal{L}_3=&\,2\,\rho^{-1}\bra{\partial_\Lambda}T^\alpha\ket{\Lambda}\bra{\partial_\Lambda}T^\gamma\,T^\beta\cM^{-1}\ket{\cJ^-_\gamma}\eta_{1\,\alpha\beta}\nonumber\\[1.5mm]
=&\,2\,\rho^{-1}\bra{\partial_\Lambda}T^\alpha\ket{\Lambda}\bra{\partial_\Lambda}\big[T^\gamma,T^\beta\big]\cM^{-1}\ket{\cJ^-_\gamma}\eta_{1\,\alpha\beta}\nonumber\\[1.5mm]
=&\,2\,\rho^{-1}\eta_{\alpha\beta} \bra{\partial_\Lambda}T^\alpha\ket{\Lambda}\bra{\partial_\Lambda}\big[\mathcal{S}_{-1}(T^\gamma),\mathcal{S}_1(T^\beta)\big]\cM^{-1}\ket{\cJ_\gamma}\nonumber\\[1.5mm]
=&\,2\,\rho^{-1}\eta_{\alpha\beta} \bra{\partial_\Lambda}T^\alpha\ket{\Lambda}\bra{\partial_\Lambda} [ T^\gamma , T^\beta ] \cM^{-1}\ket{\cJ_\gamma}\, \nonumber\\[1.5mm]
&-2\,\rho^{-1}\bra{\partial_\Lambda}T^\alpha\ket{\Lambda}\bra{\partial_\Lambda}\cM^{-1}\ket{\cJ_\alpha}+\,2\,\rho^{-1}\langle{\partial_\Lambda}|\Lambda\rangle\bra{\partial_\Lambda}T^\alpha\cM^{-1}\ket{\cJ_\alpha}\,.
\end{align}
The section constraints \eqref{eq:SC2} and \eqref{eq:SC3} were used in the first and second line, respectively, and in the last line we used that for $\alpha$ and $\beta$ restricted to $\mf{e}_9$ one has 
\be [ \mathcal{S}_{-1}(T^\alpha),\mathcal{S}_1(T^\beta)] =  [ T^\alpha , T^\beta] - \eta^{\alpha\beta} \dK - \delta_0^\alpha T^\beta - \delta_0^\beta T^\alpha \ , \ee
with the Kronecker symbol defined such that $\delta_0^\alpha \langle \mathcal{J}_\alpha |= \langle J_0|$, as well as the section constraint \eqref{eq:SC1}. 

 Combining the above non-covariant variations, we find 
\begin{align}
\Delta_\Lambda\pot\,=&\frac12 \rho^{-1}\,\langle\partial_\Lambda|\Lambda\rangle\bra{\partial_\Lambda}\big(2\,T^\alpha\cM^{-1}\ket{\cJ_\alpha}-\cM^{-1}\ket{J_0}\big)\nonumber\\[1.5mm]
&\,- \rho^{-1}\,\langle\cJ_\alpha|\Lambda\rangle\bra{\partial_\Lambda}T^\alpha\cM^{-1}\ket{\partial_\Lambda}+\frac12 \rho^{-1}\langle J_0|\Lambda\rangle\bra{\partial_\Lambda}\cM^{-1}\ket{\partial_\Lambda}\label{eq:respot}
\end{align}
which, upon using \eqref{eq:cJdef2} and \eqref{eq:defJ0}, reduces to a total derivative
\begin{equation}
\Delta_\Lambda\pot= \bra{\partial}\big(\rho^{-1}\ket{\Lambda}\bra{\partial_\Lambda}\cM^{-1}\ket{\partial_\Lambda}-\rho^{-1}\cM^{-1}\ket{\partial_\Lambda}\langle\partial_\Lambda|\Lambda\rangle\big)\,.
\end{equation}
This proves that the potential \eqref{eq:pot} is invariant, at $\tilde\rho=0$, under generalised diffeomorphisms up to total derivatives.

\subsection{The potential at $\tilde\rho\neq 0$}
\label{sec:pottilderho}
We shall now present the general expression of the potential \eqref{eq:pot} at $\tilde\rho\neq0$. In this case, $\cM\in\hat E_8\rtimes SL(2)$ and the various terms read
\begin{subequations}
\label{eq:potterm}
\begin{align}
\mathcal{L}_1=&\,\rho^{-1}\sum\limits_{n\in\mathds{Z}}\sum\limits_{k=0}^\infty\tilde\rho^k\langle J_A^{\,k-n} |\mathcal{M}^{-1}\ket{J^{\,n}_B}\eta^{AB}-2\,\rho^{-1}\bra{J_0+2\,\tilde\rho \,J_1}\cM^{-1}\ket{J_\dK}\nonumber\\[1.5mm]
&\,-2\,\rho^{-1}\bra{J_1}\cM^{-1}\ket{2\,\rho^2\,\chi-\rho^2 J_1 +\Omega^\alpha(\cM)\,\cJ_\alpha}\,,\label{eq:pot1}\\[1.5mm]
\mathcal{L}_2=&\,\rho^{-1}\bra{\mathcal{J}_\alpha}T^\beta\mathcal{M}^{-1}T^{\alpha\,\dagger}\ket{\mathcal{J}_{\beta}}\,,\label{eq:pot2}\\[1.5mm]
\mathcal{L}_3=&\,\rho\,\bra{\mathcal{J}_\alpha^-}T^\beta\mathcal{M}^{-1}T^{\alpha\,\dagger} | \mathcal{J}^-_\beta \rangle \,,\label{eq:pot3}\\[1.5mm]
\mathcal{L}_4=&\,\rho^{-1}\bra{J_0+2\,\tilde\rho \,J_1}T^\alpha\mathcal{M}^{-1}\ket{\mathcal{J}_\alpha}\,\label{eq:pot4}.
\end{align}
\end{subequations}
The currents components $\bra{\cJ_\alpha}$ defined in \eqref{eq:cJdef2} are now non-vanishing along $\mathfrak{e}_8\oleft \mathfrak{sl}(2)$, while $\bra{\cJ^-_\alpha}$ defined in \eqref{eq:shiftcur} has non-vanishing components along all Virasoro generators $L_n$ with $n\le 1$ in $\mathfrak{f}$. The $\mathfrak{sl}(2)$ components of the current are the same as for $SL(2)/SO(1,1)$ 
\begin{subequations}
\label{eq:Jsl2def}
\begin{align}
\bra{J_0}=&\,-2\,\rho^{-1}\bra{\partial_\rho}\rho+2\,\rho^{-2}\,\tilde\rho\bra{\partial_{\tilde\rho}}\tilde\rho\,,\\[1.5mm]
\bra{J_{-1}}=&\,2\,\rho^{-1}\,\tilde\rho\bra{\partial_\rho}\rho-\big(1+\rho^{-2}\tilde\rho^2\big)\bra{\partial_{\tilde\rho}}\tilde\rho\,,\\[1.5mm]
\bra{J_1}=&\,-\rho^{-2}\bra{\partial_{\tilde\rho}}\tilde\rho\,.
\end{align}
\end{subequations}
It is straightforward to check that these components satisfy the identity \eqref{eq:jrelation}. The function $\Omega^\alpha(\cM)$ defined in \eqref{eq:w(M)} is a combination of $E_9$ group cocycles that reduces to the cocycle $\omega_1^\alpha(\cM)$ when $\tilde\rho$ is set to zero. In the following, we will show that each term is invariant under rigid $\hat E_8\rtimes \mathds{R}_{L_{-1}}$  and scales  with weight one one under rigid $\mathds{R}^+_{\dL}$. Then we will show that the combination \eqref{eq:pot} is invariant under $\Lambda$ and $\Sigma$ generalised diffeomorphisms. At this point, the most striking difference with the expression of the potential at $\tilde\rho=0$ is perhaps the complexity of the expression of $\mathcal{L}_1$, and the fact that its rigid $\hat E_8$ invariance is not manifest. This is due to the absence of an invariant bilinear form over $\hat{\mathfrak{e}}_8\oleft\mathfrak{sl}(2)$.

\subsubsection{Rigid $\hat E_8\rtimes ( \mathds{R}^+_{\dL} \ltimes \mathds{R}_{L_{-1}})$ symmetry}
\label{sec:E8L-1inv}
For clarity, we will treat separately the transformations under $ E_9$ and $\mathds{R}_{L_{-1}}$. Under $g\in E_9$, the derivatives, $\cM$ and the field $\rho$ still transform as in \eqref{eq:E9actder}, \eqref{eq:E9actM} and \eqref{eq:E9actrho}, respectively. From the parameterisation of the coset element \eqref{eq:paracoset}, one finds that 
\begin{equation}
\label{eq:tilderhovar}
\tilde\rho\to\rho(g)^2\tilde\rho\,.
\end{equation}
The transformation of the current and its components still take the same form as in the $\tilde\rho=0$ case
\begin{align}
\bra{\cJ_\alpha}\otimes T^\alpha&\to\rho(g)^2 \bra{\cJ_\alpha}g\otimes g^{-1}T^\alpha g\,,\\[1.5mm]
\bra{\cJ_\alpha}&\to \rho(g)^2\,R(g)^\beta{}_\alpha\bra{\cJ_\beta}g\,,\label{eq:Jalvar}
\end{align} 
but now the current is $\mathfrak{e}_8\oleft\, \mathfrak{sl}(2)$-valued. The adjoint representation matrix $R(g)^\alpha{}_\beta$ is still defined from the generators of $\mathfrak{f}$ by \eqref{eq:Rdef'}. Note also that, using \eqref{eq:Jsl2def}, the combination appearing in $\mathcal{L}_4$ reads
\begin{equation}
\bra{J_0+2\,\tilde\rho \,J_1}=\,-2\,\rho^{-1}\bra{\partial_\rho}\rho\,,
\end{equation}
and therefore simply transforms as $\bra{J_0+2\,\tilde\rho \,J_1}\to\rho(g)^2\bra{J_0+2\,\tilde\rho \,J_1}g$ under $E_9$.

The computation of the variation of the shifted current relies on a similar reasoning as for  $\tilde\rho=0$. We start  with the infinite series of shift operators in the expression of the shifted current \eqref{eq:shiftcur}, that transforms under $E_9$ as 
\begin{equation}
\label{eq:genshift2}
\bra{\mathcal{J}_\alpha}\otimes\sum\limits_{k=0}^\infty\tilde\rho^k \mathcal{S}_{-1-k}(T^\alpha)\to\rho(g)^2\bra{\cJ_\alpha}g\,\otimes\sum\limits_{k=0}^\infty\tilde\rho^k\big(\rho(g)^{-2}\,g^{-1}\cS_{-1-k}(T^\alpha)g-\rho(g)^{2k}\,\omega^\alpha_{1+k}(g^{-1})\dK\big)\,,
\end{equation}
where we used \eqref{eq:coad}. Each cocycle $\omega^\alpha_k(g)$, together with the character $\rho(g)$, defines an extension of the adjoint representation of $E_9$ by the generator $L_{-k}$. Once again we choose the variation of the field $\bra{\chi}$ such that the shifted current transforms covariantly under $E_9$,
\begin{equation}
\label{eq:chiglotranst}
\bra{\chi}\to\bra{\chi}g+\sum\limits_{k=0}^\infty\tilde\rho^k\,\rho(g)^{2k+2}\,\omega_{1+k}^\alpha(g^{-1})\bra{\cJ_\alpha}g\,.
\end{equation}
Indeed, it follows from \eqref{eq:genshift2} and \eqref{eq:chiglotranst} that the shifted current transforms as 
\begin{equation}
\bra{\cJ_\alpha^-}\otimes T^\alpha\to\bra{\cJ_\alpha^-}g\otimes g^{-1}T^\alpha g\,,
\end{equation} 
which admits non-zero components along all Virasoro generators $L_n$ for $n\le 1$. Using the above results, it is straightforward to check that under $E_9$, the terms $\mathcal{L}_2$, $\mathcal{L}_3$ and $\mathcal{L}_4$ only scale by a factor $\rho(g)^2$. The term $\mathcal{L}_1$ is more complicated, and we will only show its invariance under infinitesimal $\mf{e}_9$ transformations. To begin with we first show invariance of $\mathcal{L}_2$, $\mathcal{L}_3$ and $\mathcal{L}_4$ under infinitesimal $E_9$ and tackle $\mathcal{L}_1$ afterwards.

Under $e^{X_- L_{-1}}\in\mathds{R}_{L_{-1}}$, we have the following transformations
\begin{align}
\cM\to&\,e^{X_- L_{1}}\cM \,e^{X_- L_{-1}}\,,\nonumber\\[1.5mm]
\bra{\partial}\to&\,\bra{\partial}e^{X_- L_{-1}}\,,\nonumber\\[1.5mm]
\rho\to&\,\rho\,,\label{eq:L-1chivar}\\[1.5mm]
\tilde\rho\to&\,\tilde\rho-X_-\,\nonumber\\[1.5mm]
\bra{\chi}\to&\bra{\chi}e^{X_- L_{-1}}\,.\nn
\end{align}
This implies
\begin{subequations}
\begin{align}
\bra{\cJ_\alpha}\otimes T^\alpha\to&\,\bra{\cJ_\alpha}e^{X_- L_{-1}}\otimes \,e^{-X_- L_{-1}}\,T^\alpha \,e^{X_- L_{-1}}\,,\label{eq:L-1varcur}\\[1.5mm]
\bra{\cJ_\alpha^-}\otimes T^\alpha\to&\,\bra{\cJ_\alpha^-}e^{X_- L_{-1}}\otimes \,e^{-X_- L_{-1}}\,T^\alpha \,e^{X_- L_{-1}}\,.\label{eq:L-1varscur}
\end{align}
\end{subequations}
The covariance of the term involving the infinite series of shift operators in the expression of the shifted current $\langle \cJ_\alpha^-|$ can be verified using 
\be   \langle \cJ_\alpha | \otimes \sum_{k=0}^\infty \tilde{\rho}^k \cS_{-1-k}(T^\alpha) =  \langle \cJ_\alpha | \otimes e^{\tilde \rho L_{-1}} \cS_{-1}( e^{-\tilde \rho L_{-1}} T^\alpha e^{\tilde \rho L_{-1}}) e^{-\tilde \rho L_{-1}}  \ .\label{eq:L-1shiftid} \,\ee
which follows from \eqref{GeneratingInfinitShift}. The $\mathds{R}_{L_{-1}}$ invariance of $\mathcal{L}_2$, $\mathcal{L}_3$ and $\mathcal{L}_4$ is then a direct consequence of \eqref{eq:L-1chivar}, \eqref{eq:L-1varcur} and \eqref{eq:L-1varscur}.

Let us finally consider the transformation of $\mathcal{L}_1$ under $E_9$ and $\mathds{R}_{L_{-1}}$. As mentioned previously, this is more easily tackled by considering the infinitesimal variations of the current components. Under the infinitesimal variation of parameter $X$ 
\begin{equation}
X=\,X_\dK \,\dK+X_0\,\dL+\sum\limits_{n\in\mathds{Z}}X^n_A\, T^A_n+X_- L_{-1}\,,\label{eq:Xgen}
\end{equation}    
one has the variations
\be \delta_X  \cM^{-1} = - X_\alpha T^\alpha \cM^{-1} - \cM^{-1}X_\alpha T^{\alpha \dagger} \ , \qquad  \delta_X \bra{\partial} = \bra{\partial} ( X_\alpha T^\alpha-X_0) \ , \label{deltaXM}  \ee
with by definition 
\begin{equation}
X_\alpha T^\alpha =\,X_\dK \,\dK+X_0\,L_0+\sum\limits_{n\in\mathds{Z}}X^n_A\, T^A_n+X_- L_{-1}\, . 
\end{equation}    
One obtains for  the current components
\begin{align}
\label{eq:infvarcur}
\delta_X\bra{J_A^{\,n}}=\,&\sum\limits_{m\in\mathds{Z}}X_B^{n-m}\bra{J_C^{\,m}}f^{BC}{}_A-n\,X_A^n\bra{J_0}+(n-1)\,X_0\bra{J_A^{\,n}}\nonumber\\
&\,-(n+1)\,X_A^{n+1}\bra{J_{-1}}-(n-1)\,X_A^{n-1}\bra{J_1}+(n+1)\,X_{-}\bra{J_A^{\,n+1}}+\bra{J_A^{\,n}}X_\alpha T^\alpha\,,\nonumber\\[1.5mm]
\delta_X\bra{J_\dK}=\,&\sum\limits_{m\in\mathds{Z}}m\,\eta^{AB}X_A^{-m}\bra{J_B^{\,m}}-X_0\bra{J_\dK}+\bra{J_\dK}X_\alpha T^\alpha\,,\nonumber\\
\delta_X\bra{J_0}=\,&-X_0\bra{J_0}+2\,X_-\bra{J_1}+\bra{J_0}X_\alpha T^\alpha\,,\nonumber\\[1.5mm]
\delta_X\bra{J_{-1}}=\,&-2\,X_0\bra{J_{-1}}+X_-\bra{J_0}+\bra{J_{-1}}X_\alpha T^\alpha\,,\nonumber\\[1.5mm]
\delta_X\bra{J_1}=\,&\bra{J_1}X_\alpha T^\alpha\, . 
\end{align}
The last term in $X_\alpha T^\alpha$ of each expression comes from the expansion of $g\in E_9$ and $e^{X_-L_{-1}}$ acting on the derivative bra of $\bra{\cJ_\alpha}$ in the $\overline{R(\Lambda_0)}_0$ representation as  in \eqref{eq:Jalvar} and \eqref{eq:L-1varcur}. These contributions trivially cancel the variation \eqref{deltaXM} of $\cM^{-1}$. All the other terms follow from the linearisation of the character $\rho(g)$ and the adjoint representation matrix $R(g)^\alpha{}_\beta$ defined in \eqref{eq:rhoX} and \eqref{eq:Rdef'}, respectively. For the infinitesimal variation of the constrained field $\bra{\chi}$ under $E_9$ and $\mathds{R}_{L_{-1}}$, we obtain from \eqref{eq:chiglotranst} and \eqref{eq:L-1chivar},
\begin{equation}
\label{eq:varchiE9}
\delta_X\bra{\chi}=\,\sum\limits_{n\in\mathds{Z}}\sum\limits_{k=0}^\infty\tilde\rho^k(n-1-k)\,\eta^{AB}X_A^{1+k-n}\bra{J_B^n}+\bra{\chi}\,X_\alpha T^\alpha\,,
\end{equation}
using the linearisation \eqref{eq:cocexp} of the cocyles $\omega^\alpha_n(g)$. With \eqref{eq:infvarcur}, we find that the infinitesimal variation of the first two term of $\mathcal{L}_1$ gives
\begin{align}
\label{eq:varL12}
\delta_X&\Big(\rho^{-1}\sum\limits_{n\in\mathds{Z}}\sum\limits_{k=0}^\infty\tilde\rho^k\bra{J_A^{k-n}}\mathcal{M}^{-1}\ket{J^{\,n}_B}\eta^{AB}-2\,\rho^{-1}\bra{J_0+2\,\tilde\rho \,J_1}\cM^{-1}\ket{J_\dK}\Big)\nonumber\\[1.5mm]
=&\,-2\,\rho^{-1}\sum\limits_{n\in\mathds{Z}}\Big(\sum\limits_{k=0}^\infty\tilde\rho^k\rho^2(k-n+1)X_A^{k+1-n} -(n+1)X_A^{-n-1}+n\,\tilde\rho X_A^{-n}\Big)\bra{J_B^{\,n}}M^{-1}\ket{J_1}\eta^{AB}\nonumber\\[1.5mm]
&\,-X_0\Big(\rho^{-1}\sum\limits_{n\in\mathds{Z}}\sum\limits_{k=0}^\infty\tilde\rho^k\bra{J_A^{k-n}}\mathcal{M}^{-1}\ket{J^{\,n}_B}\eta^{AB}-2\,\rho^{-1}\bra{J_0+2\,\tilde\rho \,J_1}\cM^{-1}\ket{J_\dK}\Big)\,.
\end{align}
where we used \eqref{eq:jrelation} to eliminate all the dependence on the components $\bra{J_0}$ and $\bra{J_{-1}}$ in the first line. Note in particular the invariance of the above combination under $\mathds{R}_{L_{-1}}$. For the remaining term in $\mathcal{L}_1$, we need to consider the variation of $\Omega^\alpha(\cM)\bra{\cJ_\alpha}$. We start from the expression
\begin{equation}
\label{eq:OmJdef}
 \Omega^\alpha(\cM)\,\bra{\cJ_\alpha}\otimes\dK=\,\bra{\cJ_\alpha}\otimes\Big(\rho^2\cM^{-1}\sum\limits_{k=0}^\infty\tilde\rho^k\mathcal{S}_{-1-k}(T^\alpha)^\dagger\cM+\tilde\rho\, T^\alpha-\mathcal{S}_1(T^\alpha)-\tilde\rho\,\delta^\alpha_\dK\dK\Big)\,,
\end{equation}
which is obtain by using \eqref{eq:w(M)} and \eqref{eq:curid}.
With \eqref{eq:L-1shiftid}, one can show that the above combination transforms as
\begin{equation}
\label{eq:L-1transOm}
\Omega^\alpha(\cM)\bra{\cJ_\alpha}\to\Omega^\alpha(\cM)\bra{\cJ_\alpha}e^{X_- L_{-1}}\,,
\end{equation}
under $\mathds{R}_{L_{-1}}$, while under $g\in E_9$ one finds
\begin{align}
\label{eq:transOmE9}
\Omega^\alpha(\cM)\bra{\cJ_\alpha}\to\,&\rho(g)^4\,\Omega^\alpha(\cM)\bra{\cJ_\alpha}g-\sum\limits_{k=0}^\infty\tilde\rho^k\rho(g)^{2k+6}\,\omega_{1+k}^\alpha(g^{-1})\bra{\cJ_\alpha}g\nonumber\\[1.5mm]
&\,+\rho(g)^2\,\omega_{-1}(g^{-1})\bra{\cJ_\alpha}g-\tilde\rho\,\rho(g)^4R(g)^\alpha{}_\dK\bra{\cJ_\alpha}g\,.
\end{align}
using \eqref{eq:coad}. By linearising \eqref{eq:L-1transOm} and \eqref{eq:transOmE9} using \eqref{eq:cocexp}, we find the following infinitesimal variation 
\begin{align}
\delta_X\Big(\Omega^\alpha(\cM)\bra{\cJ_\alpha}\Big)=\,&-\sum\limits_{n\in\mathds{Z}}\Big(\sum\limits_{k=0}^\infty\tilde\rho^k\rho^2 (n-1-k) X_B^{k+1-n}-(n+1)X_B^{-n-1}+n\,\tilde\rho X_B^{-n}\Big)\eta^{AB}\bra{\cJ_\alpha}\nonumber\\[1.5mm]
&-2\,X_0\,\Omega^\alpha(\cM)\bra{\cJ_\alpha}+\Omega^\alpha(\cM)\bra{\cJ_\alpha}X_\beta T^\beta\,.
\end{align}
With this result and \eqref{eq:varchiE9}, it is easy to verify that the last term of $\mathcal{L}_1$ is $\mathds{R}_{L_{-1}}$ invariant, while its $\hat E_8$ variation cancels out that of the first two terms in \eqref{eq:varL12}. We are then left with
\begin{equation}
\delta_X \mathcal{L}_1=\,-X_0\,\mathcal{L}_1\,,
\end{equation} 
which is the action of $\dL$ on $\mathcal{L}_1$. The full potential is then $\hat E_8\rtimes \mathds{R}_{L_{-1}}$ invariant and, as in the $\tilde\rho=\,0$ case \eqref{eq:potscale}, uniformly scales by a factor $\rho(g)^2$ under finite transformations generated~by~$\dL$.

\subsubsection{Invariance under generalised diffeomorphisms}
\label{sec:gendifftilderho}
The fields $\cM\in\hat E_8\rtimes SL(2)$  and $\rho$ still transform covariantly under generalised diffeomorphisms, {\it i.e.} as in \eqref{eq:gendiffM} and \eqref{eq:gendiffr}, respectively. According to \eqref{eq:tilderhovar}, the field $\tilde\rho$ is an $E_9$ scalar density of weight one and thus also transforms as a total derivative,
\begin{equation}
\delta_{\Lambda}\,\tilde\rho=\,\mathscr{L}_\Lambda\,\tilde\rho=\,\bra{\partial}\big(\ket{\Lambda}\tilde\rho\big)\,.
\end{equation}
In the previous section, we have shown that the full potential also transforms as an $E_9$ scalar of weight one. The $ \mathfrak{\hat e}_8\oleft\mathfrak{sl}(2)$-valued current $\bra{\cJ_\alpha}\otimes T^\alpha$ still transforms as in \eqref{eq:vargencurr} and \eqref{eq:Liecur}. In particular, its non-covariant variation still reads 
\begin{equation}
\Delta_\Lambda \bra{\cJ_\alpha}\otimes T^\alpha=\,\eta_{\alpha\beta}\bra{\partial_\Lambda}T^\alpha\ket{\Lambda}\bra{\partial_\Lambda}\otimes(T^\beta + \cM^{-1}T^{\beta\,\dagger}\cM \big)\,.
\end{equation} 
However, the bilinear form $\eta_{\alpha\beta}$ is not invariant under the extended group $\hat E_8\rtimes SL(2)$, so that the non-covariant variation of the current components, according to \eqref{eq:dreamform},
 is then 
\begin{equation}
\label{eq:difcurcomp}
\Delta_\Lambda\bra{\cJ_\alpha}= \bra{\partial_\Lambda} \Bigl( \eta_{\alpha\beta} T^\beta +\Big[\frac{\tilde\rho}{\rho^2}\,\eta_{1\,\alpha\beta}+\Big(1-2\,\frac{\tilde\rho^2}{\rho^2}\big)\,\eta_{\alpha\beta}-\tilde\rho\,\Big(1-\frac{\tilde\rho^2}{\rho^2}\Big)\eta_{-1\,\alpha\beta}\Big] \cM^{-1}  T^{\beta \dagger} \cM  \Bigr)  \ket{\Lambda}\bra{\partial_\Lambda}\;  .
\end{equation}
The Lie derivative of the current components is still given by \eqref{eq:Liecurcomp}, although the structure constant indices now take values over $\hat{\mathfrak{e}}_8\oleft \mf{sl}(2)$.

Let us now turn to the variation of the shifted current. 
Using \eqref{eq:etashiftid} and \eqref{eq:difcurcomp}, one obtains that the non-covariant variation of the shifted current  $\langle \cJ_\alpha^-|$ defined in \eqref{eq:shiftcur} for $\alpha \ne \dK$, transforms as
\bea
\label{eq:varshiftcur2}
 \Delta_\Lambda\bra{\cJ_\alpha^-}_{\alpha\neq\dK}  \hspace{-1mm}  &=& \hspace{-3mm}  \sum_{k=0}^\infty \tilde{\rho}^k \bra{\partial_\Lambda} \biggl( \eta_{-1-k\, \alpha\beta}\, T^\beta \nn\\
 && \hspace{1mm} +\Big[\frac{\tilde\rho}{\rho^2}\,\eta_{-k\,\alpha\beta}+\Big(1-2\,\frac{\tilde\rho^2}{\rho^2}\Big)\,\eta_{-1-k\, \alpha\beta}-\tilde\rho\,\Big(1-\frac{\tilde\rho^2}{\rho^2}\Big)\eta_{-2-k\,\alpha\beta}\Big] \cM^{-1}  T^{\beta \dagger} \cM  \biggr)  \ket{\Lambda}\bra{\partial_\Lambda} \nn \\
 \hspace{-1mm}   &=& \hspace{-2mm}   \bra{\partial_\Lambda} \Bigl(  \sum_{k=0}^\infty \tilde{\rho}^k \eta_{-1-k\,  \alpha\beta} T^\beta  + \Bigl[ \frac{ \tilde \rho}{\rho^2} \eta_{\alpha\beta} + \Big(1-\frac{\tilde\rho^2}{\rho^2} \Bigr) \eta_{-1\, \alpha\beta}\Big] \cM^{-1}  T^{\beta \dagger} \cM \Bigr)  \ket{\Lambda}\bra{\partial_\Lambda} \; . 
\eea
It is therefore natural to define the non-covariant variation of $\langle \chi|$ such that this transformation rule also applies for $\alpha=\dK$. We then choose
\begin{align}
\label{eq:Delchi}
\Delta_\Lambda\bra{\chi}=\,& -\bra{\partial_\Lambda}\Bigl( \sum\limits_{k=0}^\infty\tilde\rho^k L_{-1-k} + \cM^{-1} \Bigl[ \frac{\tilde{\rho}}{\rho^2} L_0 + \Bigl( 1-\frac{\tilde{\rho}^2}{\rho^2} \Bigr) L_1 \Bigr] \cM  \Bigr) \ket{\Lambda}\bra{\partial_\Lambda}  -\frac{\tilde{\rho}}{\rho^2}  \langle \partial_\Lambda | \Lambda \rangle \langle \partial_\Lambda |  \,.
\end{align}
The presence of the last term does not follow from the previous argument, but we observe that it could be removed by redefining $\langle \chi | \rightarrow \langle \chi | - \langle J_1 |$, since \mbox{$\Delta_\Lambda \langle J_1 | = - \frac{\tilde{\rho}}{\rho^2}  \langle \partial_\Lambda | \Lambda \rangle \langle \partial_\Lambda |  $}. Such a redefinition would not modify the rigid transformation rule of $\bra{\chi}$ under $\hat{E}_8 \rtimes (\mathds{R}_\dL^+ \ltimes \mathds{R}_{L_{-1}})$, but would lead to a slightly less compact expression for the potential. The Lie derivative of $\bra{\chi}$ follows from its infinitesimal variation \eqref{eq:varchiE9} under $E_9$ and reads
\begin{equation}
\mathscr{L}_\Lambda\bra{\chi}=\,\langle\partial_\chi|\Lambda\rangle\bra{\chi}+\langle\chi|\Lambda\rangle\bra{\partial_\Lambda}-\langle\partial_\Lambda|\Lambda\rangle\bra{\chi}+\sum\limits_{n\in\mathds{Z}}\sum\limits_{k=0}^\infty(n-1-k)\bra{\partial_\Lambda}T^A_{n-1-k}\ket{\Lambda}\bra{J^{\,n}_A}\,.
\end{equation}
Combining \eqref{eq:varshiftcur2} and \eqref{eq:Delchi} and using \eqref{eq:dreamform}, one finds for the shifted current 
\begin{align}
\Delta_\Lambda\bra{\cJ_\alpha^-} = \, &  \sum_{k=0}^\infty \tilde{\rho}^k \eta_{-1-k\,  \alpha\beta} \bra{\partial_\Lambda} T^\beta \ket{\Lambda}\bra{\partial_\Lambda}  - \delta_\alpha^{\dK} \frac{\tilde{\rho}}{\rho^2}  \langle \partial_\Lambda | \Lambda \rangle \langle \partial_\Lambda |  \nn\\
&\; + \Bigl( \frac{ \tilde \rho}{\rho^2} \eta_{\alpha\beta} + \Big(1-\frac{\tilde\rho^2}{\rho^2} \Bigr) \eta_{-1\, \alpha\beta}\Big)  \bra{\partial_\Lambda}   \cM^{-1}  T^{\beta \dagger} \cM  \ket{\Lambda}\bra{\partial_\Lambda} \nn\\
\Delta_\Lambda\bra{\cJ_\alpha^-}\otimes T^\alpha=\,&\sum\limits_{k=0}^\infty\tilde\rho^k\,\eta_{-1-k\,\alpha\beta}\bra{\partial_\Lambda}T^\alpha\ket{\Lambda}\bra{\partial_\Lambda}\otimes T^\beta-\frac{\tilde\rho}{\rho^2}\langle\partial_\Lambda|\Lambda\rangle\bra{\partial_\Lambda}\otimes \dK\nonumber\\
&\,+\frac{1}{\rho^2}\big(\eta_{1\,\alpha\beta}-\tilde\rho\,\eta_{\alpha\beta}\big)\bra{\partial_\Lambda}T^\beta\ket{\Lambda}\bra{\partial_\Lambda}\otimes \cM^{-1} T^{\beta\dagger}\cM\,,  \label{DeltaJminus} 
\end{align}
while its Lie derivative is still given by the expression \eqref{eq:Liescur}.

Let us now consider the variation of each term in the potential. Just as in the $\tilde\rho=\,0$ case, we will only focus on the non-covariant variations $\Delta_\Lambda \pot$ as the Lie derivative of the potential reduces to a total derivative. This is ensured by the fact that, as proved in section \ref{sec:E8L-1inv}, the potential transforms as an $E_9 $ scalar of weight one. For $\mathcal{L}_2$ and $\mathcal{L}_4$, the computation is the same as in the $\tilde\rho=\,0$ case and one finds
\begin{align}
\Delta_\Lambda\mathcal{L}_2=\,&2\,\rho^{-1}\langle{\cJ_\alpha}|\Lambda\rangle\bra{\partial_\Lambda}T^\alpha\cM^{-1}\ket{\partial_\Lambda}-2\rho^{-1}\langle{\partial_\Lambda}|\Lambda\rangle\bra{\partial_\Lambda}T^\alpha\cM^{-1}\ket{\cJ_\alpha}\label{eq:fvarL2} \nn \\[1.5mm]
&\,+2\,\rho^{-1}\bra{\partial_\Lambda}T^\alpha\ket{\Lambda}\bra{\partial_\Lambda}[ T^\gamma , T^\beta] \cM^{-1}\ket{\cJ_\gamma}\,\eta_{\alpha\beta} \,,\\[1.5mm]
\Delta_{\Lambda}\mathcal{L}_4=\,&\,-2\,\rho^{-1}\langle\partial_\Lambda|\Lambda\rangle\bra{\partial_\Lambda}T^\alpha\cM^{-1}\ket{\cJ_\alpha}\nonumber\\[1.5mm]
&\,-\rho^{-1}\langle\partial_\Lambda|\Lambda\rangle\bra{J_0+2\,\tilde\rho\,J_1}\cM^{-1}\ket{\partial_\Lambda}+\rho^{-1}\langle J_0+2\,\tilde\rho \,J_1|\Lambda\rangle\bra{\partial_\Lambda}\cM^{-1}\ket{\partial_\Lambda}\,.\label{eq:fvarL4}
\end{align} 
For $\mathcal{L}_3$, one obtains that the first infinite sum in \eqref{DeltaJminus} gives terms that all vanish according to the section constraints, while the others give
\begin{align}
\Delta_\Lambda\mathcal{L}_3=\,&2\,\rho^{-1}\bra{\partial_\Lambda}T^\alpha\ket{\Lambda}\bra{\partial_\Lambda}T^\gamma\,T^\beta\cM^{-1}\ket{\cJ^-_\gamma}(\eta_{1\,\alpha\beta}-\tilde\rho\,\eta_{\alpha\beta})\nonumber\\[1.5mm]
&\,-2\,\rho^{-1}\tilde\rho\,\langle\partial_\Lambda|\Lambda\rangle\bra{\partial_\Lambda}T^\alpha\cM^{-1}\ket{\cJ_\alpha^-}\nonumber\\[2mm]
=\,&2\,\rho^{-1}\bra{\partial_\Lambda}T^\alpha\ket{\Lambda}\bra{\partial_\Lambda}[T^\gamma,T^\beta]\,\cM^{-1}\ket{\cJ^-_\gamma}(\eta_{1\,\alpha\beta}-\tilde\rho\,\eta_{\alpha\beta})\nonumber\\[1.5mm]
&\,-2\,\rho^{-1}\sum\limits_{k=1}^\infty\tilde\rho^k\,\langle\partial_\Lambda|\Lambda\rangle\bra{\partial_\Lambda}\mathcal{S}_{-k}(T^\alpha)\,\cM^{-1}\ket{\cJ_\alpha}-2\,\rho^{-1}\tilde\rho\,\langle\partial_\Lambda|\Lambda\rangle\bra{\partial_\Lambda}\cM^{-1}\ket{\chi}\nonumber\\[1.5mm]
=\,&2\,\rho^{-1}\bra{\partial_\Lambda}T^\alpha\ket{\Lambda}\bra{\partial_\Lambda}[T^\gamma , T^\beta] \cM^{-1}\ket{\cJ_\gamma}\,\eta_{\alpha\beta}\,\nonumber\\[1.5mm]
&\,-2\,\rho^{-1}\sum\limits_{k=0}^\infty\tilde\rho^k\bra{\partial_\Lambda}\mathcal{S}_{-k}(T^\alpha)\ket{\Lambda}\bra{\partial_\Lambda}\cM^{-1}\ket{\cJ_\alpha}\nonumber\\[1.5mm]
&+\,2\,\rho^{-1}\langle{\partial_\Lambda}|\Lambda\rangle\bra{\partial_\Lambda}T^\alpha\cM^{-1}\ket{\cJ_\alpha}-2\,\rho^{-1}\tilde\rho\,\langle\partial_\Lambda|\Lambda\rangle\bra{\partial_\Lambda}\cM^{-1}\ket{\chi}\,,\label{eq:fvarL3}
\end{align}
where we used the following identity in the last step
\begin{align}
& \sum\limits_{k=0}^\infty\tilde\rho^k\bra{\partial_\Lambda}T^\alpha\ket{\Lambda} \bra{\partial_\Lambda}[\mathcal{S}_{-1-k}(T^\gamma),T^\beta]\,\cM^{-1}\ket{\cJ_\gamma}(\eta_{1\,\alpha\beta}-\tilde\rho\,\eta_{\alpha\beta})\nonumber\\
=\,&\sum\limits_{k=0}^\infty\tilde\rho^k\Big(\langle\partial_\Lambda|\Lambda\rangle\bra{\partial_\Lambda}\mathcal{S}_{-k}(T^\alpha)\cM^{-1}\ket{\cJ_\alpha}-\bra{\partial_\Lambda}\mathcal{S}_{-k}(T^\alpha)\ket{\Lambda}\bra{\partial_\Lambda}\cM^{-1}\ket{\cJ_\alpha}\Big)\nonumber\\[1.5mm]
&\,+\bra{\partial_\Lambda}T^\alpha\ket{\Lambda}\bra{\partial_\Lambda}[T^\gamma,T^\beta]\,\cM^{-1}\ket{\cJ_\gamma}\eta_{\alpha\beta}\,.
\end{align}
Let us finally consider the variation of $\mathcal{L}_1$. Using \eqref{eq:difcurcomp} one computes that
\begin{equation}
\Delta_{\Lambda}\sum\limits_{k=0}^\infty\tilde\rho^k\langle {J^{\,k-n}_A} |=\,\eta_{AB}\bra{\partial_\Lambda}\Bigl( \sum\limits_{k=0}^\infty\tilde\rho^k\, T^B_{n-k}+\frac{\tilde\rho}{\rho^2}\cM^{-1}T^{B\,\dagger}_{1+n}\,\cM+\Big(1-\frac{\tilde\rho^2}{\rho^2}\Big)\cM^{-1}T^{B\,\dagger}_n\cM \Bigr) \ket{\Lambda}\bra{\partial_\Lambda}\,, 
\end{equation}
which allows to compute the variation of the first  term in $\rho \mathcal{L}_1$ as
\begin{align}
\label{eq:vL11}
&\Delta_\Lambda\Big(\sum\limits_{k=0}^\infty\tilde\rho^k\langle {J_A^{k-n}}|\cM^{-1}\ket{J_B^n}\,\eta^{AB}\Big)\nonumber\\
&=\,2\bra{\partial_\Lambda} \Bigl( \sum\limits_{k=0}^\infty\tilde\rho^k\mathcal{S}_{-k}(T^\alpha)+\frac{\tilde\rho}{\rho^2}\cM^{-1}\mathcal{S}_1(T^\alpha)^{\dagger}\,\cM+\Big(1-\frac{\tilde\rho^2}{\rho^2}\Big)\cM^{-1}T^{\alpha\,\dagger}\cM\Bigr) \ket{\Lambda}\bra{\partial_\Lambda}\cM^{-1}  \ket{\cJ_\alpha}\nonumber\\[1.5mm]
&\;\;\;-2 \hspace{-3mm} \sum\limits_{\tiny{q=-1,0,1}} \hspace{-3mm} \bra{\partial_\Lambda} \Bigl( \sum\limits_{k=0}^\infty\tilde\rho^k\mathcal{S}_{-k}(L_q)+\frac{\tilde\rho}{\rho^2}\cM^{-1}\mathcal{S}_1(L_q)^{\dagger}\,\cM+\Big(1-\frac{\tilde\rho^2}{\rho^2}\Big)\cM^{-1}L_q^{\dagger}\,\cM \Bigr) \ket{\Lambda}\bra{\partial_\Lambda}\cM^{-1}\ket{J_q}\nonumber\\[1.5mm]
&\;\;\;-2\bra{\partial_\Lambda}\Bigl( 2-\frac{\tilde\rho^2}{\rho^2} \Bigr) \ket{\Lambda}\bra{\partial_\Lambda}\cM^{-1}\ket{J_\dK}\nonumber\\[1.5mm]
&=\,2\bra{\partial_\Lambda}\Bigl( 2\sum\limits_{k=0}^\infty\tilde\rho^k\mathcal{S}_{-k}(T^\alpha)-\frac{\tilde\rho}{\rho^2}\Omega^\alpha(\cM)\Bigr) \ket{\Lambda}\bra{\partial_\Lambda}\cM^{-1}\ket{\cJ_\alpha}-4\,\langle\partial_\Lambda|\Lambda\rangle\bra{\partial_\Lambda}\cM^{-1}\ket{J_\dK}\nonumber\\[1.5mm]
&\;\;\;-2 \hspace{-3mm} \sum\limits_{\tiny{q=-1,0,1}} \hspace{-3mm}  \bra{\partial_\Lambda} \biggl( \sum\limits_{k=0}^\infty\tilde\rho^k L_{q-k} +\cM^{-1} \Bigl[ \frac{\tilde\rho}{\rho^2} L_{-1-q}+\Big(1-\frac{\tilde\rho^2}{\rho^2}\Big) L_{-q} \Bigr]  \,\cM \biggr) \ket{\Lambda}\bra{\partial_\Lambda}\cM^{-1}\ket{J_q}\,,
\end{align}
In the first step, the loop valued currents were completed to the full $\mathfrak{\hat e}_8\oleft \mathfrak{sl}(2)$ currents by adding and subtracting explicitly the missing components. In the second step, we used \eqref{eq:Hid} and substituted $\mathcal{S}_1(T^\alpha)$ by its expression following from \eqref{eq:OmJdef}. Using once again \eqref{eq:difcurcomp}, the variation of the second term in $\rho \mathcal{L}_1$ gives
\begin{align}
\label{eq:vL12}
&\Delta_\Lambda\Big(-2\,\bra{J_0+2\,\tilde\rho \,J_1}\cM^{-1}\ket{J_\dK}\Big)\nonumber\\[1.5mm]
\,\,&=\,2\bra{\partial_\Lambda} \Bigl( L_0+\cM^{-1} \Big[\frac{\tilde\rho}{\rho^2}  L_{-1}+\Big(1-2\,\frac{\tilde\rho^2}{\rho^2}\Big)L_0 -\tilde\rho\Big(1-\frac{\tilde\rho^2}{\rho^2}\Big) L_1\Big]\cM \Bigr) \ket{\Lambda}\bra{\partial_\Lambda}\cM^{-1}\ket{J_0+2\,\tilde\rho J_1}\nonumber\\[1.5mm]
\,\,&\;\;\;\;\,+4\,\langle\partial_\Lambda|\Lambda\rangle\bra{\partial_\Lambda}\cM^{-1}\ket{J_\dK}\,.
\end{align}
For the variation of the last term in $\mathcal{L}_1$, we need that
\begin{equation}
\Delta_\Lambda\Omega^\alpha(\cM)=\,0\,,
\end{equation}
which follows from the fact $\Omega^\alpha(\cM)$ is a function of $\cM$ and not of its derivative. Using moreover that \eqref{RewriteDeltachi}  one shows that  
\begin{align}
\label{eq:Delchi2}
\Delta_\Lambda\bra{\chi}=\,&  -\sum\limits_{k=0}^\infty\tilde\rho^k\bra{\partial_\Lambda}L_{-1-k}\ket{\Lambda}\bra{\partial_\Lambda}\nonumber\\
&\,-\frac{1}{\rho^2}\bra{\partial_\Lambda} \Bigl( L_{1}+\eta_{\alpha\beta} \,H(\cM)^\alpha{}_\gamma\big(\Omega^\gamma(\cM)+\tilde\rho\,\delta^\gamma_\dK\big)\,T^\beta+\tilde\rho\,\dK \Bigr) \ket{\Lambda}\bra{\partial_\Lambda}\,,
\end{align}
Together with \eqref{eq:difcurcomp}, this yields
\begin{align}
\label{eq:vL13}
&\Delta_\Lambda\Big(-2\bra{J_1}\cM^{-1} | 2\,\rho^2\,\chi-\rho^2 J_1 +\Omega^\alpha(\cM)\,\cJ_\alpha \rangle \Big)\\[1.5mm]
&\,=2\,\frac{\tilde\rho}{\rho^2}\langle\partial_\Lambda|\Lambda\rangle\bra{J_1}\cM^{-1} | 2\,\rho^2\,\chi +\Omega^\alpha(\cM)\,\cJ_\alpha \rangle \nonumber\\[1.5mm]
&\;\;\;\;+2\bra{\partial_\Lambda} \Bigl( 2\,\rho^2\sum\limits_{k=0}^\infty\tilde\rho^k L_{-1-k}+2\,L_1+\eta_{\alpha\beta}T^\beta\big(H(\cM)^\alpha{}_\gamma\Omega^\gamma(\cM)-\Omega^\alpha(\cM)\big) \Bigr) \ket{\Lambda}\bra{\partial_\Lambda}\cM^{-1}\ket{J_1}\nonumber\\
&\;\;\;\;-4\,\tilde\rho\bra{\partial_\Lambda} \cM^{-1} \Bigl(  \frac{\tilde\rho}{\rho^2}L_{-1}+\Big(1-2\,\frac{\tilde\rho^2}{\rho^2}\Big)L_0-\tilde\rho\Big(1-\frac{\tilde\rho^2}{\rho^2}\Big)L_1 \Bigr)  \cM \ket{\Lambda}\bra{\partial_\Lambda}\cM^{-1}\ket{J_1} \nonumber
\end{align}
By combining the contributions \eqref{eq:vL11}, \eqref{eq:vL12} and \eqref{eq:vL13}, and using the relation \eqref{eq:jrelation} to eliminate $\bra{J_0}$ and $\bra{J_{-1}}$, we find 
\begin{align}
&\Delta_\Lambda\big(\rho\,\mathcal{L}_1\big)\\
&=\,4\bra{\partial_\Lambda}\sum\limits_{k=0}^\infty\tilde\rho^k\mathcal{S}_{-k}(T^\alpha)\ket{\Lambda}\bra{\partial_\Lambda}\cM^{-1}\ket{\cJ_\alpha}+4\,\tilde\rho\langle\partial_\Lambda|\Lambda\rangle\bra{\partial_\Lambda}\cM^{-1}\ket{\chi}\nonumber\\
&\;\;\;+2\bra{\partial_\Lambda}\left(L_1+\tilde\rho\,L_0+\rho^2\sum\limits_{k=0}^\infty\tilde\rho^k L_{-1-k}+\eta_{\alpha\beta}T^\beta\big(H(\cM)^\alpha{}_\gamma\Omega^\gamma(\cM)-\Omega^\alpha(\cM)\big)\right.\nonumber\\
&\left.\;\;\;\;\;\;\;\;\;-\cM^{-1}\Big[\frac{\tilde\rho}{\rho^2}L_{-2}+\Big(1-\frac{\tilde\rho^2}{\rho^2}\Big)L_{-1}+\tilde\rho\Big(1-\frac{\tilde\rho^2}{\rho^2}\Big)L_0+\rho^2\Big(1-\frac{\tilde\rho^2}{\rho^2}\Big)^2L_1\Big]\cM\right)\ket{\Lambda}\bra{\partial_\Lambda}\cM^{-1}\ket{J_1}\nonumber
\end{align}
Writing $\cM=m\,\hat g_\cM$, with $\hat g_\cM\in \hat E_8$ and $m\in SL(2)$ as in \eqref{eq:mSL2}, one computes that
\begin{align}
\cM^{-1}&\Big(\frac{\tilde\rho}{\rho^2}L_{-2}+\Bigl(1-\frac{\tilde\rho^2}{\rho^2}\Bigr)L_{-1}+\tilde\rho \Bigl(1-\frac{\tilde\rho^2}{\rho^2}\Bigr)L_0+\rho^2\Bigl(1-\frac{\tilde\rho^2}{\rho^2}\Bigr)^2 L_1\Big)\cM\nonumber\\
&=\,\hat g_\cM^{-1}\Big(L_1+\tilde\rho\,L_0+\rho^2\sum\limits_{k=0}^\infty\tilde\rho^k L_{-1-k}\Big)\hat g_\cM\,.
\end{align}
Together with the identities \eqref{eq:fid1} and \eqref{eq:fid2}, this implies that the variation of $\mathcal{L}_1$ reduces to
\begin{align}
\Delta_\Lambda\mathcal{L}_1=\,&4\,\rho^{-1}\bra{\partial_\Lambda}\sum\limits_{k=0}^\infty\tilde\rho^k\mathcal{S}_{-k}(T^\alpha)\ket{\Lambda}\bra{\partial_\Lambda}\cM^{-1}\ket{\cJ_\alpha}+4\,\frac{\tilde\rho}{\rho}\langle\partial_\Lambda|\Lambda\rangle\bra{\partial_\Lambda}\cM^{-1}\ket{\chi}\,.\label{eq:fvarL1}
\end{align}
With \eqref{eq:fvarL2}--\eqref{eq:fvarL3} and \eqref{eq:fvarL1}, one can easily check that just as in the $\tilde\rho=\,0$ case, the non-covariant variations of \eqref{eq:pot} exactly recombine into the expression \eqref{eq:respot}, such that the full potential at $\tilde\rho\neq 0$ is invariant under generalised diffeomorphisms up to total derivatives.

\subsubsection{Invariance under $\Sigma$ transformations}
We conclude our proof of the gauge invariance of the potential by considering its variation under $\Sigma$ generalised diffeomorphisms. We denote such variations simply by $\delta_\Sigma$. The fields $\cM$, $\rho$ and $\tilde\rho$ transform covariantly
\begin{subequations}
\begin{align}
\delta_\Sigma\,\cM=&\,\mathscr{L}_\Sigma\,\cM=\,\eta_{-1\,\alpha\beta} \Tr( \Sigma T^\alpha ) \big(\cM\, T^\beta+T^{\beta\,\dagger}\cM\big)\,, \label{eq:SigmaM} \\[1.5mm] 
\delta_\Sigma\,\rho=&\,\mathscr{L}_\Sigma\,\rho=\,0\,,\\[1.5mm]
\delta_\Sigma\,\tilde\rho=&\,\mathscr{L}_\Sigma\,\tilde\rho=\,\Tr (\Sigma)\,.
\end{align}
\end{subequations}
These expressions follow from the general definition \eqref{eq:Lieformal} of the Lie derivative and the transformation properties \eqref{eq:L-1chivar} of the fields under rigid $\mathds{R}_{L_{-1}}$. We then deduce that 
\begin{equation}
\delta_\Sigma\bra{\cJ_\alpha}\otimes T^\alpha=\,\mathscr{L}_\Sigma \bra{\cJ_\alpha}\otimes T^\alpha+\eta_{-1\,\alpha\beta}\,\Tr(T^\alpha\Sigma)\bra{\partial_\Sigma}\otimes\big(T^\beta+\cM^{-1}T^{\beta\,\dagger}\cM\big)\,,
\end{equation}
with 
\begin{equation}
\mathscr{L}_\Sigma\bra{\cJ_\alpha}\otimes T^\alpha=\,-\eta_{-1\,\alpha\beta}\,\Tr(T^\alpha\Sigma)\bra{\cJ_\gamma}\otimes [T^\beta,T^\gamma]\,.
\end{equation}
This implies the following variation for the current components
\begin{equation}
\delta_\Sigma\bra{\cJ_\alpha}=\,\mathscr{L}_\Sigma\bra{\cJ_\alpha}+\Big(\eta_{-1\,\alpha\beta}\,\Tr(T^\beta\Sigma)+\frac{1}{\rho^2}\big(\eta_{1\,\alpha\beta}-2\tilde\rho\,\eta_{\alpha\beta}+\tilde\rho^2\eta_{-1\,\alpha\beta}\big) \Tr (\cM^{-1} T^{\beta \dagger} \cM  \Sigma ) \Big) \bra{\partial_\Sigma}\,,\label{eq:Sigvarcurc}
\end{equation}
where we used the first identity in \eqref{eq:dreamform}. Note that this implies $\Delta_\Sigma\bra{J_0+2\,\tilde\rho J_1}=\,0$. The Lie derivative  simply reads
\begin{equation}
\mathscr{L}_\Sigma\bra{\cJ_\alpha}=\,-\eta_{-1\,\gamma\delta} \,\Tr(T^\gamma\Sigma) \,f^{\delta\beta}{}_\alpha\bra{\cJ_\beta }\,.
\end{equation}

To derive the variation of the shifted current, we start by considering the Lie derivative f the term involving the infinite series of shift operators 
\begin{align} 
&\hspace{2mm} \mathscr{L}_\Sigma \bra{\cJ_\alpha}\otimes\sum\limits_{k=0}^\infty\tilde\rho^k\,\mathcal{S}_{-1-k}(T^\alpha)\label{eq:Svarsh}\\
&=-\eta_{-1\,\alpha\beta}\,\Tr(T^\alpha\Sigma)\bra{\cJ_\gamma}\otimes\sum\limits_{k=0}^\infty\tilde\rho^k\,[T^\beta,\mathcal{S}_{-1-k}(T^\gamma)]-\sum\limits_{n\in\mathds{Z}}\sum\limits_{k=0}^\infty\tilde\rho^k(n-1-k)\,\Tr(T^A_{n-2-k}\Sigma)\bra{J_A^{\,n}}\otimes\dK\, . \nn
\end{align}
The Lie derivative of the shifted current follows from its covariance \eqref{eq:L-1varscur} under rigid $\mathds{R}_{L_{-1}}$ transformations
\begin{equation}
\mathscr{L}_\Sigma\bra{\cJ^-_\alpha}\otimes T^\alpha=\,-\eta_{-1\,\alpha\beta}\,\Tr(T^\alpha\Sigma)\bra{\cJ^-_\gamma}\otimes [T^\beta,T^\gamma]\, .
\end{equation}
Together with \eqref{eq:Svarsh}, this implies that the Lie derivative of $\bra{\chi}$ must be given by 
\begin{equation}
\mathscr{L}_\Sigma\bra{\chi}=\,\sum\limits_{n\in\mathds{Z}}\sum\limits_{k=0}^\infty\tilde\rho^k\,(n-1-k)\,\Tr(T^A_{n-2-k}\Sigma)\bra{J_A^{\,n}}\, ,
\end{equation}
and which is indeed consistent with the infinitesimal action \eqref{eq:varchiE9} of $\mathds{R}_{L_{-1}}$ on $\bra{\chi}$. In order to determine the non-covariant variation of $\langle \chi|$ we proceed as in the preceding section, and compute from \eqref{eq:Sigvarcurc} the non-covariant variation of the shifted current  $\langle \cJ_\alpha^-|$ for $\alpha \ne \dK$, 
\be
\label{eq:varshiftcur2}
 \Delta_\Sigma \bra{\cJ_\alpha^-}_{\alpha\neq\dK} = \sum\limits_{k=0}^\infty \tilde\rho^k\,\eta_{-2-k\,\alpha\beta}\,\Tr(T^\alpha\Sigma)\bra{\partial_\Sigma}+\frac{1}{\rho^2}\big(\eta_{\alpha\beta}-\tilde\rho\,\eta_{-1\,\alpha\beta}\big)\,\Tr(\cM^{-1} T^{\alpha \dagger} \cM \Sigma)\bra{\partial_\Sigma} \; . 
\ee
We define the non-covariant variation of the field $\langle \chi|$ as 
\be\label{eq:Sigvarchi}
\Delta_\Sigma\bra{\chi}=-\frac{1}{\rho^2}\Big(\rho^2\sum\limits_{k=0}^\infty\tilde\rho^k \Tr(L_{-2-k}\Sigma)+\Tr\big(\cM^{-1}(L_0-\tilde\rho\,L_1)\cM\,\Sigma\big)\Big)\bra{\partial_\Sigma}-\frac{1}{\rho^2}\bra{\partial_\Sigma}\Sigma\,,
\ee
such that the first two terms precisely reproduce the expression of the non-covariant variation \eqref{eq:varshiftcur2}, but for $\alpha =\dK$. The last term turns out to be necessary for the closure of the algebra of generalised diffeomorphisms on $\langle \chi |$. Note that this expression is consistent with the section constraint since the parameter $\Sigma$ is covariantly constrained on its right. It is important to mention, that by using \eqref{eq:Delchi} and \eqref{eq:Sigvarchi}, one can verify the closure of the gauge algebra on $\bra{\chi}$. This is proven explicitly in Appendix \ref{sec:closure}.  
With the above results, we find that the non-covariant variation of the shifted current reads
\begin{align}
\delta_{\Sigma}\bra{\cJ^-_\alpha}\otimes T^\alpha=\,&\sum\limits_{k=0}^\infty \tilde\rho^k\,\eta_{-2-k\,\alpha\beta}\,\Tr(T^\alpha\Sigma)\bra{\partial_\Sigma}\otimes T^\beta-\frac{1}{\rho^2}\,\Tr(\Sigma)\bra{\partial_\Sigma}\otimes\dK\nonumber\\
&+\frac{1}{\rho^2}\big(\eta_{\alpha\beta}-\tilde\rho\,\eta_{-1\,\alpha\beta}\big)\,\Tr(T^\alpha\Sigma)\bra{\partial_\Sigma}\otimes\cM^{-1}T^{\beta\,\dagger}\,\cM\,,\label{eq:Sigvarscur}
\end{align}
where we used \eqref{InvarBil} for the the last term.

With the various transformation properties derived above, we are now equipped to discuss the $\Sigma$ variation of each term in the potential. The rigid $\hat{E}_8 \rtimes \mathds{R}_{L_{-1}}$ invariance of the potential, proven in section \ref{sec:E8L-1inv}, directly implies that the variation generated by the Lie derivative vanishes, such that
\begin{equation}
\delta_\Sigma\pot=\,\mathscr{L}_\Sigma\pot+\Delta_\Sigma\pot=\,\Delta_\Sigma\pot\,.
\end{equation}
In the following we therefore exclusively focus on the non-covariant variations of the various terms. For $\mathcal{L}_2$ and $\mathcal{L}_4$, we obtain  
\begin{align}
\Delta_\Sigma\mathcal{L}_2&=\,2\,\rho^{-1}\,\eta_{-1\,\alpha\beta}\,\Tr(T^\alpha\Sigma)\bra{\partial_\Sigma}[T^\gamma,T^\beta]\cM^{-1}\ket{\cJ_\gamma}\,,\label{eq:varSL2}\\
\Delta_\Sigma\mathcal{L}_4&=\,0\,,\label{eq:varSL4}
\end{align}
using the section constraint \eqref{eq:SC2}. For the variation of $\mathcal{L}_3$, we first compute that 
\bea &&  \eta_{\alpha\beta} \Tr(T^\alpha \Sigma) \langle \cJ_\gamma | \otimes [ \cS_{-1-k}(T^\gamma) , T^\beta] \CR
&= &\eta_{-1 \alpha\beta}  \Tr(T^\alpha \Sigma) \langle \cJ_\gamma | \otimes [ \cS_{-k}(T^\gamma) , T^\beta]  - \Tr( \cS_{-1-k}(T^\alpha) \Sigma ) \langle \cJ_\alpha | \otimes \dK \CR 
&& + \Tr (\Sigma) \langle \cJ_\alpha | \otimes \cS_{-1-k}(T^\alpha)  - \sum_{\scriptstyle q=-1,0,1}  \eta_{q-1-k\, \alpha\beta} \Tr(T^\alpha \Sigma) \langle J_q | \otimes T^\beta \; ,  \eea
such that 
\bea& &\eta_{\alpha\beta} \Tr(T^\alpha \Sigma) \langle \cJ_\gamma^- | \otimes [ T^\gamma , T^\beta] \\
&=& \eta_{-1 \alpha\beta}  \Tr(T^\alpha \Sigma) \langle \cJ_\gamma | \otimes [ T^\gamma , T^\beta]  + \tilde{\rho} \, \eta_{-1 \alpha\beta}  \Tr(T^\alpha \Sigma) \langle \cJ^-_\gamma | \otimes [ T^\gamma , T^\beta] 
\nn\\
& &- \Tr( T^\alpha \Sigma ) \langle \cJ^-_\alpha | \otimes \dK + \Tr (\Sigma) \langle \cJ^-_\alpha | \otimes T^\alpha - \sum_{\scriptstyle q=-1,0,1}\sum_{k=0}^\infty \tilde{\rho}^k  \eta_{q-1-k\, \alpha\beta} \Tr(T^\alpha \Sigma) \langle J_q | \otimes T^\beta \; \nn .\eea
Using this formula one finds with \eqref{eq:Sigvarscur}, 
\begin{align}
\Delta_\Sigma\mathcal{L}_3=&\,2\,\rho^{-1}\big(\eta_{\alpha\beta}-\tilde\rho\,\eta_{-1\,\alpha\beta}\big)\,\Tr(T^\alpha\Sigma)\bra{\partial_\Sigma}[T^\gamma,T^\beta]\cM^{-1}\ket{\cJ^-_\gamma}-2\,\rho^{-1}\,\Tr(\Sigma)
\bra{\partial_\Sigma}T^\alpha\cM^{-1}\ket{\cJ^-_\alpha}\nonumber\\[1.5mm]
=&\,2\,\rho^{-1}\eta_{-1\,\alpha\beta}\,\Tr(T^\alpha\Sigma)\bra{\partial_\Sigma}[T^\gamma,T^\beta]\cM^{-1}\ket{\cJ_\gamma}-2\,\rho^{-1}\,\Tr(T^\alpha\Sigma)
\bra{\partial_\Sigma}\cM^{-1}\ket{\cJ^-_\alpha}\nonumber\\[1.5mm]
&\,-2\,\rho^{-1}\bra{\partial_\Sigma}\Sigma\,\cM^{-1}\ket{J_1}+2\,\rho^{-1}\,\Tr(\Sigma)\bra{\partial_\Sigma}\cM^{-1}\ket{J_1}\,,\label{eq:SL3}
\end{align}
where we used the section constraint on $\Sigma$ and in particular 
\be \eta_{\alpha\beta} \Tr( T^\alpha \Sigma) \langle \partial_\Sigma | T^\beta = \langle \partial_\Sigma | \Sigma -  \Tr( \Sigma) \langle \partial_\Sigma | \ . \ee
Let us finally consider the variation of $\rho\mathcal{L}_1$. For its first term, we obtain  
\begin{align}
&\Delta_\Sigma\Big(\sum\limits_{k=0}^\infty\tilde\rho^k\langle {J_A^{k-n}}|\cM^{-1}\ket{J_B^n}\,\eta^{AB}\Big)\nonumber\\
&\,=\,2\, \Bigl( \sum\limits_{k=0}^\infty\tilde\rho^k\,\Tr\big(\mathcal{S}_{-1-k}(T^\alpha)\Sigma\big)+\frac{1}{\rho^2}\Tr\Big(\cM^{-1}\big[\mathcal{S}_1(T^\alpha)^{\dagger}-\tilde\rho \,T^{\alpha\,\dagger}+\tilde\rho\,\delta^\alpha_{\dK}\big]\cM\,\Sigma\Big)\Bigr) \bra{\partial_\Sigma}\cM^{-1}  \ket{\cJ_\alpha}\nonumber\\[1.5mm]
&\;\;\;-2 \hspace{-3mm} \sum\limits_{\tiny{q=-1,0,1}} \hspace{-1mm}\Bigl( \sum\limits_{k=0}^\infty\tilde\rho^k\,\Tr \big(L_{q-1-k}\Sigma\big)+\frac{1}{\rho^2}\Tr\Big(\cM^{-1}\big[L_{-1-q}-\tilde\rho\,L_{-q}\big]\cM^{-1}\,\Sigma\Big)\Bigr)\bra{\partial_\Sigma}\cM^{-1}\ket{J_q}\nonumber\\[1.5mm]
&\,=\,4\,\Tr(T^\alpha\Sigma)\bra{\partial_\Sigma}\cM^{-1}\ket{\cJ_\alpha^-}-\frac{2}{\rho^2}\bra{\partial_\Sigma}\cM^{-1}\ket{2\,\rho^2\chi+\Omega^\alpha(\cM)\cJ_\alpha}\nonumber\\[1.5mm]
&\;\;\;-2 \hspace{-3mm} \sum\limits_{\tiny{q=-1,0,1}} \hspace{-1mm}\Bigl( \sum\limits_{k=0}^\infty\tilde\rho^k\,\Tr \big(L_{q-1-k}\Sigma\big)+\frac{1}{\rho^2}\Tr\Big(\cM^{-1}\big[L_{-1-q}-\tilde\rho\,L_{-q}\big]\cM\,\Sigma\Big)\Bigr)\bra{\partial_\Sigma}\cM^{-1}\ket{J_q}\,.\label{eq:varSL11}
\end{align}
In the first step we used \eqref{eq:Sigvarcurc} and subsequently completed the loop valued currents to their full $\mathfrak{\hat e}_8\oleft\mathfrak{sl}(2)$-valued expressions. In the second step, we used \eqref{eq:OmJdef} to eliminate the term containing $\mathcal{S}_1(T^\alpha)$. Note that we wrote the result explicitly in terms of the shifted current and that the $\bra{\chi}$ contributions in fact cancel. \newline
For the variation of the second term in $\rho\mathcal{L}_1$, we get 
\begin{align}
&\Delta_\Sigma\Big(-2\bra{J_0+2\,\tilde\rho \,J_1}\cM^{-1}\ket{J_\dK}\Big)\nonumber\\
&\;\;\;\;=\,2\,\Big(\Tr(L_{-1}\Sigma)+\frac{1}{\rho^2}\Tr\Big(\cM^{-1}\big[L_{1}-2\,\tilde\rho\,L_0+\tilde\rho^2L_{1}\big]\cM\,\Sigma\Big)\Big)\bra{\partial_\Sigma}\cM^{-1}\ket{J_0+2\,\tilde\rho J_1}\,.\label{eq:varSL12}
\end{align}
using \eqref{eq:Sigvarscur}. For computing the variation of the last term in $\rho\mathcal{L}_1$, we note that 
\begin{equation}
\Delta_\Sigma\Omega^\alpha(\cM)=\,0\,,
\end{equation}
since $\Omega^\alpha(\cM)$ does not depend on derivatives of $\cM$. Together with \eqref{eq:Sigvarchi}, this leads to 
\begin{align}
\Delta_\Sigma&\Big(-2\bra{J_1}\cM^{-1}\ket{2\,\rho^2\,\chi-\rho^2 J_1 +\Omega^\alpha(\cM)\,\cJ_\alpha}\Big)\nonumber\\[1.5mm]
&=\,\frac{2}{\rho^2}\Tr(\Sigma)\bra{\partial_\Sigma}\cM^{-1}\ket{2\,\rho^2(\chi-J_1)+\Omega^\alpha(\cM)\cJ_\alpha}+4\bra{\partial_\Sigma}\Sigma\,\cM^{-1}\ket{J_1}\nonumber\\
&\;\;\;+4\Big(\rho^2\sum\limits_{k=0}^\infty\tilde\rho^k \Tr(L_{-2-k}\Sigma)+\Tr\big(\cM^{-1}\big[L_0-\tilde\rho\,L_1\big]\cM\,\Sigma\big)\Big)\bra{\partial_\Sigma}\cM^{-1}\ket{J_1}\nonumber\\
&\;\;\;-2\,\eta_{-1\,\alpha\beta}\big(\Omega^\alpha(\cM)+\Omega^\gamma(\cM)\,H(\cM)^\alpha{}_\gamma\big)\,\Tr(T^\beta\Sigma)\bra{\partial_\Sigma}\cM^{-1}\ket{J_1}\,.\label{eq:varSL13}
\end{align}
Recombining the results \eqref{eq:varSL11}, \eqref{eq:varSL12}, \eqref{eq:varSL13} and eliminating the components $\bra{J_0}$ and $\bra{J_{-1}}$ using \eqref{eq:jrelation}, we reach 
\begin{align}
&\Delta_\Sigma\big(\rho\,\mathcal{L}_1)\label{eq:varSL14}\\[1.5mm]
&=\,4\,\Tr(T^\alpha\Sigma)
\bra{\partial_\Sigma}\cM^{-1}\ket{\cJ^-_\alpha}+4\,\bra{\partial_\Sigma}\Sigma\,\cM^{-1}\ket{J_1}-4\,\Tr(\Sigma)\bra{\partial_\Sigma}\cM^{-1}\ket{J_1}\nonumber\\[2mm]
&\;\;-2\,\eta_{-1\,\alpha\beta}\big(\Omega^\alpha(\cM)+\Omega^\gamma(\cM)\,H(\cM)^\alpha{}_\gamma\big)\,\Tr(T^\beta\Sigma)\bra{\partial_\Sigma}\cM^{-1}\ket{J_1}\nonumber\\
&\;\;+2\,\Tr\Big(\Big[\rho^2\sum\limits_{k=0}^\infty\tilde\rho^kL_{-2-k}+\tilde\rho\,L_{-1}-L_0\nonumber\\
&\;\;\;\;\;\;\;\;\;\;\;\;\;\;\;\;\;-\frac{1}{\rho^2}\cM^{-1}\big(L_{-2}-3\,\tilde\rho\,L_{-1}-\big(\rho^2-3\,\tilde\rho^2\big)L_0+\tilde\rho\big(\rho^2-\tilde\rho^2\big)L_1\big)\cM\Big]\Sigma\Big)\bra{\partial_\Sigma}\cM^{-1}\ket{J_1}\,.\nonumber
\end{align}
Using once again the decomposition $\cM=\,m\,\hat g_{\cM}$, with $m\in SL(2)$ as in \eqref{eq:mSL2}, one finds the following intermediary result
\begin{equation}
-\frac{1}{\rho^2}\cM^{-1}\big(L_{-2}-3\,\tilde\rho\,L_{-1}-\big(\rho^2-3\,\tilde\rho^2\big)L_0+\tilde\rho\big(\rho^2-\tilde\rho^2\big)L_1\big)\cM=\,\hat g_{\cM}^{-1}\big(L_0-\tilde\rho\,L_{-1}-\rho^2\sum\limits_{k=0}^\infty\tilde\rho^kL_{-2-k}\big)\hat g_{\cM}\,,\nonumber
\end{equation}
which, together with \eqref{eq:Sid1} and \eqref{eq:Sid2}, allow to reduce the expression \eqref{eq:varSL14} to
\begin{align}
\Delta_\Sigma\,\mathcal{L}_1=\,4\,\rho^{-1}\,\Big(\Tr(T^\alpha\Sigma)
\bra{\partial_\Sigma}\cM^{-1}\ket{\cJ^-_\alpha}+\bra{\partial_\Sigma}\Sigma\,\cM^{-1}\ket{J_1}-\Tr(\Sigma)\bra{\partial_\Sigma}\cM^{-1}\ket{J_1}\Big)
\end{align}
With the above result and the variations \eqref{eq:varSL2}, \eqref{eq:varSL4} and \eqref{eq:varSL4}, it is straightforward to verify that the non-covariant variation of the potential \eqref{eq:pot} vanishes, thus proving its invariance under $\Sigma$ generalised diffeomorphisms.

\subsection{The potential in the unendlichbein formalism}
\label{BorelGauge}

In contrast to finite-dimensional Lie groups, care has to be taken when defining the Lie group from the algebra in the affine case. As the Lie algebra has infinitely many generators, the formal exponential of Lie algebra elements does not manifestly form a group or it may introduce formal infinite sum expressions whose well-definedness needs to be established. For instance, the current $\cJ_M=\cM^{-1}\partial_M \cM$ suffers from this problem as $\cM$ depends on an infinity of dual potentials $Y_n^A$ and applying Baker--Campbell--Hausdorff-type identities for evaluating the algebra-valued $\cJ_M$ leads naively to infinite sums within each component along the loop algebra.

In this section, we explain how to make sense of this infinity and that our potential~\eqref{eq:pot} is well-defined. As a preliminary step, we recall different definitions of an affine Kac--Moody group. The `minimal Kac--Moody group' is defined as the closure of the group generated by the one-parameter subgroups of the real roots~\cite{KacPeterson} that can also be interpreted using the Tits group functor~\cite{Tits:1987}. This corresponds to the definition of the loop group  $\hat{E}_8$ as the group of rational functions in $E_8$ of the spectral parameter $w$, that are meromorphic on $P^1(\mathds{C})$. This group can be completed with respect to a topology of an associated building~\cite{Carbone:2003}. This `completed Kac--Moody group' is then defined as the group of meromorphic functions in $E_8$ of the spectral parameter $w\in \mathds{C}$. It corresponds to choosing one standard Borel subalgebra (in our case the negative Borel associated with negative powers of $w$) and allowing infinite analytic power series of $w$ in that direction while keeping only a finite number of powers for the other direction (for us positive powers of $w$). 

One can then write a representative $\hat{V}$ of the coset space $\hat{E}_8/ K(\hat{E}_8)$ for the completed group $\hat{E}_8$ in Borel gauge, using the Iwasawa decomposition. Putting explicit coordinates on the affine Borel group is subtle and is best done using building theory~\cite{DeMedts:2009}. Here, we will choose coordinates formally  through exponentiation as we did in~\eqref{eq:coshatE8} and we can also extend the coset representative to include the $\reals_{\dL}^+ \ltimes \reals_{L_{-1}}$ part for the axio-dilaton to have a group element $\cV = v \hat{V}$, which we consider in the $R(\Lambda_0)_0$ representation as usual. In this way of writing $\cV$ one can see  that when acting on an element of a lowest weight module, only finite expressions arise. 

The Hermitian conjugation $\cV^\dagger$ does not preserve the Kac--Moody group completed in one direction as it interchanges the two standard Borel subgroups, so $\cM=\cV^\dagger\cV$ is only well-defined when $\cV$ is in the minimal group. This is another way of seeing that the definition of $\cJ_M$ requires qualification. 

The advantage of working in the Borel gauge~\eqref{eq:coshatE8} is that the Maurer--Cartan derivative $\partial_M \cV \cV^{-1}$ \textit{is} well-defined as it requires only finitely many commutators to determine the $\hat{\mf{e}}_8\oleft \mf{sl}(2)$ element at each (negative) power of the spectral parameter. Similarly, the coset component of the Maurer--Cartan form 
\begin{align}
\cP_M = \frac12 \bigl( \partial_M\cV  \cV^{-1} +( \partial_M \cV \cV^{-1})^\dagger  \bigr)\, ,\label{eq:Maurer}
\end{align}
is well-defined. By construction, one can write the $\hat{\mathfrak{e}}_8 \oleft \mathfrak{sl}(2) $ current as
\begin{align}
\cJ_M = 2 \cV^{-1} \cP_M \cV \ , 
\end{align}
and this expression makes sense in the completed group. In the Fock space notation, this definition of the current takes the form
 \begin{align} 
 \langle \cJ_\alpha | \otimes T^\alpha = 2\ \langle \cP_\alpha | \otimes \cV^{-1} T^\alpha \cV \ , \quad \langle \cJ_\alpha | = 2 \,R(\cV)^\beta{}_\alpha \langle \cP_\beta |=2 \,R(v)^\beta{}_\gamma \,R(\hat{V})^\gamma{}_\alpha \langle \cP_\beta | \,, \label{eq:cvcur}
 \end{align} 
 where we have define $R(\cV)^\alpha{}_\beta T^\beta = \cV^{-1} T^\alpha \cV$ in analogy with~\eqref{eq:Rdef} in the appendix.
 In particular one has 
 \be \langle J_0 | = 2 \langle P_0 |  - 4 \frac{\tilde{\rho}}{\rho} \langle P_1| \; , \qquad  \langle J_1 | = 2  \rho^{-1}  \langle P_1 | \ . \ee
 For the shifted current, one computes that 
 \begin{align} 
 \label{BorelGaugeMoveCurrent} 
 & \hspace{5mm} \langle  \cJ_\alpha | \otimes \sum_{k=0}^\infty \tilde{\rho}^k \cS_{-1-k}(T^\alpha) = 2\ \langle \cP_\alpha | \otimes \sum_{k=0}^\infty \tilde{\rho}^k \cS_{-1-k}( \hat{V}^{-1} v^{-1}  T^\alpha v \hat{V} )\CR
 &= 2  \ \langle \cP_\alpha | \otimes \sum_{k=0}^\infty \tilde{\rho}^k R(v)^\alpha{}_\beta \bigl( \hat{V}^{-1} \cS_{-1-k}( T^\beta)   \hat{V}  - \omega^\beta_{1+k}(\hat{V}^{-1}) \dK \bigr) \CR
 &= 2  \ \langle \cP_\alpha | \otimes \biggl( \hat{V}^{-1} e^{\tilde{\rho} L_{-1}  }\cS_{-1}( \rho^{L_0}  T^\alpha \rho^{-L_0}   )  e^{-\tilde{\rho} L_{-1}  } \hat{V}  - \sum_{k=0}^\infty \tilde{\rho}^k R(v)^\alpha{}_\beta \omega^\beta_{1+k}(\hat{V}^{-1}) \dK \biggr)  \CR
 &= 2  \ \langle \cP_\alpha | \otimes \biggl( \rho^{-1} \cV^{-1} \cS_{-1}( T^\alpha  )  \cV  - \Bigl( \omega_1^\alpha(\rho^{L_0}) + \sum_{k=0}^\infty \tilde{\rho}^k R(v)^\alpha{}_\beta \omega^\beta_{1+k}(\hat{V}^{-1}) \Bigr) \dK \biggr) \; ,   \end{align}
using \eqref{eq:coad} and \eqref{GeneratingInfinitShift}. This motivates the change of variable 
\be  \langle \chi | = \rho^{-1} \langle \tilde{\chi} | + 2  \Bigl( \omega_1^\alpha(\rho^{L_0}) + \sum_{k=0}^\infty \tilde{\rho}^k R(v)^\alpha{}_\beta \omega^\beta_{1+k}(\hat{V}^{-1}) \Bigr) \langle \cP_\alpha |\ . \label{BorelGaugeMoveChi} \ee
Note in particular that $\langle \tilde{\chi} |$ is a constrained field, since both $\langle \chi |$ and $\langle \cP_\alpha |$ are. Using this formula one obtains by construction that 
\be \rho  \,\langle \cJ_\alpha^- | \otimes T^\alpha =  2  \langle \cP_\alpha | \otimes \cV^{-1} \cS_{-1}(T^\alpha) \cV  + \langle \tilde{\chi} | \otimes \dK   \ .\label{eq:cvscur} \ee
With the above results, it becomes straightforward to re-express $\mathcal{L}_2$, $\mathcal{L}_3$ and $\mathcal{L}_4$ in terms of the Maurer-Cartan form. We find 
\begin{align} 
\cL_2  &= 4 \rho^{-1}  \langle \cP_\alpha | \cV^{-1} T^\beta T^{\alpha} \cV^{-1\dagger} | \cP_\beta \rangle \; , \CR
\cL_3 &= 4 \rho^{-1}  \langle \cP_\alpha | \cV^{-1}\cS_{-1}(T^\beta) \cS_1( T^\alpha) \cV^{-1 \dagger}  | \cP_\beta \rangle + 4 \langle \cP_\alpha | \cV^{-1} \cS_1(T^\alpha) \cV^{-1 \dagger } | \tilde \chi \rangle + \langle \tilde \chi  | \cV^{-1} \cV^{-1\dagger } | \tilde \chi  \rangle \; ,  \CR
\cL_4 &= 4  \rho^{-1} \langle P_0 | \cV^{-1} T^\alpha \cV^{-1\dagger} | \cP_\alpha \rangle\; , 
\end{align}
Instead of working out explicitly the expression of $\mathcal{L}_1$ step by step, we argue what the result should be based on the computation of the rigid $\hat E_8  \rtimes ( \mathds{R}^+_d \ltimes \mathds{R}_{L_{-1}})$ invariance of $\mathcal{L}_1$. To this purpose, let us first observe that the changes of variable \eqref{eq:cvcur}, \eqref{BorelGaugeMoveChi} and \eqref{eq:cvscur} essentially take the same form as the rigid transformations of the currents  $\langle \cJ_\alpha|$, $\bra{\cJ_\alpha^-}$ and the field $\langle \chi |$ under $\hat{E}_8 \rtimes ( \mathds{R}^+_d \ltimes \mathds{R}_{L_{-1}})$ that were presented in Section \ref{sec:E8L-1inv}, but now with $\mathcal{V}$ playing the r\^ole of the group element associated to the transformation. An important point is that this comparison only holds if one disregards the constrained $\overline{R(\Lambda_0)_{-1}}$ vector index of the currents and $\bra{\chi}$, whose associated transformation cancels that of $\cM^{-1}$ when considering the variation of $\mathcal{L}_1$. The analogy drawn above and the rigid invariance of $\mathcal{L}_1$ imply that the explicit dependence on $\cV$, apart from the $\cM^{-1}$ contracting the derivatives, is eliminated by the substitution \eqref{BorelGaugeMoveCurrent} and \eqref{BorelGaugeMoveChi} that induces  $\cM\to \cV^{-1\dagger} \cM \cV^{-1}=\id$. In the end, this change of variables simply amounts to cancel $\tilde{\rho}$ and  $\Omega^\alpha(\cM)$ through $\Omega^\alpha(\mathds{1})=0$, such that the new expression of $\mathcal{L}_1$ in terms of $\langle \cP_\alpha |$ and $\langle \tilde{\chi }|$ only depends implicitly on $\rho$, $\tilde{\rho}$ and $\hat{V}$ through $\langle \cP_\alpha |$ and the operator $\rho^{-\frac{1}2}  \cV^{-1}$
\be \cL_1 = 4\,\rho^{-1}\sum\limits_{n\in\mathds{Z}}\langle P_A^{\,-n} |\mathcal{M}^{-1}\ket{P^{\,n}_B}\eta^{AB}-8\,\rho^{-1}\bra{P_0}\cM^{-1}\ket{P_\dK} -8 \,\rho^{-1}\bra{P_1}\cM^{-1}\ket{\tilde{\chi}- P_1 } \ . \ee

Combining these terms together one obtains the potential \eqref{eq:pot} in the following simple form
\begin{align}
\nonumber  \rho \pot =\ &  \eta^{\alpha\beta}  \langle \cP_\alpha | \cV^{-1}  \cV^{-1\dagger} | \cP_\beta \rangle  + \frac12\langle \tilde \chi-2 P_1  | \cV^{-1} \cV^{-1\dagger } | \tilde \chi -2 P_1 \rangle  - 2 \langle \cP_\alpha | \cV^{-1} T^\beta T^{\alpha} \cV^{-1\dagger} | \cP_\beta \rangle \\  \nonumber
&+ 2  \langle \cP_\alpha | \cV^{-1}\cS_{-1}(T^\beta) \cS_1( T^\alpha) \cV^{-1 \dagger}  | \cP_\beta \rangle + 2 \langle \cP_\alpha | \cV^{-1} \cS_1(T^\alpha) \cV^{-1 \dagger } | \tilde \chi \rangle \\[1ex]
&+ 2 \langle P_0 | \cV^{-1} T^\alpha \cV^{-1\dagger} | \cP_\alpha \rangle \ .\label{eq:potM}  
\end{align} 
Any solution to the section constraint only has a finite number of non-trivial components for the derivative $\bra{\partial}$. This means that $\langle \partial | \cV^{-1}$ only involves finite sums and is regular in the Borel gauge. Moreover, all potentially infinite sums of the dual potentials $Y_n^A$ cancel in the potential. This cancellation can be associated with the invariance under $\delta_\Sigma$ generalised diffeomorphisms as these can be used to gauge away almost all $Y_n^A$. We shall see these facts more explicitly in the following section where we work out the potential in an $E_8$ decomposition. In summary, using the completed Kac--Moody group and a Borel gauge representative, the $E_9$ exceptional field theory potential~\eqref{eq:pot} is completely well-defined.


\section{Reduction to $E_8$ and consistency with supergravity}
\label{sec:redE8}

An inherent property of $E_n$ exceptional field theories for $n\leq8$, is that they reduce to eleven-dimensional supergravity or type IIB supergravity upon choosing the appropriate solution to the section condition. In this section, we shall partially demonstrate this property for the part of the $E_9$ exceptional field theory dynamics encoded in the $E_9$ scalar potential. Our strategy will consist in proving that the $E_8$ exceptional field theory with two external isometries is embedded in the scalar potential of $E_9$ exceptional field theory. In other words, we will show explicitly that when the infinite number of $E_9$ internal generalised coordinates are truncated to those of $E_8$ exceptional field theory, the potential~\eqref{eq:potterm0} reproduces all the terms of the $E_8$ exceptional field theory Lagrangian for field configurations that do not depend on the two external coordinates. As a corollary, this implies that our potential encodes the dynamics of eleven-dimensional supergravity and type IIB supergravity with two external isometries.

\subsection{$E_8$ section and exceptional field theory}

In the present section, we are interested in relating the $E_9$ exceptional field theory to $E_8$ exceptional field theory~\cite{Hohm:2014fxa} in $3+248=2+(1+248)$ dimensions where the $248$ directions are subject to an analogous $E_8$ section constraint. As indicated in the decomposition, we require $1+248$ directions to emerge from the coordinates $\ket{Y}$ in $R(\Lambda_0)_{-1}$ of $E_9$. Similar to~\eqref{eq:Vdec1} we have a decomposition of the coordinates according to
\begin{align}
 | Y\rangle = \Bigl(  \varphi + \eta_{AB} y^A T_{-1}^B + \sum_{n=2}^\infty y_{A_1\dots A_{n}} T_{-1}^{A_1} \dots T_{-1}^{A_n} \Bigr)  | 0\rangle 
\end{align}
of the coordinates. It was shown in \cite{Bossard:2017aae} that for any hyperplane solution to the $E_9$ section constraint~\eqref{eq:SC} there exists an $E_9$ element that rotates this hyperplane to one lying completely along the directions $\varphi$ and $y^A$, {\it i.e.}, the lowest two pieces in the $E_8$ graded decomposition of $R(\Lambda_0)_{-1}$  corresponding to $1+248$ directions. Moreover, the remnant of the $E_9$ section constraint~\eqref{eq:SC} implies that fields depend on the $248$ directions $y^A$ in such a way that they satisfy the $E_8$ section constraint~\cite[Eq.~(1.1)]{Hohm:2014fxa}. In practice, this solution to the section constraint is implemented by only considering the corresponding of the form
\begin{align}
\label{SectionSol}
\langle \partial | = \langle 0| ( \partial_\varphi + T_1^A \partial_A ) \ , 
\end{align}
where $\partial_A = \frac{\partial\; }{\partial y^A}$ satisfies the $E_8$ section constraint.

The direction $\varphi$ has an interpretation as the third external coordinate and $y^A$ as the internal coordinates in the $E_8$ exceptional field theory. In this section, we will show that the potential \eqref{eq:pot} introduced in this paper is indeed consistent with the action of $E_8$ exceptional field theory in that it reproduces all its terms with no external derivatives with respect to the two directions $x^\mu=(t,x)$. The Lagrangian of $E_8$ exceptional field theory is of the schematic form
\begin{align}
\label{eq:E8act}
\cL_{E_8} = \sqrt{-g} \hat{R}  + \frac1{240}\sqrt{-g} g^{mn} \eta^{AB} J_{m A} J_{n B} -\sqrt{-g} V(M_0,g)  +  \cL_{\rm CS}\,,
\end{align}
where $\hat{R}$ denotes the (improved) Ricci scalar of the 3 external directions with metric $g_{mn}$, where $m=0,1,2$. $M_0$ are the scalar fields parameterising $E_8$ (see section~\ref{sec:2dmax}) and $J_{m A}$ are the components of the $E_8$ covariant current $J_m = M_0^{-1} D_m M_0$ where the covariant derivative $D_m$ featuring $J_m$ and $\hat{R}$ is covariantised with respect to the two gauge fields $A_m^A$ and $B_{m A}$ of the theory, where $A=1,\ldots, 248$ labels the adjoint of $E_8$, and the gauge fields also appear in the Chern--Simons term. The vector field $B_{m A}$ is constrained in its $E_8$ index. The potential term $V(M_0,g)$ can be expressed through the $E_8$ internal current $M_0^{-1} \partial_A M_0$ and is a combination of terms similar to~\eqref{eq:potterm0}. The exact form of the various terms of~\eqref{eq:E8act} is given in~\cite{Hohm:2014fxa}. 

As we shall show in detail in Section~\ref{sec:comp}, all terms except for the topological Chern--Simons term give contributions when restricting to derivatives along $\varphi$ and the $248$ internal coordinates $y^A$ according to~\eqref{SectionSol}. First we parameterise the $E_8$ fields in a way that facilitates the comparison.
For the metric $g_{mn}$ on the three-dimensional external space we shall consider the (static) ansatz 
\begin{align}
\label{eq:met3DStat}
ds^2 = e^{2\sigma}(-dt^2+ dx^2) + \rho^2d\varphi^2\,.
\end{align}
Compared to~\eqref{eq:met3D} there is no Kaluza--Klein vector $A_{\mu}^{\scriptscriptstyle (3)}$ since we disregard all external form fields in this paper. In other words, the only components of the two vector fields that will appear are those along $\varphi$. For simplicity we shall write the remaining components of the vector fields $A^A= A_\varphi^A$ and $B_{A} = B_{\varphi\, A}$, without writing explicitly their $\varphi$ index. 

\subsection{parameterising $\cM$ and decomposition of the potential}

We shall now decompose the potential~\eqref{eq:pot} of $E_9$ exceptional field theory in the $E_8$ solution~\eqref{SectionSol} to the section constraint. Moreover, we shall see explicitly that even though $R(\Lambda_0)_{0}$ is an infinite-dimensional representation, the total potential only gives rise to finitely many terms as we explained in section~\ref{BorelGauge}. In particular, we use the formulation in terms of the coset representative $\cV$ rather than $\cM$.

First, as explained earlier, we can work at $\tilde\rho=0$ without loss of generality by gauge-fixing partially the invariance of the potential under $\Sigma$ transformations and this simplifies the analysis in this section as we only have to analyse~\eqref{eq:potterm0}. We now demonstrate this gauge-fixing explicitly using~\eqref{SectionSol} for solution to the section condition. On~\eqref{SectionSol}, the constrained parameter $\Sigma\sim \ket{\Sigma}\bra{\pi_\Sigma}$ can be parameterised  as 
\begin{align}
\Sigma =\ & \Bigl( \sigma_0- \eta^{AB} \Sigma_{A,B} + \sum_{n=1}^\infty \sigma_{A_1\dots A_n} T_{-1}^{A_1} \dots  T_{-1}^{A_n} \Bigr) |0\rangle \langle 0 | \\[-1ex]
&+  \Bigl( \Sigma_A +  \Sigma_{A,B} T_{-1}^B+\sum_{n=2}^\infty \Sigma_{A,A_1\dots A_n} T_{-1}^{A_1} \dots  T_{-1}^{A_n} \Bigr) |0\rangle \langle 0 | T_1^A \ , \nn{}
\end{align}
where the coefficients have to be projected to the irreducible representations appearing on level $n$ in~\eqref{eq:R0e8}. Moreover, it is clear from the structure of the generalised Lie derivative~\eqref{eq:Lie} that many of the components of $\Sigma$ have a trivial action on $\cM$. 

The trace of $\Sigma$ is a finite expression and given by
\begin{align}
 \Tr\, \Sigma = \sigma_0 \,.  
\end{align}
Considering for simplicity a $\Sigma$ gauge transformation with parameter $\Sigma  = \sigma_0 | 0 \rangle \langle 0 |$, one  obtains according to~\eqref{eq:SigmaM} and~\eqref{eq:Sigvarchi} the variations 
\begin{align}
\delta_{\sigma_0} \tilde\rho &= \sigma_0\,,\nn\\
\delta _{\sigma_0} \cM &= - \sigma_0 ( \cM L_{-1} + L_1 \cM )  \,,\\
\delta_{\sigma_0}  \langle \chi | &=- \frac{1}{\rho^2}\Bigl( 1 +\langle 0 | \cM^{-1} (L_0-\tilde{\rho} L_1) \cM | 0 \rangle  \Bigr)  \langle0|  ( \partial_\varphi  + T_1^A \partial_A ) \sigma_0  \nn\ . 
 \end{align}
The first transformation shows that we can shift the field $\tilde\rho$ by a gauge parameter when exponentiated. Therefore, we can use a finite gauge transformation to set $\tilde{\rho}=0$. As is evident from the other two equations, this will have a non-trivial effect on the dual potentials and $\chi$. Moreover, setting $\tilde\rho=0$ can be done while preserving a residual gauge invariance under traceless $\Sigma$ transformations satisfying $\sigma_0 = 0$. By a similar reasoning one can consider a more general class of $\Sigma$ parameters and find that all the higher level potentials $Y_{n}^A$ for $n\ge 2$ are also pure gauge, as follows from $\eta_{-1}$ in~\eqref{eq:SigmaM}, see also~\cite[Eq.~(4.36)]{Bossard:2017aae}. By contrast, $Y_{1}^A$ cannot be completely gauged away but transforms as it should in $D=3$ under gauge transformations~\cite{Hohm:2014fxa}.

For the rest of this section, we shall then work with the potential~\eqref{eq:potterm0} at $\tilde\rho=0$. For vanishing axion $\tilde\rho$,  the matrix $\cM_{MN}$ belongs to $E_9$ and can be parameterised by a suitable $E_9/K(E_9)$ coset space representative in Borel gauge. The latter follows from \eqref{eq:coshatE8}, and reads 
\begin{align}
\label{eq:cosE9}
\cV = \rho^{-L_0} e^{-\sigma} V_0 \prod_{n=1}^\infty \exp( Y_{n}^A \eta_{AB} T_{-n}^B ) \ .
\end{align}
The potential at $\tilde\rho=\,0$ can be expressed in terms of the $\mathfrak{e}_9$-valued Maurer--Cartan form associated to the coset representative \eqref{eq:cosE9}. This is most easily obtained by taking the expression \eqref{eq:potM} and setting $\tilde\rho=\,0$. This gives
\begin{align} 
\label{eq:pot2E8}
\rho \pot(M,\tilde\chi) &=  \eta^{\alpha\beta} \langle \cP_\alpha | \cV^{-1} \cV^{-1\dagger} | \cP_\beta \rangle - 2 \langle \cP_\alpha | \cV^{-1} T^\beta T^\alpha  \cV^{-1 \dagger}  | \cP_\beta \rangle  \nn\\[1.5mm]
&\quad+2  \langle \cP_\alpha | \cV^{-1}\cS_{-1}(T^\beta) \cS_1( T^\alpha) \cV^{-1 \dagger}  | \cP_\beta \rangle+ 2 \langle \cP_\alpha | \cV^{-1} \cS_1(T^\alpha) \cV^{-1 \dagger } | \tilde \chi \rangle\nn\\
&\quad + \frac12\langle \tilde \chi | \cV^{-1} \cV^{-1\dagger } | \tilde \chi \rangle +2 \langle \cP_0 | \cV^{-1}  T^\alpha \cV^{-1\dagger} | \cP_\alpha \rangle  \ , \ 
\end{align}
where $\bra{P_\alpha}$ and $\bra{\tilde\chi}$ are given by \eqref{eq:Maurer} and \eqref{BorelGaugeMoveChi} for $\tilde\rho=\,0$, respectively. The (negative) Borel gauge representative~\eqref{eq:cosE9} was chosen such that $\langle \cP_\alpha | \cV^{-1}$ gives rise to a finite expansion. To see this explicitly, we first note that the solution~\eqref{SectionSol} of the section constraint implies for $\cP_M$ that one has the decomposition $\langle \cP_\alpha | = \langle 0 | ( \cP_{\varphi,\alpha} +  \cP_{A,\alpha} T^A_1 )$. Multiplying then by $\cV^{-1}$ from the right and using~\eqref{eq:cosE9} one obtains the finite expression
\begin{align} 
\langle \cP_\alpha | \cV^{-1} &= \langle 0 | ( \cP_{\varphi,\alpha}  + \cP_{A,\alpha} T^A_1 ) \prod_{n=1}^\infty \exp( - Y_{n B} T_{-n}^B )  \rho^{L_0} e^{\sigma} V_0^{-1}  \CR
&= \langle 0 | ( \cP_{\varphi,\alpha}   -  \eta^{AB}  Y_{1 A} \cP_{B,\alpha} + \cP_{A,\alpha} T^A_1 )  \rho^{L_0} e^{\sigma} V_0^{-1} \CR
&=e^{\sigma}    \langle 0 | ( \cP_{\varphi,\alpha}   -  \eta^{AB}  Y_{1 A} \cP_{B,\alpha} + \rho  \cP_{A,\alpha}V_0 T^{A}_1 V_0^{-1} )  \ .  
\end{align}
Similarly, the scalar field $\bra{\tilde\chi}$ from~\eqref{BorelGaugeMoveChi} satisfies the section constraint and can thus be parameterised as
\begin{align}
\langle\tilde{\chi}  | = \langle 0 | ( \tilde{\chi}_{\varphi} +  \tilde{\chi}_{A} T^A_1 ) \ . 
\end{align}

As another preparatory step we need to introduce indices for the local $K(E_8)=Spin(16)/\ints_2$ subgroup appearing in the coset space $E_8/K(E_8)$ represented by $V_0$. We do this by writing $\uA$ for the adjoint of $E_{8}$ transforming under the local $K(E_8)$ subgroup and make the definitions 
\begin{equation}
\label{eq:hatP}
\begin{split}  \hat{\cP}_{\varphi,\alpha} &\equiv \cP_{\varphi,\alpha}   -  \eta^{AB}  Y_{1 A} \cP_{B,\alpha} \ , \\
\hat{\chi}_{\varphi} &\equiv \tilde{\chi}_{\varphi}   -  \eta^{AB}  Y_{1 A} \tilde{\chi}_{B} \ , 
\end{split}\hspace{5mm}\begin{split}
\hat{\cP}_{\underline{A},\alpha} T_n^{\uA} &\equiv   \cP_{A,\alpha} V_0  T^A_n V_0^{-1}   \ ,\\
\hat{\chi}_{\underline{A}} T_n^{\uA} &\equiv    \tilde{\chi}_{A} V_0  T^A_n V_0^{-1}   \ .
\end{split}
\end{equation}
More generally, we shall consider the notation that an underlined index is related to a normal one through $X_{\underline{A}}T^{\uA} = X_A V_0  T^A V_0^{-1}$  and in particular 
\begin{align}
 \label{Switch} \cP_{A,\alpha}  \cP_{B,\beta} \langle 0 | V_0  T^A_n V_0^{-1}(V_0  T^B_n V_0^{-1})^\dagger |0\rangle &= n\, M_0^{AB}  \cP_{A,\alpha}  \cP_{B,\beta} \CR
 =\hat\cP_{\uA,\alpha}  \hat\cP_{\uB,\beta} \langle 0 |   T^\uA_n ( T^\uB_n )^\dagger |0\rangle &= n\, \delta^{\uA\uB} \hat\cP_{\uA,\alpha}  \hat\cP_{\uB,\beta} \ . 
 \end{align}
Since $\cP_{M,\alpha} T^\alpha$ belongs to the coset component it satisfies the Hermiticity property $\cP_{M,\alpha} T^\alpha =(\cP_{M,\alpha} T^{\alpha})^\dagger $ and thus the components~\eqref{eq:hatP} can be decomposed as 
\begin{align} 
\hat{\cP}_{\varphi,\alpha} T^\alpha  &= P_{\varphi,0} L_0 + P_{\varphi,\dK} \dK+ P_{\varphi,\underline{A}} T^{\underline{A}}  +\frac12 \sum_{n=1}^\infty P_{\varphi,}{}_{\underline{A}}^n \bigl( T_{-n}^{\underline{A}} + (T_{-n}^{\underline{A}})^\dagger \bigr) \ , \CR
\hat{\cP}_{\underline{A},\alpha} T^\alpha  &= P_{\underline{A},0} L_0 + P_{\underline{A},\dK} \dK+ P_{\underline{A},\underline{B}} T^{\underline{B}}   +\frac12 \sum_{n=1}^\infty P_{\underline{A},}{}_{\underline{B}}^n \bigl( T_{-n}^{\underline{B}} + (T_{-n}^{\underline{B}})^\dagger \bigr) \ . 
\end{align}
Since $(P_{\varphi,\uA}T^\uA)^\dagger = P_{\varphi,\uA} T^\uA$ we did not explicitly symmetrise this $\mf{e}_8$-valued component of $\hat{\cP}$.

Using this notation, one obtains for the $\cL_1$ part of~\eqref{eq:pot2E8}
\begin{align}  
\eta^{\alpha\beta} \langle \cP_\alpha | \cV^{-1} \cV^{-1\dagger} | \cP_\beta \rangle = e^{2\sigma}  \eta^{\alpha\beta}  \Bigl(  \hat{\cP}_{\varphi,\alpha}  \hat{\cP}_{\varphi,\beta} +\rho^{2} M_0^{AB} {\cP}_{A,\alpha}  {\cP}_{B,\beta} \Bigr) \ , 
\end{align}
for the $\cL_2$ part
\begin{align}
 \langle \cP_\alpha | \cV^{-1} T^\beta T^\alpha  \cV^{-1 \dagger}  | \cP_\beta \rangle &=  e^{2\sigma}  \langle 0|   \Bigl(\hat{\cP}_{\varphi,\beta} T^\beta  \hat{\cP}_{\varphi,\alpha} T^\alpha   + \rho^{2}  T_1^{\underline{C}}   \hat{\cP}_{\underline{D},\beta} T^\beta  \hat{\cP}_{\underline{C},\alpha} T^\alpha  (T_{1}^{\underline{D}})^\dagger  \Bigr) | 0\rangle \CR
&\quad +2 \rho e^{2\sigma}  \langle 0|  T_1^{\underline{C}}  \hat{\cP}_{\varphi,\beta} T^\beta  \hat{\cP}_{\underline{C},\alpha} T^\alpha  | 0\rangle  \ ,  \qquad 
\end{align}
and similarly for $\cL_3$ 
\bea && 2  \langle \cP_\alpha | \cV^{-1}\cS_{-1}(T^\beta) \cS_1( T^\alpha) \cV^{-1 \dagger}  | \cP_\beta \rangle + 2 \langle \cP_\alpha | \cV^{-1} \cS_1(T^\alpha) \cV^{-1 \dagger } | \tilde \chi \rangle + \frac12\langle \tilde \chi | \cV^{-1} \cV^{-1\dagger } | \tilde \chi \rangle \CR
&=& \frac12e^{2\sigma} \langle 0|   (2\hat{\cP}_{\varphi,\beta} \cS_{-1}(T^\beta) + \hat{\chi}_\varphi)  ( 2\hat{\cP}_{\varphi,\alpha} \cS_{-1}(T^\alpha)^\dagger + \hat{\chi}_\varphi)    |0\rangle \CR
&&  +\frac12e^{2\sigma} \rho^2  \langle 0|  T_1^{\underline{C}}  ( \hat{\cP}_{\underline{D},\beta} \cS_{-1}(2T^\beta)  + \hat{\chi}_{\underline{D}} ) ( 2 \hat{\cP}_{\underline{C},\alpha} \cS_{-1}(T^\alpha)^\dagger  + \hat{\chi}_{\underline{C}}) (T_{1}^{\underline{D}})^\dagger   |0\rangle \CR
&& +e^{2\sigma} \rho \langle 0|  T_1^{\underline{C}}  ( 2\hat{\cP}_{\varphi,\beta} \cS_{-1}(T^\beta ) + \hat{\chi}_\varphi)(2 \hat{\cP}_{\underline{C},\alpha} \cS_{-1}(T^\alpha)^\dagger + \hat{\chi}_{\underline{C}})  | 0\rangle  \ . \eea

We now start to collect the different pieces in the potential term $\frac14 \cL_1-\frac12\cL_2+\frac12\cL_3$ in order to match them with the corresponding terms in the $E_8$ exceptional field theory action. We begin with the terms bilinear in $\cP_{\varphi,\alpha}$. These are, after removing the overall $e^{2\sigma}$ factor,
\begin{align}
\label{eq:bil1}
 &\quad  \eta^{\alpha\beta} \hat{\cP}_{\varphi,\alpha}  \hat{\cP}_{\varphi,\beta}  -2  \langle 0|   \hat{\cP}_{\varphi,\beta} T^\beta  \hat{\cP}_{\varphi,\alpha} T^\alpha  |0\rangle + \frac12\langle 0|   (2\hat{\cP}_{\varphi,\beta} \cS_{-1}(T^\beta) + \hat{\chi}_\varphi)  ( 2\hat{\cP}_{\varphi,\alpha} \cS_{-1}(T^\alpha)^\dagger + \hat{\chi}_\varphi)    |0\rangle \CR
&=   \delta^{\underline{A}\underline{B}} \hat{P}_{\varphi,\underline{A}}  \hat{P}_{\varphi,\underline{B}} -2 \hat{P}_{\varphi,0} \hat{P}_{\varphi,\dK} + \frac12 \sum_{n=1}^\infty  \delta^{\underline{A}\underline{B}}   \hat{P}_{\varphi,}{}_{\underline{A}}^n\hat{P}_{\varphi,}{}_{\underline{B}}^n \CR
&\quad -2 \hat{P}_{\varphi,\dK}\hat{P}_{\varphi,\dK} - \frac12\sum_{n=1}^\infty n  \delta^{\underline{A}\underline{B}}  \hat{P}_{\varphi,}{}_{\underline{A}}^n\hat{P}_{\varphi,}{}_{\underline{B}}^n+ \frac12\hat{\chi}_\varphi\hat{\chi}_\varphi +  \frac12\sum_{n=1}^\infty n \delta^{\underline{A}\underline{B}}  \hat{P}_{\varphi,}{}_{\underline{A}}^{n+1}\hat{P}_{\varphi,}{}_{\underline{B}}^{n+1}  \CR
&=  \delta^{\underline{A}\underline{B}} \hat{P}_{\varphi,\underline{A}}  \hat{P}_{\varphi,\underline{B}} - 2(\hat{P}_{\varphi,0}+ \hat{P}_{\varphi,\dK} ) \hat{P}_{\varphi,\dK} +  \frac12\hat{\chi}_\varphi\hat{\chi}_\varphi  \ .    
\end{align}
Note that all the higher potential field strengths $\hat{P}_{\varphi,}{}_{\underline{A}}^{n}$ cancel. 

Next we consider all terms bilinear in $\cP_{\uC,\alpha}$. The $\cL_1$ terms are simply $2  \eta^{\alpha\beta}  M_0^{CD} {\cP}_{C,\alpha}  {\cP}_{D,\beta}$, while the ones from $\cL_2$ are
\begin{align} 
&\quad 4\langle 0 | T_1^{\underline{C}} \cP_{\underline{D},\beta} T^\beta \cP_{\underline{C},\alpha} T^\alpha  (T_1^{\underline{D}})^\dagger | 0\rangle \CR
&=  \langle 0 | \Bigl[ 2( P_{\underline{D},0} + P_{\underline{D},\dK})  T_1^{\underline{C}} +2 P_{\underline{D},\underline{B}} f^{\uC\uB}{}_\uE T_1^\uE +P_{\underline{D},}{}^1_{\underline{B}} \eta^{\uC\uB} + \sum_{n=1}^\infty  P_{\underline{D},}{}^n_{\underline{B}}  T_1^{\underline{C}} (T_{-n}^\uB)^\dagger \Bigr]\CR
& \qquad  \Bigl[ 2( P_{\underline{C},0} + P_{\underline{C},\dK})  (T_1^{\underline{D}} )^\dagger + 2 P_{\underline{C},\underline{A}} f^{\uD\uA}{}_\uF( T_1^\uF)^\dagger +P_{\underline{C},}{}^1_{\underline{A}} \eta^{\uD\uA} + \sum_{n=1}^\infty  P_{\underline{C},}{}^n_{\underline{A}}  T_{-n}^A (T_1^{\underline{D}})^\dagger \Bigr]  | 0\rangle \CR
&=  4 \delta^{\underline{C}\underline{D}} ( P_{\underline{D},0} + P_{\underline{D},\dK}) ( P_{\underline{C},0} + P_{\underline{C},\dK}) +  4 P_{\underline{D},\underline{B}} P_{\underline{C},\underline{A}} \delta^{\underline{E}\underline{F}} f^{\underline{C}\underline{B}}{}_{\underline{E}} f^{\underline{D}\underline{A}}{}_{\underline{F}}\CR
&\quad + 8  ( P_{\underline{D},0} + P_{\underline{D},\dK}) P_{\uC,\uA} \delta^{\uC\uE} f^{\uD\uA}{}_\uE + ( \eta^{\uC\uB} \eta^{\uD\uA} + \eta^{\uB\uD} \eta^{\uA\uC})  P_{\uD,}{}^1_\uB P_{\uC,}{}^1_\uA \CR
& \qquad + \sum_{n=1}^\infty ( n \delta^{\uA\uB} \delta^{\uC\uD} + \delta^{\uB\uE} \delta^{\uD\uF} f_{\uG\uE}{}^\uA f^{\uC\uG}{}_\uF ) P_{\uD,}{}^n_\uB P_{\uC,}{}^n_\uA \ .  
\end{align}
The terms quadratic in  $\cP_{\uC,\alpha}$ coming from $\cL_3$ are
\begin{align} 
\label{eq:bil2}
&\quad  \langle 0|  T_1^{\underline{C}}  ( 2\hat{\cP}_{\underline{D},\beta} \cS_{-1}(T^\beta)  + \hat{\chi}_{\underline{D}} ) (  2\hat{\cP}_{\underline{C},\alpha} \cS_{-1}(T^\alpha)^\dagger  + \hat{\chi}_{\underline{C}}) (T_{1}^{\underline{D}})^\dagger   |0\rangle \CR
&=  \langle 0 | \Bigl[ \hat{\chi}_\uD T_1^{\underline{C}} + P_{\underline{D},}{}^1_{\underline{B}} \delta^{\uB\uF} f_{\uF\uE}{}^\uC T_1^\uE +2 P_{\underline{D},\underline{B}} \eta^{\uC\uB} + \sum_{n=1}^\infty  P_{\underline{D},}{}^{n+1}_{\underline{B}}  T_1^{\underline{C}} (T_{-n}^\uB)^\dagger \Bigr]\CR
& \qquad  \Bigl[ \hat{\chi}_\uC  (T_1^{\underline{D}} )^\dagger + P_{\underline{C},}{}^1_{\underline{A}}  \delta^{\uA\uG} f_{\uG\uH}{}^\uD( T_1^\uH)^\dagger +2P_{\underline{C},\underline{A}} \eta^{\uD\uA} + \sum_{n=1}^\infty  P_{\underline{C},}{}^{n+1}_{\underline{A}}  T_{-n}^A (T_1^{\underline{D}})^\dagger \Bigr]  | 0\rangle \CR
&=  \delta^{\underline{C}\underline{D}}  \hat{\chi}_\uC \hat{\chi}_\uD +4 \eta^{\uC\uB} \eta^{\uD\uA}   P_{\underline{D},\underline{B}} P_{\underline{C},\underline{A}} + 2f^{\uC\uA}{}_\uB \delta^{\uB\uD} \hat{\chi}_\uD P_{\uC,}{}^1_\uA  + \delta^{\uE\uF} f^{\uD\uB}{}_\uE f^{\uC\uA}{}_\uF P_{\uD,}{}^1_\uB P_{\uC,}{}^1_\uA  \CR
&\quad + \eta^{\uB\uD} \eta^{\uA\uC}  P_{\uD,}{}^2_\uB P_{\uC,}{}^2_\uA  + \sum_{n=1}^\infty \bigl( (n-1) \delta^{\uA\uB} \delta^{\uC\uD} + \delta^{\uB\uE} \delta^{\uD\uF} f_{\uG\uE}{}^\uA f^{\uC\uG}{}_\uF \bigr) P_{\uD,}{}^n_\uB P_{\uC,}{}^n_\uA \ .  
\end{align}
In rewriting the final expression we have used,\footnote{The structure constants $f_{\uA\uB}{}^\uC$ are given by $f_{\uA\uB}{}^\uC = \eta_{\uA\uD} \eta_{\uB\uE} \eta^{\uC\uF} f^{\uD\uE}{}_{\uF} = -\delta_{\uA\uD} \delta_{\uB\uE} \delta^{\uC\uF} f^{\uD\uE}{}_{\uF}  $ and one uses the Jacobi identity to derive this identity.}
\begin{align} 
\delta^{\uB\uE} \delta^{\uD\uF} f_{\uG\uE}{}^\uC f^{\uA\uG}{}_\uF = \delta^{\uB\uE} \delta^{\uD\uF} f_{\uG\uE}{}^\uA f^{\uC\uG}{}_\uF + \delta^{\uE\uF} f^{\uD\uB}{}_\uE f^{\uC\uA}{}_\uF \ . 
\end{align}
Combining the terms bilinear in $\cP_{\uC,\alpha}$ in $\frac14 \cL_1-\frac12\cL_2+\frac12\cL_3$ determined above then gives the following somewhat lengthy expression
\begin{align} 
&\quad   \eta^{\alpha\beta}  M_0^{CD} {\cP}_{C,\alpha}  {\cP}_{D,\beta}  - 2 \langle 0|  T_1^{\underline{C}}   \hat{\cP}_{\underline{D},\beta} T^\beta  \hat{\cP}_{\underline{C},\alpha} T^\alpha  (T_{1}^{\underline{D}})^\dagger   |0\rangle \CR
& \hspace{30mm}  + \frac12 \langle 0|  T_1^{\underline{C}}  (2 \hat{\cP}_{\underline{D},\beta} \cS_{-1}(T^\beta)  + \hat{\chi}_{\underline{D}} ) (  2\hat{\cP}_{\underline{C},\alpha} \cS_{-1}(T^\alpha)^\dagger  + \hat{\chi}_{\underline{C}}) (T_{1}^{\underline{D}})^\dagger   |0\rangle \CR
&= M_0^{CD} \delta^{\underline{A}\underline{B}}  {P}_{C,\underline{A}}  {P}_{D,\underline{B}} -2  M_0^{CD}  {P}_{C,0} {P}_{D,\dK} + \frac12\delta^{\underline{C}\underline{D}}  \sum_{n=1}^\infty \delta^{\underline{A}\underline{B}}  {P}_{\underline{C},}{}_{\underline{A}}^n {P}_{\underline{D},}{}_{\underline{B}}^n \CR
&\quad  - 2 \delta^{\underline{C}\underline{D}} ( P_{\underline{D},0} + P_{\underline{D},\dK}) ( P_{\underline{C},0} + P_{\underline{C},\dK}) -  2P_{\underline{D},\underline{B}} P_{\underline{C},\underline{A}} \delta^{\underline{E}\underline{F}} f^{\underline{C}\underline{B}}{}_{\underline{E}} f^{\underline{D}\underline{A}}{}_{\underline{F}}\CR
&\quad - 4  ( P_{\underline{D},0} + P_{\underline{D},\dK}) P_{\uC,\uA} \delta^{\uC\uE} f^{\uD\uA}{}_\uE - \frac12( \eta^{\uC\uB} \eta^{\uD\uA} + \eta^{\uB\uD} \eta^{\uA\uC})  P_{\uD,}{}^1_\uB P_{\uC,}{}^1_\uA \CR
& \qquad - \frac12\sum_{n=1}^\infty ( n \delta^{\uA\uB} \delta^{\uC\uD} + \delta^{\uB\uE} \delta^{\uD\uF} f_{\uG\uE}{}^\uA f^{\uC\uG}{}_\uF ) P_{\uD,}{}^n_\uB P_{\uC,}{}^n_\uA  \CR
&\quad + \frac12\delta^{\underline{C}\underline{D}}  \hat{\chi}_\uC \hat{\chi}_\uD +2\eta^{\uC\uB} \eta^{\uD\uA}   P_{\underline{D},\underline{B}} P_{\underline{C},\underline{A}} + f^{\uC\uA}{}_\uB \delta^{\uB\uD} \hat{\chi}_\uD P_{\uC,}{}^1_\uA  + \frac12\delta^{\uE\uF} f^{\uD\uB}{}_\uE f^{\uC\uA}{}_\uF P_{\uD,}{}^1_\uB P_{\uC,}{}^1_\uA  \CR
&\quad + \frac12 \eta^{\uB\uD} \eta^{\uA\uC}  P_{\uD,}{}^2_\uB P_{\uC,}{}^2_\uA  + \frac12\sum_{n=1}^\infty \bigl( (n-1) \delta^{\uA\uB} \delta^{\uC\uD} + \delta^{\uB\uE} \delta^{\uD\uF} f_{\uG\uE}{}^\uA f^{\uC\uG}{}_\uF \bigr) P_{\uD,}{}^n_\uB P_{\uC,}{}^n_\uA \CR
 &=  M_0^{CD} \eta^{AB}  {P}_{C,A}  {P}_{D,B} -2 M_0^{EF}  f^{CB}{}_{E}  f^{DA}{}_{F} P_{D,B}P_{C,A} +2 \eta^{CB} \eta^{DA} P_{D,B}P_{C,A}    \CR
&\quad  - \frac12 M_0^{CD}  \bigl( 4 P_{C,0} P_{D,0} + 12 P_{C,0} P_{D,\dK} + 4 P_{C,\dK} P_{D,\dK}-\hat{\chi}_{C} \hat{\chi}_{D} \bigr) \CR
&\quad  -\frac12(\eta^{CB} \eta^{DA} + \eta^{BD} \eta^{AC} )P_{D,}{}^1_{B}P_{C,}{}^1_{A}  + \frac12M_0^{EF} f^{DB}{}_E f^{CA}{}_F P_{D,}{}^1_{B}P_{C,}{}^1_{A} \CR
&\quad  - 4  M_0^{AB} f^{CD}{}_A P_{C,D} (P_{B,0} +P_{B,\dK})   + f^{AB}{}_{D} M_0^{DC} P_{A,}{}^1_{B} \hat{\chi}_{C} + \frac12\eta^{BD} \eta^{AC}  P_{D,}{}^2_B P_{C,}{}^2_A\ .   
\end{align}
Because the structure constant $f^{AB}{}_C$ and the Killing form $\eta^{AB}$ are $E_8$ invariant, trading local $K(E_8)$ indices $\uA,\uB,\dots$ for $E_8$ indices $A,B,\dots$ by conjugation with $V_0$ amounts in practice to simply replacing $\delta^{\uA\uB}$ by $M_0^{AB}$ according to \eqref{Switch}. Once again all the higher level scalar field strengths $P_{A,}{}^n_B$ cancel out for $n>2$ for the final expression involving all bilinears in $\cP_{\uC,\alpha}$. 

Finally, we collect all the terms in $\cP_{\varphi,\beta} \cP_{\uC, \alpha}$. These have no contribution from $\cL_1$ and the terms arising in  $-\frac12\cL_2+\frac12\cL_3$  are
\begin{align} 
\label{eq:bil3}
&\quad-2\langle 0|  T_1^{\underline{C}}  \hat{\cP}_{\varphi,\beta} T^\beta  \hat{\cP}_{\underline{C},\alpha} T^\alpha  | 0\rangle + \frac12\langle 0|  T_1^{\underline{C}}  ( 2\hat{\cP}_{\varphi,\beta} \cS_{-1}(T^\beta ) + \hat{\chi}_\varphi)(2 \hat{P}_{\underline{C},\alpha} \cS_{-1}(T^\alpha)^\dagger + \hat{\chi}_{\underline{C}})  | 0\rangle \\
&=- \eta^{\underline{A}\underline{B}} P_{\varphi,}{}^1_{\underline{A}} P_{\underline{B},\dK} + 2f^{\underline{A}\underline{B}\underline{C}} P_{\varphi,\underline{A}} P_{\underline{B},}{}^1_{\underline{C}}- ( P_{\varphi,0}+ P_{\varphi,\dK}) \eta^{\underline{A}\underline{B}} P_{\underline{A},}{}^1_{\underline{B}} +\frac12\sum_{n=1}^\infty \delta^{\underline{C}\underline{D}}  f^{\underline{A}\underline{B}}{}_{\underline{D}} P_{\varphi,}{}^n_{\underline{C}}   P_{\underline{A},}{}^{n+1}_{\underline{B}} \CR
&\quad + \eta^{\underline{A}\underline{B}} P_{\varphi,\underline{A}} \hat{\chi}_{\underline{B}}+ \frac12\hat{\chi}_\varphi \eta^{\underline{A}\underline{B}} P_{\underline{A},}{}^2_{\underline{B}} - \frac12\sum_{n=1}^\infty \delta^{\underline{C}\underline{D}}  f^{\underline{A}\underline{B}}{}_{\underline{D}} P_{\varphi,}{}^n_{\underline{C}}   P_{\underline{A},}{}^{n+1}_{\underline{B}} \CR 
&=- \eta^{AB} P_{\varphi,}{}^1_{A} P_{B,\dK} + f^{ABC} P_{\varphi,A} P_{B,}{}^1_{C}- ( P_{\varphi,0}+ P_{\varphi,\dK}) \eta^{AB} P_{A,}{}^1_{B}   + \eta^{AB} P_{\varphi,A} \hat{\chi}_{B}+ \frac12 \hat{\chi}_\varphi \eta^{\underline{A}\underline{B}} P_{\underline{A},}{}^2_{\underline{B}}  \ . \nn\
\end{align}
Finally we compute the expression of $\frac12\cL_4$ that gives 
\begin{align} 
\label{eq:bil4}
2 \langle \cP_0 | \cV^{-1}  T^\alpha \cV^{-1\dagger} | \cP_\alpha \rangle &= e^{2\sigma} \Bigl( 2 P_{\varphi,0} P_{\varphi,\dK} +  \rho \eta^{AB} P_{A,0} P_{\varphi,}{}_B^1 +  \rho P_{\varphi,0} \eta^{AB} P_{A,}{}^1_B \\ 
&\quad+ 2\rho^2 M_0^{AB} P_{A,0} ( P_{B,0} + P_{B,\dK}) + 2 \rho^2 M_0^{AB} f^{CD}{}_B P_{A,0} P_{C,D} \Bigr) \nn \ . \end{align}

Having collected and simplified all the terms appearing in the potential, we now need to explain how the various components relate to the quantities of $E_8$ exceptional field theory. 
First, we identify the dual potential $Y_1^A$ with the three-dimensional vector field along the $\varphi$ direction $Y_1^A = A^A$. This is natural as the vector fields in $D=3$ are dual to the scalar fields and after reduction to two dimensions the relevant part of this duality equation becomes exactly~\eqref{eq:dual1}. Similarly, the $\varphi$ component of the $E_8$ constrained vector field reduces to a constrained scalar in two dimensions, that is we impose $\tilde{\chi}_A = \rho^{-1} B_A$. Evaluating the components of $\cP_\varphi$ one obtains
\begin{align}
\begin{split} P_{\varphi,0}  &= -  \rho^{-1} ( \partial_\varphi  \rho - A^A \partial_A \rho ) \equiv  -  \rho^{-1} D \rho \; , \\
P_{\varphi,\dK}  &=  -   ( \partial_\varphi  \sigma - A^A \partial_A \sigma ) \equiv  -D \sigma \; ,  \\
P_{\varphi,}{}^1_A & = \rho^{-1} \eta_{AB} ( \partial_\varphi  A^B - A^C \partial_C A^B ) \equiv  \rho^{-1} \eta_{AB} D A^B\; , 
\end{split}
\end{align}
where we have introduced the notation $D$ for $\partial_\varphi-A^A\partial_A$. Note that $D$ as introduced here is \textit{not} the full covariant derivative $D_\varphi$ that defines $J_{m A}$ and $\hat{R}$ in~\eqref{eq:E8act}  \cite{Hohm:2014fxa}, but only includes the transport term. 

The components of $\cP_A$ become similarly
\begin{align}
\begin{split} P_{A,0}  &= - \rho^{-1} \partial_A \rho  \ ,  \\
2P_{A,B} &= \eta_{BC} J_{A,}{}^C \ , \\
P_{A,}{}^2_B &= \rho^{-2} \eta_{BC} \bigl( \partial_A Y^C_2 + \tfrac12 f_{DE}{}^C A^D \partial_A A^E \bigr)  \ . 
\end{split}\hspace{10mm} 
\begin{split} 
P_{A,\dK}  &=    -\partial_A \sigma \; ,  \\
P_{A,}{}^1_B &= \rho^{-1} \eta_{BC} \partial_A A^C \ , \\
&
\end{split}
\end{align}
Here, $J_{A,}{}^C$ denotes the internal component of the $\mf{e}_8$ current defined from $M_0$ as 
\be M_{0}^{-1 CD} \partial_A M_{0 BD} = J_{A,}{}^D f_{DB}{}^C \ . \ee

With these identifications we can now rewrite the full potential~\eqref{eq:pot2E8} using also the rearrangements~\eqref{eq:bil1}, \eqref{eq:bil2}, \eqref{eq:bil3} and~\eqref{eq:bil4} for the various bilinears. 
The result is the following long expression
\begin{multline}  
\label{eq:pot2E8a}
e^{-2\sigma} \pot = -2 \rho^{-1} (D \sigma )^2  + D A^A  (  \partial_A \rho^{-1} +2 \rho^{-1} \partial_A \sigma) + ( - D \rho^{-1} + 2 \rho^{-1} D\sigma)  \partial_A A^A   \\
+  \rho^{-1}  \eta^{AB} P_{\varphi,A} P_{\varphi,B}   -2 \rho^{-1} f_{AB}{}^{C} \eta^{AD} P_{\varphi,D}   \partial_C A^B + \frac12\rho^{-3} \bigl(\rho \hat{\chi}_\varphi+  \partial_A Y^A_2 + \tfrac12 f_{AB}{}^C A^A \partial_C A^B \bigr)^2\\
 -\frac12\rho^{-1} ( \partial_A A^B \partial_B A^A+ \partial_A A^A \partial_B A^B ) +\frac12 \rho^{-1}   M_0^{AB} f_{AC}{}^D \partial_D A^C f_{BE}{}^F \partial_F A^E \\
+\frac14\rho \Bigl( M_0^{CD} \eta_{AB}  {J}_{C,}{}^{A}  {J}_{D,}{}^{B} -2M_0^{EF}  f_{BE}{}^C   f_{AF}{}^D  J_{D,}{}^{B}J_{C,}{}^{A} + 2 J_{A,}{}^{B}J_{B,}{}^{A} \Bigr)   \\
 -2 M_0^{AB} \partial_A \sigma  \bigl(  \rho \partial_B \sigma + 2 \partial_B \rho  \bigr)  - M_0^{AB} f_{AD}{}^C J_{C,}{}^{D} (\partial_B \rho + 2 \rho \partial_B \sigma ) \\
 +  2 \rho^{-1} \eta^{AB} P_{\varphi,B}  B_{A} +\frac12 \rho^{-1}  M_0^{AB} B_A B_B -  \rho^{-1}  M_0^{AB} f_{AC}{}^D \partial_{D} A^C B_{B}\,.
\end{multline} 

\subsection{Comparison with $E_8$ exceptional theory}
\label{sec:comp}

The above form of the potential still does not look very similar to the standard action of $E_8$ exceptional field theory sketched in~\eqref{eq:E8act}, and in particular it still contains the dual potential $Y_2^A$ and the constrained scalar $\hat\chi_\varphi$ that were not considered in  \cite{Hohm:2014fxa}. In order to recognise the standard terms we now expand them out.

The kinetic term for the scalar fields reduces to an expression in terms of the covariant current along the $\varphi$ direction
\begin{align}
\label{E8current} J^A = 2\eta^{AB} P_{\varphi,B}  - ( M_0^{AB} + \eta^{AB} ) ( f_{BC}{}^D \partial_D A^C -B_B) \ , 
\end{align}
with $2 P_{\varphi,A} T^A = {M}_0^{-1} ( \partial_\varphi  - A^A \partial_A ) {M_0}=M_0^{-1} DM_0$.
This is the only non-trivial surviving part of the kinetic term and becomes explicitly
\begin{multline}  
\label{eq:JJ}
\frac{1}{2} \eta_{AB} J^A J^B = 2 \eta^{AB} P_{\varphi,A} P_{\varphi,B} -4  f_{AB}{}^{C} \eta^{AD} P_{\varphi,D}  \partial_C A^B + \partial_A A^B \partial_B A^A+ (\partial_A A^A)^2 + M_0^{AB} B_A B_B \\+ 4 B_A \eta^{AB} P_{\varphi,B}  - 2 M_0^{AB} f_{AC}{}^D \partial_{D} A^C B_{B} + M_0^{AB} f_{AC}{}^D \partial_D A^C f_{BE}{}^F \partial_F A^E \ , 
\end{multline}
where we can already anticipate how several of the terms in~\eqref{eq:pot2E8a} above simplify.

With the metric  \eqref{eq:met3DStat}, one computes that
\begin{align} 
g^{-1} \partial_A g &= 2\rho^{-1}( 2\rho  \partial_A \sigma + \partial_A \rho ) \ , \CR
g^{-1} \partial_A g g^{-1} \partial_B g+  \partial_A g^{\mu\nu}   \partial_B g_{\mu\nu} &= 8 \rho^{-1} \partial_{(A} \sigma ( \rho \partial_{B)} \sigma + 2 \partial_{B)} \rho )\ , 
\end{align}
and dropping the dependence in $t$ and $x$ one obtains for the improved Ricci scalar that 
\begin{align}
e^{2\sigma} \rho \hat{R}- d( \varepsilon_{abc} e^a \wedge \omega^{bc}) +\partial_A ( A^A e^a \wedge \omega^{bd})  = - 2 \rho^{-1} e^{2\sigma} \bigl( D\sigma - \partial_A A^A \bigr)^2  \ . 
\end{align}
The two total derivatives were introduced to write the Einstein--Hilbert Lagrangian in terms of the generalised anholonomies. Using integration by part one obtains from this
\begin{align} 
\label{eq:Rhat}
&\quad e^{2\sigma} \Bigl( 2 D A^A  (  \partial_A \rho^{-1} +2 \rho^{-1} \partial_A \sigma) +2 ( - D \rho^{-1} + 2 \rho^{-1} D\sigma)  \partial_A A^A \Bigr) \CR
&=  2 \rho^{-1} e^{2\sigma} \Bigl( 4 D \sigma \partial_A A^A + \partial_A A^B \partial_B A^A -  ( \partial_A A^A)^2 \Bigr) \CR
& \quad  + 2 \partial_A \Bigl( \rho^{-1} e^{2\sigma} ( D A^A +A^A \partial_B A^B ) \Bigr) - 2 \partial_\varphi  \Bigl( \rho^{-1} e^{2\sigma}  \partial_A A^A \Bigr)\ .   
\end{align}

Using~\eqref{eq:JJ} and~\eqref{eq:Rhat} to rewrite~\eqref{eq:pot2E8a} one obtains that
\begin{multline} 
\label{eq:pot2E8b}
 -e^{-2\sigma} \pot =  \rho \hat{R} - \frac14 \rho^{-1}  \eta_{AB} J^A J^B  +\frac12 \partial_A \mathcal{B}^A + \frac12\partial_\varphi  \mathcal{B}^0 \\
-\rho \Bigl( \frac14 M_0^{CD} \eta_{AB}  {J}_{C,}{}^{A}  {J}_{D,}{}^{B} -\frac12M_0^{EF}  f_{BE}{}^C   f_{AF}{}^D  J_{D,}{}^{B}J_{C,}{}^{A} +  \frac12J_{A,}{}^{B}J_{B,}{}^{A} \Bigr)   \\
+\frac12M_0^{AB} f_{AD}{}^C J_{C,}{}^{D}g^{-1} \partial_B g+\frac{1}{4} M_0^{AB} \bigl( g^{-1} \partial_A g g^{-1} \partial_B g+  \partial_A g^{\mu\nu}   \partial_B g_{\mu\nu} \bigr)    \\
 - \frac12 \rho^{-1} \bigl( \hat{\chi}_\varphi+ \rho^{-1} (  \partial_A Y^A_2 + \tfrac12 f_{AB}{}^C A^A \partial_C A^B) \bigr)^2 \ .
\end{multline} 
Here,  $\mathcal{B}^A$ and $\mathcal{B}^0$ are boundary terms introduced by the partial integrations. The three first lines of~\eqref{eq:pot2E8b} reproduce indeed the Lagrangian~\eqref{eq:E8act}, when neglecting the dependence in the two-dimensional external coordinates $t$ and $x$, such that the topological term does not contribute. In particular, the second and third line correspond to the potential of $E_8$ exceptional field theory written in terms of currents and is structurally the same as~\eqref{eq:potterm0}. The last line exhibits that $\hat{\chi}_\varphi$ is an auxiliary field that can be integrated out without affecting the other fields, and after its elimination the Lagrangian does not depend on the higher level potential $Y^A_2$.
The fields $\langle \chi|$ is eventually fixed to 
\begin{align}
\langle \chi | = \rho^{-2} \langle 0 | \Bigl( - \partial_A Y^A_2 - \tfrac12 f_{AB}{}^C A^A \partial_C A^B  + B_A A^A + B_A T_1^A \Bigr) + 2 \omega_1^\alpha(\cV^{-1}) \langle P_\alpha | \ . 
\end{align}
This removes all dependence on the unwanted fields so that we obtain a perfect match with all terms of $E_8$ exceptional field theory that can be reproduced from the $E_9$ potential constructed in this paper.

It was shown in~\cite{Hohm:2013pua,Blair:2013gqa}, and more generally in~\cite{Bossard:2015foa}, that the only two inequivalent maximal hyperplane solutions to the section constraint of $E_n$ exceptional field theories for $n\leq 8$, correspond to eleven-dimensional supergravity and type IIB supergravity. For $E_8$ exceptional field theory, a partial dictionary with eleven-dimensional supergravity was provided explicitly in~\cite{Godazgar:2013rja,Hohm:2014fxa}. The above results then imply that, after appropriately solving the section constraint, our $E_9$ exceptional field theory potential reduces to the eleven-dimensional or type IIB supergravity Lagrangians for field configurations that do not depend on the two external coordinates $t$ and $x$.

\section{Conclusions}
\label{sec:conc}

In this paper, we have constructed the potential of $E_9$ exceptional field theory as an invariant under $E_9$ generalised diffeomorphisms. This potential  is constructed out of (at most) two internal derivatives acting on the scalar fields and is the first example of such an invariant potential
for an infinite-dimensional duality group and an infinite-dimensional coordinate representation.
The potential consists of four terms, separately invariant under rigid $\hat E_8 \rtimes \mathds{R}_{L_{-1}}$ transformations and transforming homogeneously under $\mathds R^+_\dL$, where \mbox{$E_9 = \hat E_8\rtimes \reals^+_\dL$}. 
Invariance under generalised diffeomorphisms (up to a total derivative) is only achieved by conspiring cancellations among the variations of the different terms. Another key new feature of $E_9$ exceptional field theory is the appearance of a covariantly constrained field $\chi_M$ already in the scalar sector. This constrained scalar field also enters crucially in the potential by forming an indecomposable representation together with the (non-central) components of the $\mathfrak{e}_9$ current. 

Because of the complicated representation theory of $E_9$ and of its extension by $\reals_{L_{-1}}$, which admit indecomposable (but not irreducible) representations,
it is not known whether there are only a finite number of terms invariant under the rigid symmetries of the theory that could in principle contribute to the potential. 
It is therefore difficult to state whether our result could be uniquely determined by requiring invariance under generalised diffeomorphisms.
This is however not necessary for our purpose, as we also require that the dynamics of $D=11$ and type IIB supergravity are reproduced upon solving the section constraint, and have proved that this is the case by mapping our expression to the potential of $E_8$ exceptional field theory.
This is sufficient to guarantee uniqueness of our result.

The result of this paper is the first building block for the full $E_9$ exceptional field theory.
Specifically, it represents the truncation of $E_9$ exceptional field theory to scalar fields and
vanishing external derivatives. The full theory will combine the scalar fields introduced in this paper
with gauge fields $\{A_\mu{}^M, B_{\mu}{}^M{}_N\}$, 
transforming in the representations of the gauge parameters of the
generalised diffeomorphisms \eqref{eq:Lie}. These gauge fields will covariantise external derivatives
but also couple separately via a topological (Wess--Zumino-like) term.
As customary in all even dimensions, the full theory will presumably admit its most compact formulation
in terms of a pseudo-action supplemented by certain first-order duality equations,
in this case for the scalar fields. The latter would define the extension of the linear system
underlying two-dimensional maximal supergravity 
to the full exceptional field theory --- after solving the section constraints thus to
full $D=11$ and type IIB supergravity.
In particular, these equations should provide first order equations for the constrained 
scalar field $\chi_M$, confirming that this is not an additional propagating degree of freedom,
but rather is determined by the physical fields of the theory.
The precise match with two-dimensional supergravity will require the identification of 
the dictionary among the components of  our matrix ${\cal M}^{MN}$ and the
infinite tower of dual scalar potentials encoded in the various
formulations of the linear system~\cite{Belinsky:1971nt,Maison:1978es,Nicolai:1987kz,Julia:1996nu,Bernard:1997et}.

As already discussed in the introduction, an immediate application of $E_9$ exceptional field theory
will be its reduction by means of a generalised Scherk--Schwarz ansatz \cite{Bossard:2017aae}, 
which together with the dictionary to supergravity fields would exhibit the structure of
the yet elusive scalar potential of gauged maximal $D=2$ supergravity~\cite{Samtleben:2007an}
without the need to resort to the fermionic sector and supersymmetry of the theory. A notable aspect of the gauged maximal $D=2$ supergravities studied in~\cite{Samtleben:2007an} is the ubiquity of the gauging of the $L_{-1}$ generator that also featured in our construction and the generalised Lie derivative. The constrained scalar field $\chi_M$ is also indispensable in the generalised Scherk--Schwarz ansatz for such gaugings.

It would also be very desirable to reformulate our potential in terms of the manifestly covariant components of a suitably defined internal Ricci tensor, analogous to the structures identified for the lower-rank groups
\cite{Coimbra:2011ky,Cederwall:2015ica,Baguet:2016jph}.
However, such a formulation would first require the identification of (the unambiguous components of) an appropriate internal $K(E_9)$ spin connection, which at the moment seems a formidable task given the non-reductiveness of $K(E_9)$ and the fact that it does not admit highest weight representations.

Our work can also be considered as a step towards understanding the $E_{11}$ conjecture~\cite{West:2001as,West:2014eza,Tumanov:2016abm} as well as the $E_{10}$ conjecture~\cite{Damour:2002cu}.
The advantage of the group $E_9$ considered here is that it admits an explicit realisation as a vertex operator algebra which allows to define explicitly the full non-linear theory.
One can nevertheless expect that there exists an $E_{11}$ exceptional field theory that would include all the others by considering specific partial solutions of its section constraint. 
The latter does not appear explicitly in the formulation of~\cite{Tumanov:2016abm} but played a crucial role in a different linearised system extending $E_{11}$~\cite{Bossard:2017wxl}.

While we have focussed on the construction of the potential of $E_9$ exceptional field theory, our expressions and proof of invariance are equally valid for any affine Kac--Moody group based on a finite-dimensional simple Lie group $G$, in which case the rigid symmetry group of the potential is $\hat G\rtimes (\reals_\dL^+\ltimes\reals_{L_-1})$, with $\hat G$ the centrally extended loop group over $G$. The expressions for the generalised Lie derivative are entirely analogous as proved in \cite{Bossard:2017aae}.
Our result can then also be seen as the extension of the expressions for extended field theory potentials in \cite{Cederwall:2017fjm} to the case of affine Kac--Moody groups with scalar fields in indecomposable representations.
An especially interesting case is the affine group over $SO(8,n)$ governing two-dimensional half-maximal supergravity.
Extended field theories for the duality groups of half-maximal supergravities in four and three dimensions have been recently formulated \cite{Ciceri:2016hup,Hohm:2017wtr} and capture ten-dimensional heterotic and six-dimensional $(2,0)$ supergravities as solutions of the section constraint. 
The potential constructed in this paper corresponds to ten-dimensional $(1,0)$ and six-dimensional $(2,0)$ supergravities coupled to abelian supermultiplets.
The introduction of gauge interactions for these half-maximal extended field theories requires a deformation of the generalised diffeomorphisms, of the potential and of the full dynamics \cite{Hohm:2011ex,Ciceri:2016hup} which would also be interesting to pursue.
Along these lines,
a further interesting development would be the construction of an `$X$-deformation' of our potential (and later of the full $E_9$ exceptional field theory) that would also reproduce the dynamics of massive type IIA supergravity upon solving the section constraint, in analogy with the higher-dimensional cases \cite{Ciceri:2016dmd}.

\subsection*{Acknowledgements}

We would like to thank Martin Cederwall, Blagoje Oblak and Jakob Palmkvist for useful discussions. This work was partially supported by a PHC PROCOPE, projet No 37667ZL, and by DAAD PPP grant 57316852 (XSUGRA). The work of GB was partially supported by the ANR grant Black-dS-String (ANR-16-CE31-0004). The work of GI was supported by  STFC consolidated grant ST/P000754/1.

\appendix
\section{Properties of the cocycle}
\label{app:cocycle}

\subsection{$E_9$ group 1-cocycles in the co-adjoint}

The group $E_9$ acts on its Lie algebra $\mf{e}_9$ by conjugation and we   aim to extend this group action on the extra $\mf{vir}$ generators $L_m$, $m\neq0$. To this end let 
\begin{align}
X = X_{\sf K} \dK +  X_0 L_0 + \sum_{n\in\ints} X_A^n\, T^A_n = X_\alpha T^\alpha
\end{align}
be an element of $\mf{e}_9$ in the $R(\Lambda_0)_0$ representation, where $\dL=L_0$. As in this section we are only concerned with the adjoint representation of $\mf{e}_9$ and its extension by $\langle L_m\rangle$ for fixed $m\neq0$, we are allowed to ignore the distinction between $\dL$ and $L_0$ throughout our discussion. 
Notice also that compared to \eqref{eq:Xgen}, there is no $L_{-1}$ component here.
The non-trivial commutator between the Virasoro generators and $\mf{e}_9$ is given by
\begin{align}
\label{eq:ad1}
\lb X, L_{m}  \rb &= -m X_0 L_{m} + \sum_{n\in \ints} n X_A^n T_{n +m}^A
= -m X_0 L_{m} + \sum_{n\in \ints} (n-m)  X_A^{n-m} T_n^A\,.
\end{align}
We see that $L_m$ transforms under $\mf{e}_9$ by a rescaling proportional to the derivation component of $X$, plus extra elements in the loop algebra $\hat{\mf{e}}_8$.
The same happens for a finite transformation $g\in E_9$, where we define
\begin{align}
\label{eq:E9L1}
g^{-1} L_{m} g 
&\equiv  \rho(g)^{-2m} L_{m} - \omega^\alpha_{-m} (g) \,\eta_{\alpha\beta}T^\beta\,,
\end{align}
where $\rho(g)^2$ is the component of $g$ along the one-dimensional subgroup generated by the derivation:
\begin{equation}
\label{eq:rhogdef}
g \equiv \rho(g)^{-2L_0}\hat g\,,\quad \hat g\in \hat E_8\,.
\end{equation}
By construction and for fixed $m$, the algebra $\mf{e}_9\oleft \langle L_{m}\rangle$ with commutation relations  \eqref{eq:ad1} defines a representation of $E_9$ under the adjoint action. For $X\in\mf{e}_9$ and $e_{m}\in\reals$, no sum over $m$, one gets 
\begin{align}
\label{gOnExtension}
g^{-1}( X - e_{m} L_{m} ) g = \left(g^{-1}X\,g + e_{m} \eta_{\alpha\beta} \omega^\alpha_{-m}(g) T^\beta \right) - \rho(g)^{-2m} e_{m} L_{m}\,,
\end{align}
such that there is a non-trivial effect of the $L_{m}$ component on the $\mf{e}_9$ component $X$, {\it i.e.}, the representation matrices are block triangular and the representation is indecomposable. One can understand this representation to be built out of two $E_9$  representations, the adjoint representation $\mf{e}_9$ and the one-dimensional representation $\rho(g)^{-2m}$ mentioned above~\eqref{eq:Vdec1}, linked together by the non-trivial map from $E_9$ to $\mf{e}_9$ defined by $g\mapsto \eta_{\alpha\beta} \omega^\alpha_{-m}(g) T^\beta$. The $\omega^\alpha_{-m}$ are the components of a map from $E_9$ to the co-adjoint representation $\mf{e}_9^*$ that one calls a group 1-cocycle. The $\omega^\alpha_{-m}$ determine in this sense the extension of the adjoint $E_9$ representation $\mf{e}_9$ to the indecomposable $E_9$ representation $\mf{e}_9\oleft \langle L_{m}\rangle$. For the above formula \eqref{gOnExtension} to define an action of $E_9$, {\it i.e.}, acting twice being compatible with the group multiplication, the map $\omega^\alpha_{-m}$ must satisfy the 1-cocyle condition\footnote{In the standard mathematical definition it is $\rho^{2m}(g) \omega^\alpha_{-m}(g)$ that defines an $E_9$ 1-cocycle in the co-adjoint representation.}
 \begin{align}
\label{eq:cp}
\omega^{\alpha}_{-m} (g_1 g_2) = R(g^{-1}_2)^\alpha{}_\beta\, \omega^\beta_{-m} (g_1) + \rho(g_1)^{- 2m} \omega^\alpha_{-m} (g_2)\,,
\end{align}
for any $g_1, g_2\in E_9$ and where
\begin{align}
\label{eq:Rdef}
g^{-1} T^\alpha g = R(g)^\alpha{}_\beta T^{\beta}
\end{align} 
defines representation matrices $R(g)^\alpha{}_\beta$ of the adjoint $E_9$ action: $R(g_1)^\alpha{}_\gamma R(g_2)^\gamma{}_\beta = R(g_1g_2)^\alpha{}_\beta$. Note also that the invariance of $\eta^{\alpha\beta}$ on $\mf{e}_9$ implies $\eta_{\beta\gamma} \,R(g^{-1})^\gamma{}_\delta \eta^{\delta\alpha} = R(g)^\alpha{}_\beta$. 
We will discuss the proof of \eqref{eq:cp} momentarily. If the 1-cocycle was trivial, {\it i.e.~}if there existed a co-adjoint vector $v^\alpha$ such that 
\be \omega^\alpha_{-m}(g) \overset{?}{=} R^\alpha{}_\beta(g^{-1}) v^\beta -  \rho^{-2m}(g) v^\alpha \ , \ee
then the representation $\mf{e}_9\oleft \langle L_{m}\rangle$ would decompose into the direct sum of $\mf{e}_9$ and the one-dimensional representation $\rho(g)^{-2m}$, but $\omega^\alpha_{-m}$ is a non-trivial cocycle for all $m$ and $\mf{e}_9\oleft \langle L_{m}\rangle$ is indecomposable.

The dual of the extended representation can be constructed in the usual way. Denoting the basis dual to $\eta_{\alpha\beta}T^\beta$ and $L_{m}$ by $\Lambda^\alpha$ and $\Lambda^{m}$ one finds for the action of $E_9$ that (no sum over $m$)
\begin{align}
\label{eq:dualext}
g^{-1} \left( j_\alpha \Lambda^\alpha + \chi_{m} \Lambda^{m}\right) g = j_\alpha R(g)^\alpha{}_\beta \Lambda^\beta +\left( \rho(g)^{2m}\chi_{m} - j_\alpha \,\rho(g)^{2m}R(g)^\alpha{}_\beta\, \omega_{-m}^\beta(g)\right) \Lambda^m\,.
\end{align}
from which one can read off the transformation of the coefficients $j_\alpha$ and $\chi_{m}$.
Using \eqref{eq:cp} we find 
\begin{align}
\omega^\alpha_{-m}(g) = - \rho(g)^{-2m} R(g^{-1})^\alpha{}_\beta \omega^\beta_{-m}(g^{-1})
\end{align} 
so that~\eqref{eq:dualext} can be rewritten succinctly as follows
\begin{align}
\label{eq:coadext}
(j_\alpha,\, \chi_m ) \to \left( j_\beta R(g)^\beta{}_\alpha \,,\ \rho(g)^{2m} \chi_{m} + \omega^\alpha_{-m}(g^{-1}) j_\alpha\right)\,.
\end{align}
Setting $m=-1$, this is the transformation we have for the currents and the field $\chi$ at $\tilde\rho=0$ \eqref{eq:chiglotrans}.

An important observation is that $\omega^\alpha_{-m}(g)$ can be defined in terms of the shift operators \eqref{eq:Sm}.
First, we notice that the bilinear forms \eqref{eq:etam} transform under $E_9$ only by a rescaling
\begin{equation}
\eta_{m\,\alpha\beta}\,g^{-1}T^\alpha g\,\otimes\, g^{-1}T^\beta g =
\rho(g)^{-2m} \eta_{m\,\alpha\beta}T^\alpha\otimes T^\beta \,.
\end{equation}
Then, using \eqref{eq:etashiftid} for $n=0$ and conjugating by $g$, we obtain for $m\neq 0$,
\begin{equation}
\label{eq:coad}
\omega^\alpha_{-m} (g) \dK \equiv \rho(g)^{-2m} g\,  \cS_{m}(T^\alpha) g^{-1} - \cS_{m} ( g \,T^\alpha g^{-1} ) \,,
\end{equation}
which is therefore equivalent to \eqref{eq:E9L1} and allows to straightforwardly prove the cocycle condition \eqref{eq:cp}.
Crucially, \eqref{eq:coad} holds for any $T^\alpha \in \mf{f}$ (with $g\in E_9$)
and for this reason we will take it as our definition of $\omega^\alpha_{-m} (g)$.
The cocycle condition \eqref{eq:cp} is still satisfied by this more general definition, where \eqref{eq:Rdef} also generalises to any $T^\alpha \in \mathfrak f$ and  $\omega^\alpha_{-m} (g)$ is an $E_9$ group 1-cocyle in the conjugate representation $\mf{f}^*$.

For the computation of Section \ref{sec:pot}, it is useful to present the expansion of the loop components of the cocycle. Up to linear order in the components of $X\in\mathfrak{e}_9$ of $g=e^X$, we find using \eqref{eq:coad} that\footnote{One computes that $[X,\cS_m(T_n^A)] +m X_0 \cS_m(T_n^A)- \cS_m([X,T_n^A]) = -(m+n) \eta^{AB} X_B^{-m-n}$, while $[X,\cS_m(L_n)] +m X_0 \cS_m(L_n)- \cS_m([X,L_n]) = 0$.}
\begin{equation}
\label{eq:cocexp}
\omega_{-m\, n}^{\hspace{4.9mm} A}(g)= -(m+n)\eta^{AB}\,X_B^{-m-n}+\mathcal{O}(X^2) \ . 
\end{equation}
The expansion of the $L_n$ components of the cocycle $\omega_{-m\, n}(g)=\mathcal{O}(X^2)$ only starts at quadratic order in $X$ while the component along $\dK$ simply vanishes. Another particularly important expression will be the cocycle associated with the Hermitian coset representative $\cM$ at $\tilde\rho=0$. A convenient expression equivalent to \eqref{eq:coad} is
\begin{equation}
\label{eq:omegaM}
\omega_m^\alpha(\cM)\dK = \rho^{2m}\cM^{-1}\cS_m\left(T^{\alpha\,\dagger}\right)\cM - \cS_m\left(\cM^{-1}T^{\alpha\,\dagger}\cM \right)\,\quad\text{when }\tilde\rho=0\,.
\end{equation}

Finally, it is instructive to write explicit expressions for $\omega^\alpha_{-m} (g)$ for $g=e^X$.
Here we restrict to $m=\pm1$ only.
Going back to \eqref{eq:ad1}, it is useful to rewrite the 
second term as
\begin{align}
\label{eq:rew}
-\sum_{n\in \ints} (n\pm 1)  X_A^{n\pm 1} T_n^A =  \left\langle \lb L_0, X\rb , T^\alpha\ \right\rangle_{\pm 1} \eta_{\alpha\beta}T^\beta \,,
\end{align}
where the sum on $\alpha$ runs over all $E_9$ generators with indices raised and lowered with the standard $\eta_{\alpha\beta}=\eta_{0\, \alpha\beta}$ of~\eqref{eq:etam} and the shifted form on the loop generators in $\mf{e}_9$ is defined as
\begin{align}
\left\langle T^A_m , T^B_n \right\rangle_{\pm 1} = \eta^{AB} \delta_{m,-n\pm 1}
\end{align}
and agrees with the restriction to the loop part of the bilinear form $\eta_{\pm 1\, \alpha\beta}$.  To illustrate~\eqref{eq:rew} further, we write out the right-hand side explicitly
\begin{align}
- \left\langle \lb L_0, X\rb , T^\alpha \right\rangle_{\pm 1} \eta_{\alpha\beta}T^\beta  
&= \left\langle \lb L_0, X\rb , \dK \right\rangle_{\pm 1} \! L_0 +\left\langle \lb L_0, X\rb , L_0 \right\rangle_{\pm 1} \! \dK- \sum_{n\in\ints} \left\langle \lb L_0, X\rb , T^A_{-n} \right\rangle_{\pm 1} \! T^B_n \eta_{AB} \nn\\
&= \sum_{m,n\in\ints} m \,X^m_C \left\langle  T^C_m , T^A_{-n} \right\rangle_{\pm 1} \! T^B_n \eta_{AB} \nn\\
&= \sum_{n\in \ints} (n\pm 1) X^{n\pm 1}_A T^A_n\,,
\end{align}
where we have used that $\eta_{\pm 1}$ does not pair $\dK$ and $L_0$ non-trivially with anything in $\mf{e}_9$.

Using~\eqref{eq:rew}, we can thus restate~\eqref{eq:ad1} as
\begin{align}
\ad_X L_{\pm 1} = \mp X_0 L_{\pm 1} -  \left\langle \lb L_0, X\rb , T^\alpha \right\rangle_{\mp 1} \eta_{\alpha\beta}T^\beta 
\end{align}
and we have also introduced the notation $\ad_X L_{\pm 1} =\lb X, L_{\pm 1}\rb$ for the commutator between $\mf{e}_9$ and $L_{\pm 1}$. This is the action we aim to exponentiate.

By induction one can show for any $k\geq 0$ that
\begin{align}
\ad_X^k L_{\pm 1} = (\mp X_0)^k L_{\pm 1} - \biggl\langle \lb L_0, X\rb , \sum_{\ell=0}^{k-1} (\mp X_0)^\ell (-\ad_X)^{k-1-\ell} \,T^\alpha \biggr\rangle_{\mp 1} \eta_{\alpha\beta}T^\beta \,,
\end{align}
which can be exponentiated to $g=e^X$ easily since $X_0$ is central in the representation. This leads to 
\begin{align}
g^{-1} L_{\pm 1} g &= e^{\pm X_0} L_{\pm 1} -\left\langle \lb L_0, X \rb , \frac{e^{\ad_X} - e^{\pm X_0}}{-\ad_X \pm X_0} T^\alpha \right\rangle_{\mp 1} \eta_{\alpha\beta}T^\beta \,,
\end{align}
which compared to \eqref{eq:E9L1} gives
\begin{subequations}
\begin{align}
\rho(g)&=\,e^{-\frac{X_0}{2}}\,, \label{eq:rhoX}\\
\omega^\alpha_\pm (g) &= \left\langle \lb L_0, X \rb , \frac{e^{\ad_X} - e^{\mp X_0}}{-\ad_X \mp X_0} T^\alpha\right\rangle_{\pm 1}\,,\label{eq:co9}
\end{align}
\end{subequations}
This explicit expression of the cocycle is only valid for $T^\alpha\in\mathfrak{e}_9$.

\subsection{Generalisation to $\hat{E}_8\rtimes SL(2)$}
\label{sec:coSL2}

In the previous discussion $g$ was an element of $E_9$. When the axion $\tilde\rho\neq0$, $\cM$ is no longer an element of $E_9$ and instead belongs to the group $\hat{E}_8\rtimes SL(2)$. We thus require an extension of some of the formulas above to this case. 
We shall now give a generalisation of \eqref{eq:omegaM} that can be expressed as an infinite power series in $\tilde\rho$ and reduces to the previous formula when $\tilde\rho=0$.

As in~\eqref{eq:cM_co} we shall decompose the Hermitian $\cM$ as
\begin{align}
\cM = m \hat{g}_\cM = \hat{g}_\cM^\dagger m
\end{align}
with Hermitian $m\in SL(2)$ and non-Hermitian $\hat{g}_\cM$ in $\hat{E}_8$. The generalisation of the cocycle that has the needed properties is
\begin{align}
\label{eq:coc}
\Omega^\alpha(\cM) \dK= \hat{g}_\cM^{-1} \cS_{+1} \left( m^{-1} T^{\alpha\,\dagger} m \right) \hat{g}_\cM - \cS_{+1} \left(\cM^{-1} T^{\alpha\,\dagger} \cM\right)\,.
\end{align}
Compared with~\eqref{eq:omegaM}, we see that we have made a choice in the split between $m$ and $\hat{g}_\cM$ and that the character factor $\rho$ does not appear explicitly anymore. This is natural since this part is contained in the action of $m\in SL(2)$. We have also defined this only for $\cS_{+1}$.
Notice that because of the presence of $m$ inside the first shift operator, this expression does not satisfy \eqref{eq:cp} and is therefore not a group cocycle.
However, when $\tilde\rho=0$ we have $m\in\mathds R^+_\dL$ and $\Omega^\alpha(\cM)$ reduces to $\omega_1^\alpha(\cM)$.

Since the $T^\alpha$ that belong to $\hat{\mf{e}}_8$ can be represented by $\mf{e}_8$ elements that depend on a spectral parameter $w$ via $T_m^A  = w^m T^A $, leading to meromorphic functions of $w$, and $SL(2)$ acts on these generators by M\"{o}bius transformations of $w$, it is convenient to work out the conjugation by $m$ in this picture. More explicitly, the form of $SL(2)$ generators as differential operators in $w$ is
\begin{align}
L_{+1} = - w^2\partial_w \,,\quad L_0 = - w \partial_w \,,\quad L_{-1} = - \partial_w\,,
\end{align}
such that their exponentiated action on any function $f(w)$ is given by M\"obius transformations leading to\footnote{The exponentiation of the individual transformations is
\begin{equation*}
e^{\tilde\rho L_{+1}} f(w) e^{-\tilde\rho L_{+1}} = f\left(\frac{w}{1+\tilde\rho w}\right)\,,\quad
\rho^{2L_0} f(w) \rho^{-2L_0} = f(\rho^{-2} w)\,,\quad
e^{\tilde\rho L_{-1}} f(w) e^{-\tilde\rho L_{-1}} = f(w-\tilde\rho)\,.
\end{equation*}}
\begin{align}
m^{-1} f(w) m = f\left( \frac{\rho^{-2} (w-\tilde\rho)}{1+\tilde\rho \rho^{-2}(w-\tilde\rho)}\right)\,,
\end{align}
which when combined with the shift operator $\cS_{+1}$ that multiplies by $w$ leads to
\begin{align}
m\, \cS_{+1} \left( m^{-1} f(w) m \right) m^{-1} = \left(\rho^2 \frac{w}{1-\tilde\rho w} + \tilde\rho\right) f(w)
&= \rho^2\sum_{k\geq 0}  \tilde\rho^k w^{k+1} f(w) +\tilde\rho f(w)\nn\\
&= \rho^2 \sum_{k \geq 0} \tilde\rho^k \cS_{1+k}( f(w) )+\tilde\rho f(w)\,.
\end{align}
Inserting this into~\eqref{eq:coc} leads to 
\begin{align}
\label{eq:w(M)}
\Omega^\alpha(\cM)\dK = \cM^{-1} \Big( \rho^2 \sum_{k\geq 0} \tilde\rho^k \cS_{1+k}( T^{\alpha\,\dagger}) + \tilde\rho \,T^{\alpha\,\dagger}\Big) \cM - \cS_{+1} \left(\cM^{-1} T^{\alpha\,\dagger}\cM \right) -\tilde\rho \,\delta^\alpha_\dK \dK\,,
\end{align}
where the last term is due to the fact that $\Omega^\dK(\cM) =0$ in~\eqref{eq:coc} but the expansion using the M\"obius transformations above generates a spurious term.

\subsection{Useful identities}

Here we collect some useful identities for the generalisation of the cocycle $\Omega^\alpha(\cM)$ discussed above and the effect of $SL(2)$ conjugations on expressions appearing in the derivation of the potential.

Using the same argument as above based on M\"obius transformations, one works out the conjugation under $L_{-1}$ of the shift operator as
\begin{align}
e^{\tilde \rho L_{-1}} \cS_{-1}( e^{-\tilde \rho L_{-1}} f(w) e^{\tilde \rho L_{-1}}) e^{-\tilde \rho L_{-1}} = \frac{1}{w-\tilde{\rho}} f(w) 
\end{align}
such that
\begin{align}
  e^{\tilde \rho L_{-1}} \cS_{-1}( e^{-\tilde \rho L_{-1}} T^\alpha e^{\tilde \rho L_{-1}}) e^{-\tilde \rho L_{-1}}  = \sum_{k=0}^\infty \tilde{\rho}^k \cS_{-1-k}(T^\alpha) \ .  \label{GeneratingInfinitShift} 
\end{align}
It is sometimes convenient to rewrite the geometric series of shifted generators appearing in $\cJ^-_\alpha$ in this way.

The $\hat{E}_8$ invariant bilinear form $\eta_{n \, \alpha\beta}$ is not invariant under the action of $SL(2)$. To compute the effect of a conjugation with the Virasoro generator $L_q$ for $q= -1,0,1$, it is useful to first show that at the Lie algebra level
\begin{align}
 \eta_{n {\alpha\beta}} [ L_q , T^\alpha ] \otimes T^\beta +  \eta_{n {\alpha\beta}} T^\alpha \otimes [ L_q , T^\beta ]  = (q-n) \eta_{n+q \alpha\beta} T^\alpha \otimes T^\beta \ .  
\end{align}
Using this formula, one computes that for $m\in SL(2)$ one has 
\begin{align} \label{eq:dreamform} 
 \eta_{-1 \alpha\beta} m^{-1} T^{\alpha \dagger} m \otimes T^\beta &=  \Bigl( \frac{1}{\rho^2} \eta_{1 \alpha\beta} - 2 \frac{\tilde{\rho}}{\rho^2} \eta_{ \alpha\beta} +  \frac{\tilde{\rho}^2}{\rho^2} \eta_{-1  \alpha\beta}  \Bigr) T^\alpha \otimes m^{-1} T^{\beta \dagger} m  \\
\eta_{\alpha\beta} m^{-1} T^{\alpha \dagger} m \otimes T^\beta  &=  \Bigl(\frac{\tilde{\rho}}{\rho^2} \eta_{1 \alpha\beta} + \Bigl( 1-2\frac{\tilde{\rho}^2}{\rho^2} \Bigr)  \eta_{ \alpha\beta}  - \tilde{\rho}   \Bigl( 1-\frac{\tilde{\rho}^2}{\rho^2} \Bigr) \eta_{-1  \alpha\beta}   \Bigr) T^\alpha \otimes m^{-1} T^{\beta \dagger} m  \nn \\
\eta_{1 \alpha\beta} m^{-1} T^{\alpha \dagger} m \otimes T^\beta \hspace{-1mm} &= \hspace{-1mm}  \Bigl(  \frac{\tilde{\rho}^2}{\rho^2} \eta_{1 \alpha\beta} + 2\tilde{\rho} \Bigl( 1-\frac{\tilde{\rho}^2}{\rho^2} \Bigr)  \eta_{ \alpha\beta}  + \rho^2  \Bigl( 1-\frac{\tilde{\rho}^2}{\rho^2} \Bigr)^2 \eta_{-1  \alpha\beta}  \Bigr) T^\alpha \otimes m^{-1} T^{\beta \dagger} m \nn  \\
\eta_{2 \alpha\beta} m^{-1} T^{\alpha \dagger} m \otimes T^\beta \hspace{-1mm} &= \hspace{-1mm}  \Bigl( \frac{\tilde{\rho}^3}{\rho^2} \eta_{1 \alpha\beta} + \tilde{\rho}^2 \Bigl( 3-2\frac{\tilde{\rho}^2}{\rho^2} \Bigr)  \eta_{ \alpha\beta}  + z \Bigl( 3\rho^2 - 3 {\tilde{\rho}}^2 + \frac{{\tilde{\rho}}^4}{\rho^2} \Bigr) \eta_{-1  \alpha\beta}  \nn \\
& \hspace{45mm}  + \rho^4 \sum_{k=0}^\infty \tilde{\rho}^k   \eta_{-2-k\, \alpha\beta} \Bigr) T^\alpha \otimes m^{-1} T^{\beta \dagger} m  \  . \nn
 \end{align}
Note in particular that 
\begin{align}
 \bigl( \eta_{\alpha\beta} - \tilde{\rho}\,  \eta_{-1 \alpha\beta} \bigr) m^{-1} T^{\alpha \dagger} m \otimes T^\beta = \bigl( \eta_{\alpha\beta} - \tilde{\rho} \, \eta_{-1 \alpha\beta} \bigr) T^\alpha \otimes m^{-1} T^{\beta \dagger} m \ . \label{InvarBil}  
\end{align}

These formulas can be used to obtain some properties of $\Omega^\alpha(\cM)$. One computes using the definition \eqref{eq:coc} that
\bea  \eta_{\alpha\beta} \Omega^\alpha(\cM) \dK \otimes T^\beta &=&\eta_{\alpha\beta}  \Bigl(  \hat{g}_\cM^{-1} \mathcal{S}_1( m^{-1} T^{\alpha \dagger} m) \hat{g}_\cM - \mathcal{S}_1( \hat{g}_\cM^{-1}  m^{-1} T^{\alpha \dagger} m \hat{g}_\cM) \Bigr) \otimes T^\beta  \\
&=& \Bigl( \frac{\tilde{\rho}}{\rho^2} \eta_{1 \alpha\beta} + \Bigl( 1-2\frac{\tilde{\rho}^2}{\rho^2} \Bigr)  \eta_{ \alpha\beta}  - \tilde{\rho}   \Bigl( 1-\frac{\tilde{\rho}^2}{\rho^2} \Bigr) \eta_{-1  \alpha\beta}  \Bigr)  \CR
&& \times \Bigl( \hat{g}_\cM^{-1} \mathcal{S}_1(  T^\alpha ) \hat{g}_\cM  \otimes m^{-1} T^{\beta \dagger} m - \mathcal{S}_1(  T^\alpha )\otimes  \hat{g}_\cM^{-1} m^{-1} T^{\beta \dagger} m \hat{g}_\cM  \Bigr) \CR
&=& \dK \otimes \Bigl( \tilde{\rho} ( L_0 - \hat{g}_\cM^{-1} L_0 \hat{g}_\cM ) + \rho^2 \sum_{k=0}^\infty \tilde{\rho}^k (L_{-1-k} - \hat{g}_\cM^{-1} L_{-1-k} \hat{g}_\cM)\Bigr) \; ,  \nn  \eea
where we used \eqref{eq:dreamform} in the first step and \eqref{eq:etashiftid} in the second. One therefore obtains  the useful identity
\be \label{eq:fid1}  \eta_{\alpha\beta} \Omega^\alpha(\cM)  T^\beta  = \tilde{\rho} ( L_0 - \hat{g}_\cM^{-1} L_0 \hat{g}_\cM ) + \rho^2 \sum_{k=0}^\infty \tilde{\rho}^k (L_{-1-k} - \hat{g}_\cM^{-1} L_{-1-k} \hat{g}_\cM)  \ . \ee

 We want also to compute $\eta_{\alpha\beta} H(\cM)^{\alpha}{}_\gamma \Omega^\gamma(\cM) T^\beta$ with 
 $H(\cM)^\alpha{}_\beta$  defined in \eqref{eq:Hdef}. Using instead formula \eqref{eq:w(M)}, one computes that
 \bea  && \eta_{\alpha\beta} H(\cM)^{\alpha}{}_\gamma \Omega^\gamma(\cM) \dK \otimes T^\beta \CR
 &=& \eta_{\alpha\beta} \Bigl( \cM^{-1} \rho^2  \sum_{k=0}^\infty \tilde{\rho}^k \mathcal{S}_{-1-k}(\cM^{-1} T^{\alpha \dagger} \cM)^\dagger \cM+ \tilde \rho T^\alpha - \mathcal{S}_1(T^\alpha) \Bigr) \otimes T^\beta \CR
 && \hspace{20mm}+ \tilde \rho \, \dK \otimes \Bigl(\frac{\tilde{\rho}}{\rho^2} \cM^{-1} L_{-1} \cM + \Bigl( 1-2\frac{\tilde{\rho}^2}{\rho^2} \Bigr) \cM^{-1} L_0 \cM   - \tilde{\rho}   \Bigl( 1-\frac{\tilde{\rho}^2}{\rho^2}  \Bigr) \cM^{-1} L_1 \cM  \Bigr) \CR
 &=& \dK \otimes \bigl( \hat{g}_\cM^{-1} L_1 \hat{g}_\cM - L_1 \bigr)  \eea
 so one obtains the very simple result that 
 \be  \label{eq:fid2} \eta_{\alpha\beta} H(\cM)^{\alpha}{}_\gamma \Omega^\gamma(\cM) T^\beta = \hat{g}_\cM^{-1} L_1 \hat{g}_\cM - L_1   \ . \ee
Using this formula and reinserting the $m$ matrix using \eqref{eq:dreamform} one obtains that 
\be L_1 + \eta_{\alpha\beta} H(\cM)^{\alpha}{}_\gamma \Omega^\gamma(\cM) T^\beta  = \cM^{-1} \Bigl( \rho^2 \Bigl( 1-\frac{\tilde{\rho}^2}{\rho^2}\Bigr)^2 L_1 + 2 \tilde{\rho} \Bigl( 1-\frac{\tilde{\rho}^2}{\rho^2} \Bigr) L_0 + \frac{\tilde{\rho}^2}{\rho^2} L_{-1} \Bigr) \cM \ .  \ee
Using again \eqref{eq:dreamform} one computes that 
\be \eta_{\alpha\beta} H(\cM)^\alpha{}_\dK T^\beta =  \cM^{-1} \Bigl( \tilde\rho \Bigl( 1-\frac{\tilde{\rho}^2}{\rho^2}\Bigr) L_1 - \Bigl( 1-2\frac{\tilde{\rho}^2}{\rho^2} \Bigr) L_0 - \frac{\tilde{\rho}}{\rho^2} L_{-1} \Bigr) \cM \; , \ee
such that 
\be L_1 + \eta_{\alpha\beta} H(\cM)^{\alpha}{}_\gamma ( \Omega^\gamma(\cM) + \tilde\rho \delta^\gamma_\dK) T^\beta =\cM^{-1} \Bigl(  (\rho^2 - \tilde{\rho}^2 ) L_1 + \tilde \rho L_0 \Bigr) \cM \; . \label{RewriteDeltachi}  \ee

By a similar reasoning, one also obtains 
\begin{equation}
\eta_{-1\,\alpha\beta}\,\Omega^\alpha(\cM)\,T^\beta=\,\tilde\rho\,L_{-1}+\rho^2\sum\limits_{k=0}^\infty\tilde\rho^k L_{-2-k}-\hat g_{\cM}^{-1}\Big(\tilde\rho\,L_{-1}+\rho^2\sum\limits_{k=0}^\infty\tilde\rho^k L_{-2-k}\Big)\hat g_\cM\,,\label{eq:Sid1}
\end{equation}
and
\begin{equation}
\eta_{-1\,\alpha\beta}\,\Omega^\gamma(\cM)\,H(\cM)^\alpha{}_\gamma\,T^\beta=\,\hat g_\cM^{-1}\,L_0\,\hat g_\cM-L_0\, \label{eq:Sid2}\,.
\end{equation}

\section{Gauge algebra closure on $\bra{\chi}$}
\label{sec:closure}

The closure of the algebra of generalised diffeomorphisms on the $\hat{E}_8 \rtimes ( \mathds{R}^+_\dL \ltimes \mathds{R}_{L_{-1}})$ scalars $\cM$ follows from its closure on a vector field $|V\rangle$ \eqref{eq:Lie},  which was derived in \cite{Bossard:2017aae}. The field $\langle \chi |$ does not transform under generalised diffeomorphisms simply as a generalised Lie derivative, and the mixing with the $\hat{\mf{e}}_8 \oleft \mf{sl}(2)$ current makes it non-obvious that the algebra closes on this field. In this appendix we show that this is indeed the case. 

Because the algebra closes on $\cM$ and therefore on $\langle \cJ_\alpha |$, one can check the closure of the algebra on any linear combination of $\langle \chi |$ and $\langle \cJ_\alpha |$. Since the transformation of $\langle \chi |$ is not manifestly covariant, it is indeed convenient to check the closure of the algebra on the combination 
\begin{align}
 \langle \xi | \equiv \langle \chi | + \sum_{k=0}^\infty \tilde{\rho}^k \langle \cJ_\alpha  | \cS_{-1-k}( T^\alpha) + \frac{\tilde{\rho}}{\rho^2} \bigl( \langle \cJ_\alpha | T^\alpha  + \rho^{-1}  \langle \partial \rho  | \bigr)   \ ,  
\end{align}
that transforms as
\begin{multline}\delta \langle \xi | = \langle \partial_{\xi} | \Lambda \rangle \langle \xi | + \eta_{\alpha\beta} \langle \partial_\Lambda | T^\alpha | \Lambda \rangle \langle \xi| T^\beta + \frac{1}{\rho^2} \eta_{1\alpha\beta} \langle \partial_\Lambda | T^\alpha | \Lambda \rangle \langle \partial_\Lambda | \cM^{-1} T^{\beta \dagger} \cM \\
+ \eta_{-1\alpha\beta} \tr ( T^\alpha \Sigma ) \langle \xi | T^\beta +  \frac{1}{\rho^2} \eta_{\alpha\beta} \tr( T^\alpha \Sigma)  \langle \partial_\Sigma | \cM^{-1} T^{\beta \dagger} \cM  \\ - \frac{1}{\rho^2} \langle \partial_\Sigma | \Sigma + \frac{1}{\rho^2} \tr(\Sigma) \bigl( \langle \cJ_\alpha | T^\alpha  + \rho^{-1}  \langle \partial \rho  |  \bigr)  \ . \end{multline} 
To compute the closure it is convenient to use the BRST formalism, for which $\Lambda$ and $\Sigma$ are understood as anticommuting ghost fields, with their own variation defined according to \eqref{ClosureGD} as
\begin{align}
 \label{BRST} \delta | \Lambda\rangle &=\frac{1}{2} \Bigl( \langle \partial_{\Lambda_1} | \Lambda_2 \rangle | \Lambda_1\rangle -\eta_{\alpha\beta} \langle \partial_{\Lambda_2} | T^\alpha | \Lambda_2 \rangle T^\beta | \Lambda_1 \rangle - \langle \partial_{\Lambda_2} | \Lambda_2 \rangle | \Lambda_1 \rangle \Bigr) \; , \\
\delta \Sigma &= \langle \partial_\Sigma | \Lambda \rangle  \Sigma - \eta_{\alpha\beta} \langle \partial_\Lambda | T^\alpha | \Lambda \rangle T^\beta \Sigma - \Sigma | \Lambda \rangle \langle \partial_\Lambda | \CR
&\quad +\frac14  \eta_{1\, \alpha\beta}  \Bigl(  \langle\partial_{\Lambda_2} | T^\alpha | \Lambda_2 \rangle T^\beta | \Lambda_1 \rangle +  \langle\partial_{\Lambda_2} | T^\alpha  | \Lambda_1 \rangle T^\beta | \Lambda_2 \rangle  \Bigr)  \langle \partial_{\Lambda_2} | - \frac12 \eta_{-1 \alpha \beta} \tr ( T^\alpha \Sigma)  T^\beta \Sigma  \  ,  \nonumber 
\end{align}
for which the labels on $| \Lambda_1\rangle$ and $|\Lambda_2\rangle$ only indicate on which $|\Lambda\rangle$ the derivative acts, despite the fact that they are the same anticommuting ghost $|\Lambda\rangle$. For example, in index notation one has
\begin{align}
 \delta \Lambda^M = \frac{1}{2} \Bigl(\Lambda^N \partial_N \Lambda^M  -\eta_{\alpha\beta} (T^\alpha)^P{}_N \partial_P \Lambda^N  \;  (T^\beta)^M{}_Q \Lambda^Q - \partial_N \Lambda^N \; \Lambda^M  \Bigr) \ . 
\end{align}
In this notation, the closure on the algebra on a vector field $|V\rangle$ is equivalent to the property that $\delta^2 | V \rangle = 0$. Note that $\delta^2 | \Lambda \rangle \ne 0$, but it gives a trivial generalised diffeomorphism, whereas the definition of a truly nilpotent operator requires the introduction of an infinite chain of ghosts for ghosts generating an $L_\infty $ algebra structure \cite{Berman:2012vc,Hohm:2017pnh,Cederwall:2018aab}.  

In the BRST formulation it is easier to check that $\delta^2 \langle   \xi |$ indeed vanishes. Here, we shall only give some of the steps for the terms quadratic in $|\Lambda\rangle$. The parts of the transformations corresponding to the Lie derivative of an ordinary vector field work as in \cite{Bossard:2017aae}, whereas the other give the following contributions 
\begin{align}
 \delta^2 \langle \xi | &= \eta_{1\alpha\beta} \langle \partial_1 | T^\alpha | \Lambda_1 \rangle \langle \partial_1 | \Lambda_2 \rangle \langle \partial_1 | \cM^{-1} T^{\beta \dagger}  \cM \CR
& - \frac14 \eta_{\alpha\beta} \eta_{1\gamma\delta} \Bigl( \langle \partial_2  | [ T^\alpha  , T^\gamma] | \Lambda_1 \rangle  \langle \partial_2 | T^\delta | \Lambda_2 \rangle + \langle \partial_2  | [ T^\alpha  , T^\gamma] | \Lambda_2 \rangle  \langle \partial_2 | T^\delta | \Lambda_1 \rangle  \Bigr) \langle \partial_1 + \partial_2 | \cM^{-1} T^{\beta \dagger}  \cM \CR 
&\quad+ \frac14 \eta_{1\alpha\beta} \Bigl(   \langle \partial_1 | T^\alpha | \Lambda_2 \rangle \langle \partial_2 | T^\beta | \Lambda_1 \rangle + \langle \partial_1 | T^\alpha | \Lambda_1 \rangle \langle \partial_2 | T^\beta | \Lambda_2 \rangle \Bigr) \langle \partial_2 | \CR
&\quad+\eta_{1\alpha\beta}  \langle \partial_1 | T^\alpha | \Lambda_1 \rangle \Bigl( \langle \partial_2 | \Lambda_2 \rangle \langle \partial_1 | + \langle \partial_1 | \Lambda_2 \rangle \langle \partial_2 | \Bigr)  \cM^{-1} T^{\beta \dagger}  \cM \CR 
&\quad +  \eta_{1\alpha\beta}  \eta_{\gamma\delta}    \langle \partial_1 | T^\alpha | \Lambda_1 \rangle  \langle \partial_2 | T^\gamma | \Lambda_2 \rangle \langle \partial_1 | \cM^{-1} [T^\beta , T^\delta ]^\dagger \cM \CR
&\quad +  \eta_{1\alpha\beta}  \eta_{\gamma\delta}    \langle \partial_1 + \partial_2 | [ T^\alpha , T^\gamma] | \Lambda_1 \rangle  \langle \partial_2 | T^\gamma | \Lambda_2 \rangle \langle \partial_1 + \partial_2 | \cM^{-1}   T^{\beta \dagger}  \cM\CR
&\quad  + \frac12 \eta_{1\alpha\beta} \Bigl( \langle \partial_2 | T^\alpha | \Lambda_1 \rangle \langle \partial_1 | \Lambda_2 \rangle -  \langle \partial_1 | T^\alpha | \Lambda_1 \rangle \langle \partial_2 | \Lambda_2 \rangle   \Bigr) \langle \partial_1 + \partial_2 | \cM^{-1}   T^{\beta \dagger}  \cM\CR
&\quad + \frac12 \eta_{1\alpha\beta}  \langle \partial_1  + \partial_2 | T^\alpha | \Lambda_1 \rangle \langle \partial_2-\partial_1  | \Lambda_2 \rangle  \langle \partial_1 + \partial_2 |\CR
&= 0 \  ,  
\end{align}
where $\langle \partial_1  |  =\langle \partial_{\Lambda_1} |$ and $\langle \partial_2  |  =\langle \partial_{\Lambda_2} |$ for short, and  one uses that all the terms involving commutators simplify according to  
\bea && \eta_{1\alpha\beta} \eta_{\gamma\delta} \Bigl( - \langle \partial_2 | [ T^\alpha , T^\gamma ] | \Lambda_1 \rangle \langle \partial_2 | -  \langle \partial_1 | [ T^\alpha , T^\gamma ] | \Lambda_1 \rangle \langle \partial_1 | + 2 \langle \partial_1 + \partial_2 | [ T^\beta, T^\gamma ] | \Lambda_1 \rangle \langle \partial_2 | \Bigr) T^\delta | \Lambda_2 \rangle \CR
&& \hspace{100mm} \times  \langle \partial_1 + \partial_2 |  \cM^{-1}   T^{\beta \dagger}  \cM\CR
&& -4\eta_{1\alpha\beta} \eta_{\gamma\delta}  \langle \partial_1 | [ T^\alpha , T^\gamma ] | \Lambda_1 \rangle \langle \partial_2 | T^\delta | \Lambda_2 \rangle  \langle \partial_1 |  \cM^{-1}   T^{\beta \dagger}  \cM\CR 
&=&\eta_{1\alpha\beta}  \langle \partial_1 | T^\alpha | \Lambda_1 \rangle \langle \partial_2 | T^\beta | \Lambda_2 \rangle \langle \partial_1 - \partial_2 | \ .   \eea

\end{document}